\documentclass[12pt,notitlepage,a4paper]{article}
\pdfoutput=1
\usepackage{color,graphicx}
\usepackage{cite}
\usepackage{amssymb}
\usepackage{delarray,amsmath}
\usepackage[latin1]{inputenc}
\usepackage[american]{babel}
\usepackage[table]{xcolor}
\usepackage{cite}
\usepackage{tikz-cd}

\newcommand{\be}{\begin{equation}}
\newcommand{\ee}{\end{equation}}
\newcommand{\beq}{\begin{equation}}
\newcommand{\eeq}{\end{equation}}
\newcommand{\beqa}{\begin{eqnarray}}
\newcommand{\eeqa}{\end{eqnarray}}

\newcommand{\bear}{\begin{eqnarray}}
\newcommand{\eear}{\end{eqnarray}}
\newcommand{\pd}{\partial}
\numberwithin{equation}{section}

\newfont{\namefont}{cmr10}
\newfont{\addfont}{cmti7 scaled 1440}
\newfont{\boldmathfont}{cmbx10}
\newfont{\headfontb}{cmbx10 scaled 1728}

\begin{document}
\baselineskip=15.5pt
\pagestyle{plain}
\setcounter{page}{1}

\begin{center}
\vspace{0.1in}

\renewcommand{\thefootnote}{\fnsymbol{footnote}}

\begin{center}
\Large \bf    Floquet Scalar  Dynamics in  Global AdS 
\end{center}
\vskip 0.1truein
\begin{center}
\bf{Anxo Biasi${}^1$\footnote{anxo.biasi@gmail.com}, Pablo Carracedo,${}^2$\footnote{pablo.enrique.carracedo.garcia@xunta.gal}
Javier Mas,${}^1$\footnote{javier.mas@usc.es} \\ Daniele Musso${}^1$\footnote{daniele.musso@usc.es} and 
Alexandre Serantes${}^3$\footnote{alexandre.serantes@icts.res.in}}\\
\end{center}
\vspace{0.5mm}

\begin{center}\it{
${}^1$Departamento de  F\'\i sica de Part\'\i  culas \\
Universidade de Santiago de Compostela \\
and \\
Instituto Galego de F\'\i sica de Altas Enerx\'\i as (IGFAE)\\
E-15782 Santiago de Compostela, Spain}
\end{center}

\begin{center}\it{
${}^2$Meteo-Galicia, Santiago de Compostela, Spain}
\end{center}

\begin{center}\it{
${}^3$}International Centre for Theoretical Sciences-TIFR, \\
Survey No. 151, Shivakote, Hesaraghatta Hobli, \\
Bengaluru North, India 560 089
\end{center}

\setcounter{footnote}{0}
\renewcommand{\thefootnote}{\arabic{footnote}}

\vspace{0.4in}

\begin{abstract}
\noindent
We study periodically driven scalar fields and the resulting geometries with global AdS asymptotics.
These solutions describe the strongly coupled dynamics of dual finite-size quantum systems under a periodic driving which we interpret as Floquet condensates.  
They span a continuous two-parameter space that extends the linearized solutions  on AdS.
We map the regions of stability in the solution space. 
In a significant portion of the unstable subspace, two very different endpoints are reached depending upon the sign of the perturbation. 
Collapse into a black hole occurs for one sign. For the opposite sign instead one attains a regular solution with periodic modulation. 
We also construct quenches where the driving frequency and amplitude are continuously varied.
Quasistatic quenches can interpolate between pure AdS and sourced solutions with time periodic vev. 
By suitably choosing the quasistatic path one can obtain boson stars dual to Floquet condensates at zero driving field.
We characterize the adiabaticity of the quenching processes. Besides, we speculate on the possible connections of this framework with time crystals.

\smallskip
\end{abstract}
\end{center}

\newpage

\tableofcontents

\newpage

\section{Introduction and main results}

The physics of periodically driven many-body systems is a fascinating chapter in the study of out-of-equilibrium dynamics \cite{kohn2001periodic,lazarides2014periodic}.
Their behavior can differ substantially from that of their static counterparts. 
Floquet systems have been investigated at criticality with the tools of conformal field theory (CFT) (see \cite{Berdanier:2017kmd} and references therein). 
The naive expectation that they can only increase indefinitely their energy because of the driving can be avoided in some regions of the driving parameter space.
This was confirmed by explicit examples that attain a steady state at finite temperature \cite{khemani2016phase,Berdanier:2017kmd}.
Even though the actual saturation mechanism is not completely understood, the fact that periodic driving can radically alter the stability of equilibrium points is well known from nonlinear dynamical systems, the Kapitza pendulum being a prominent example \cite{Citro2015}.

The AdS/CFT correspondence maps equilibration processes onto the dynamical evolution of a dual gravitational system. 
The holographic dictionary defines the sources of the field theory in terms of asymptotic modes of the bulk fields. 
It is then possible to study a driven holographic system coupled to an external source. 
In the cases of interest here, the source corresponds to the leading asymptotic mode of a bulk scalar field in an asymptotically global AdS spacetime. 

Periodic driving has been scarcely studied in the context of AdS/CFT.
The analysis performed in \cite{rangamani2015driven} comes closest in spirit to the present study (see also  
\cite{Li:2013fhw,Basu:2013vva,Hashimoto:2016ize,auzzi2013periodically,Kinoshita:2017uch}, all in the probe approximation). 
The main difference is that their initial states are already mixed, while here we consider the driving of pure states. 
Their geometrical ansatz is embedded in the Poincar\'e patch, with an arbitrarily small planar horizon to cap the IR. 
They find a monotonous growth of the black hole horizon as a general outcome of the driving. 
Such growth leads to an infinite temperature final state. 
The result is, on the one side, compatible with the already mentioned natural expectation and, on the other side, 
it comes out as a consequence of the ingoing boundary conditions at the horizon. 
The only remaining freedom may be in the rate of growth of the energy density and entropy. 

The authors in \cite{rangamani2015driven} find three different phases depending upon the range of values acquired by the dimensionless combination 
$\chi = \phi_b \omega_b$, namely the product of the amplitude and the frequency of the driving at the boundary.
Small values of $\chi$ correspond to a dissipation dominated regime, where the horizon absorbs all the energy entering the system from the boundary. 
Here the scalar field synchronizes with the driving but does not change its amplitude.  
Raising $\chi$ they find two other phases in which the scalar field stops being ``transparent", and it may absorb part of the energy input and increase its amplitude.

The picture of \cite{auzzi2013periodically,rangamani2015driven} should be compared with the results we find here. 
The fact that our driving does not necessarily imply decoherence (i.e. a collapse into a black hole) is by no means trivial and it only occurs in certain regions of parameter space. 
Charting this region, characterizing its features, and studying its stability is the main target in the present paper.

We have examined both  complex and real massless scalar fields. 
While a priori their dynamics is quite different, remarkably, the final results of the real and complex cases exhibit strong similarities, 
both in the structure of the phase space and in the order of magnitude of specific numerical values like stability thresholds.

Finding the manifold of driven and periodic solutions is the first step. 
Since such solutions could in principle be all unstable, a thorough inspection of their stability is in order both at the linear and nonlinear level. 
Linearly stable solutions proved to exist and be robust against rather large fluctuations. 
For unstable solutions, we follow numerically the evolution in the search for the final endpoint at large times. 
The picture that emerges is rather interesting: in significant portions of the unstable subspace, the solutions lie at the boundary 
of two radically different long time behaviors.
On one side, fluctuations drive the geometry to a collapse into a black hole, the dual theory loses coherence and thermalizes as expected. 
On the other side, the geometry is regular forever. It however exhibits a periodic modulation that has the form of a 
relaxation oscillation\footnote{We are adopting the usual terminology of dynamical systems like, for example, the inverted pendulum.} 
where the solution periodically bounces back and stays for a long time close to the initial unstable solution. 
From the dual field theory point of view, the quantum state remains pure but exhibits an emergent pulsating modulation.

The framework at hand allows us to address the important question of the adiabatic preparation of Floquet condensates \cite{Heinisch2015,Gertjerenken2015,Heinisch2016}.
By slowly varying the amplitude and/or   frequency  of the driving, one can study if the system keeps up with the source and follows a trajectory on the manifold of periodic solutions. This is naively expected in the limit of slow quenches, as a natural dual counterpart of the quantum adiabatic theorem. However, a careful analysis reveals subtleties whenever the driving frequency approaches that of a normal mode of AdS.
Similarly, we study the quenching of the system from AdS through a cyclic protocol starting and ending with a vanishing driving. 
The possibility that such cycle ends on a periodic solution with vanishing source (but not just AdS) could perhaps relate to the open 
question of spontaneous breaking of continuous time translations \cite{Wilczek:2012jt,Sacha:2017fqe}. 

The structure of the paper is as follows.
In Section \ref{comp} we study the case of the driven complex field. 
We characterize in detail the space of stationary solutions and focus on their linear and nonlinear stability properties.
In Section \ref{quench}  we analyze the response of the system upon a continuously varying  driving. 
This unravels interesting structures like the presence of critical values of the quench parameters separating classes 
of qualitatively different time evolutions (ending or not in a black hole, for instance). 
In Section \ref{type1} we show that the unstable solutions are, in fact, attractor solutions in the sense of Type I gravitational phase 
transitions studied in \cite{Choptuik:1996yg,Bizon:1998qd,Hawley:2000dt}.
In Section \ref{BSlego} we describe two alternative quenching protocols to construct a boson star starting from AdS. 
One is quasistatic, while the other involves an unstable solution as an intermediate step. 
Section \ref{postcol} is devoted to the late-time evolution of the collapsing black hole solutions and establishes contact with the results of the analysis done in \cite{rangamani2015driven}. In particular 
 we are able to pin down the three regimes regimes found in that paper.
In Section \ref{real} we comment the analysis for the real scalar field, stressing the fact that it constitutes a nontrivial extension both 
at the conceptual and technical level.  
We summarize in Section \ref{futuro} accounting for and interpreting the results as well as indicating research directions for the future. 
Some technical material is collected in the appendices.

\section{Periodically driven complex scalar field}
\label{comp}

Our case study involves the simplest possible setup, namely a  complex scalar field in global AdS$_4$, 
\be
\label{com_act}
S= \frac{1}{2\kappa^2} \int d^{4}x \sqrt{-g}\left( R - 2\Lambda\right)   -\int d^{4}x\sqrt{-g}\left(   \pd_\mu\phi  \pd^\mu\phi^*  - m^2 \phi \phi^* \right) \ ,
\ee
with   $\Lambda = -3/l^2$ for AdS$_4$. We will set $\kappa^2 = 8\pi G=1$, $l=1$. The action is invariant under global $U(1)$  transformations
$\phi\to e^{i\alpha}\phi$.
Our ansatz for the metric is  
\be
ds^2 = \frac{1}{\cos^2 x}\left( - f e^{-2\delta} dt^2+ f^{-1} dx^2 + \sin^2 x \, d \Omega_{2}^2\right) \ , \label{line1}
\ee
where  $x\in [0,\pi/2)$ is the radial coordinate. 
We will examine the space of  solutions adapted to the following time-periodic ansatz
\be
\phi(t,x) = \rho(x) e^{i\omega_b t} \, , \label{bsansatz}
\ee
\begin{figure}[h!]
\begin{center}
\includegraphics[scale=0.23]{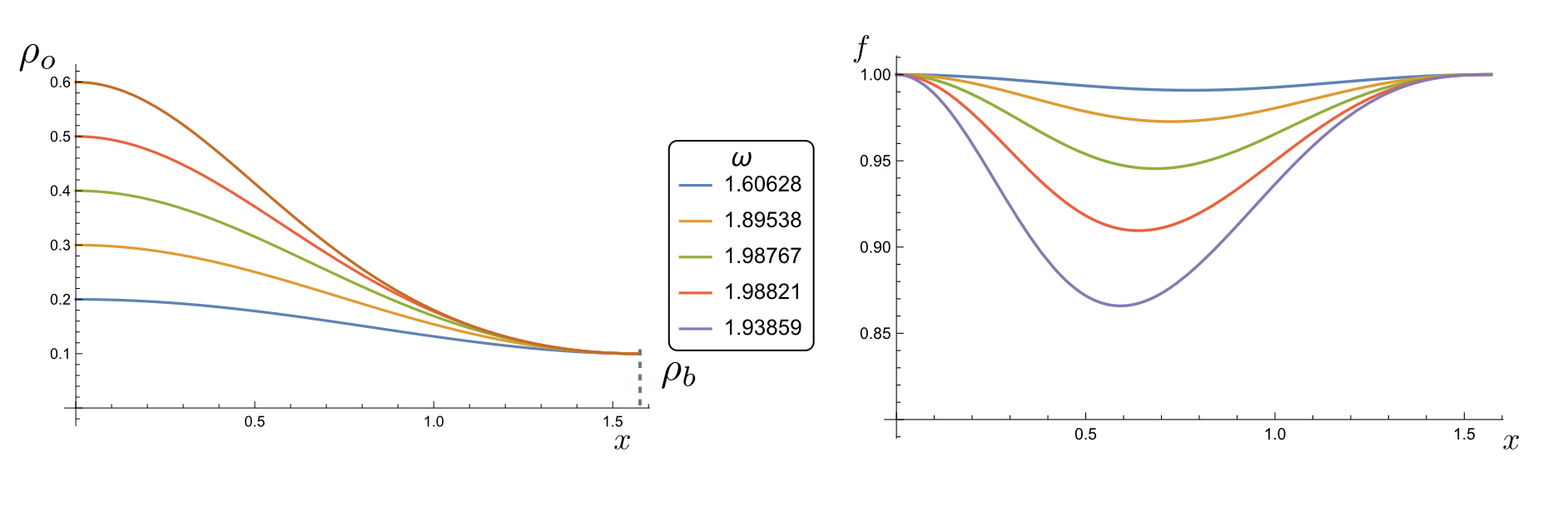}
\caption{\small Set of profiles for a range of values of $\omega_b$ holding $\rho_b = |\phi(\pi/2)|$ fixed. }
\label{fig:fullplot0}
\end{center}
\end{figure}
with $\rho(x)$ a real scalar function.
The space of solutions contains just three functions $f, \delta$ and $\rho$; they are determined by two parameters, 
namely, the frequency $\omega_b$ and a boundary value for $\rho$.

Introducing the ansatz \eqref{bsansatz} in the Einstein-Klein Gordon equations of motion, gives a set of 
$\omega$-dependent static equations for $\rho(x)$  which can be solved numerically by standard methods (see \eqref{drivenbosoneq}).
Each solution of this  system  gives rise to a static geometry where the scalar field rotates in the complex 
plane with angular velocity $\omega_b$ while keeping  constant its modulus $\rho(x)$.
The boundary values of $\rho(x)$ are respectively $\rho_o = \rho(0)$ at the origin and $\rho_b$ at the boundary; 
$\rho_b$ is defined through the asymptotic Taylor expansion
$$
\rho(x) = \rho_b (x-\pi/2)^{3-\Delta} + ...\ ,
$$
with $\Delta = 3/2 + \sqrt{9/4+m^2}$ being the conformal dimension of the dual scalar operator $\mathcal O$.
For non-vanishing $\rho_b$, the dual state is interpreted as being driven by a time-periodic source $\rho_b e^{i\omega_b t}$. 
We abbreviate these  {\em sourced periodic solutions} as SPS. In this paper we analyze the case of a massless scalar, $m^2=0$, in detail. We shall also comment on
partial results obtained in the case $m^2 = -2$, where everything seems to follow the same pattern so far. 
In the massless case, the unsourced solutions with  
$\rho_b=0$ are termed {\em boson stars} -BS- in (at least part of) the literature \cite{Astefanesei:2003qy, Buchel:2013uba} and we shall 
adhere in what follows to this name.

\begin{figure}[h!]
\begin{center}
\includegraphics[scale=0.36]{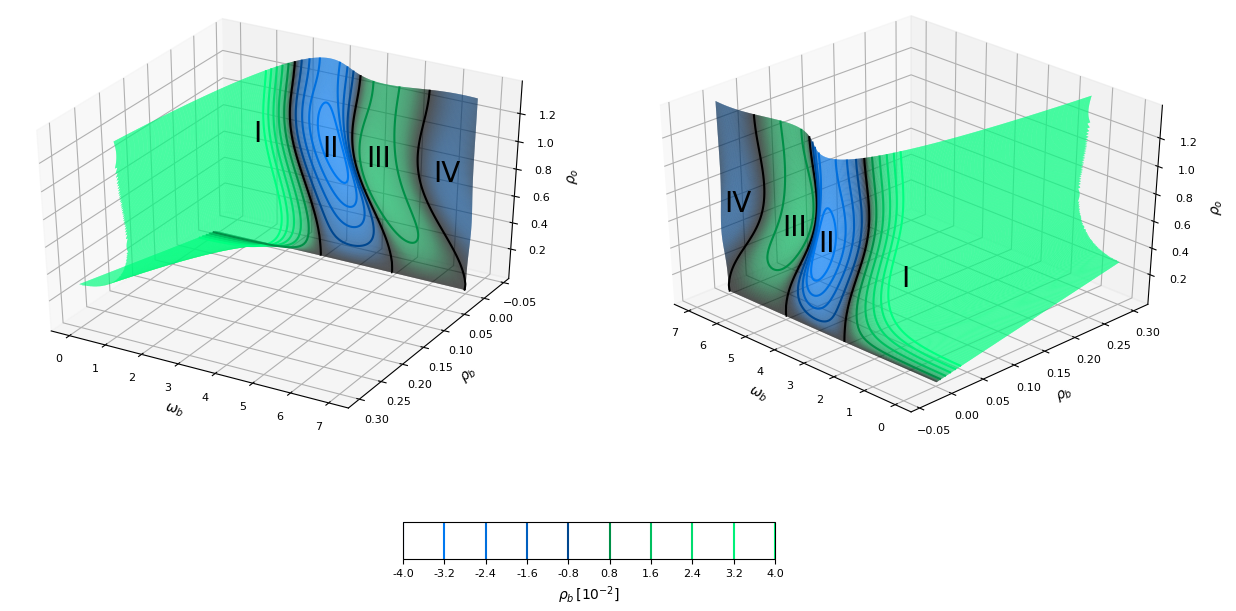}
\caption{\small The surface of static geometries corresponding to a periodically driven complex scalar field. 
The surface intersects the {\em boson star} plane $\rho_b=0$ at a set of curves as shown in the plot. In making this plot, we have adopted the phase convention that  makes $\rho_o$ a positive number (see Fig.\,\ref{fig:fig3}). Hence
the sign of $\rho_b$ is nothing but the relative sgn$(\rho_o/\rho_b)$, which is correlated with the number of nodes of $\phi(x)$. }
\label{fig:fullplot}
\end{center}
\end{figure}
\vspace{-0.5cm}

Figure \ref{fig:fullplot} will be an essential tool to understand the results of our analysis. 
The surface represents  the complete  set of static geometries corresponding to SPS's in the three dimensional space spanned by $(\omega_b, \rho_o,\rho_b)$ (see Fig.\,\ref{fig:fullplot0}  for a
visualization of these three parameters).
The SPS surface cuts at $\omega_b=0$ on a line of configurations  where the scalar takes a constant radial profile, in particular $\rho_o=\rho_b$. 
Boson stars lie at the intersection of the SPS surface with the zero source plane $\rho_b=0$. 
The boson star curves start from the bottom plane $\rho_o=0$ at the values $\omega_n = 3  + 2n$, given by the spectrum of normal frequencies of 
the massless scalar in AdS$_4$.  
As the value of $\rho_o$ is increased, the nonlinear dressing of the linearized solution shifts the value of the frequency and gives rise to the 
wiggly curves depicted in Fig.\,\ref{fig:fullplot}. 
The lower portion of this curve that bends towards smaller values of $\omega_b$ was already constructed in \cite{Buchel:2013uba}. 
The different sectors indicated with  I,II,III,... correspond to SPS whose radial profiles $\rho(x)$ have $0,1,2...$ nodes respectively.

\begin{figure}[h!]
\begin{center}
\includegraphics[scale=0.4]{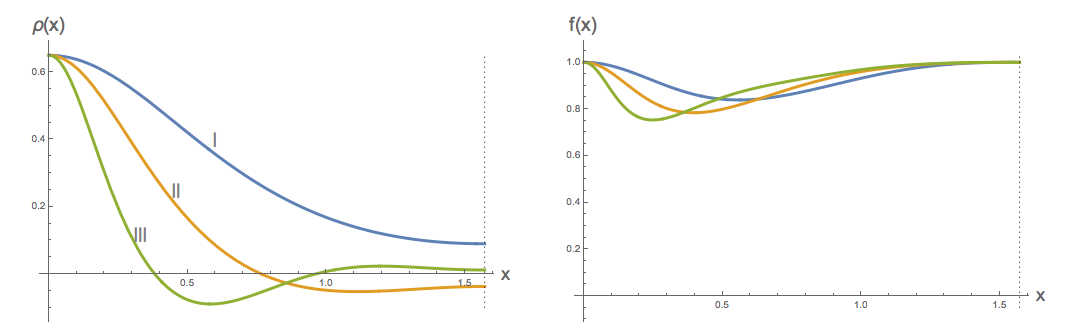}
\caption{\small  Radial profiles for higher modes in regions $I,II$ and $III$ in Fig.\,\ref{fig:fullplot} having the same value of the scalar field at the origin, $\rho_o=0.65$ and different 
$\omega_b=1.9, 2.96$ and $5.04$ from top to bottom. As the number of nodes increases, the energy  density tends to  concentrate deeper in the bulk. The reader should understant that this profile is rotating in the transverse plane. The relative sign  $\rho_b$ and $\rho_o$ correlates with the even or odd number of nodes.}
\label{fig:fig3}
\end{center}
\end{figure}
\vspace{-0.5cm}

In the case of a driving by a relevant scalar field with $m^2=-2$,  the surface remains qualitatively the same.
The unsourced solutions analogous to the boson stars branch instead from the spectrum of linearized fluctuations which is 
now $\omega_n = \Delta + 2n$ with $\Delta = 1,2$ for the alternative/standard quantization of the scalar field.

Given the external driving force, the existence of regular solutions constitutes by itself a nontrivial result, the natural expectation being that  
the system would get increasingly excited. The total mass should  grow monotonically and, eventually, lead to a collapse, signalling thermalization in the dual field theory side.
Before closing this section let us comment why this may be avoided here. The rate at which the total mass changes with time is controlled by the 
diffeomorphism Ward identity
$$
\nabla_{\mu} \langle  T^{\mu 0}\rangle  =  \langle {\mathcal O}(t)\rangle \nabla^0 {\mathcal J} \,.
$$
For the case at hand, this identity boils down to the following equation
\be
\dot m(t) = - 2{\rm Re}\left( \dot\phi^*_b(t) \langle {\mathcal O}(t)\rangle\right) \, . \label{WardId}
\ee
 In the undriven case $\dot\phi_b(t)=0$, the r.h.s. is identically zero, and the system is isolated. For $\dot\phi_b(t)\neq 0$, the magnitude and sign of the product on the right hand side is unforeseeable. The product of the two factors can be either positive or negative. This implies that, for a variable source, energy can flow  both in and out. The source function $\phi_b(t)$ is known. In contrast, the 1-point function $\langle {\mathcal O}(t)\rangle$ is a teleological quantity (in the radial direction), like event horizons are (in time). It can only be extracted from the boundary behaviour of the solution once this has been computed down to the origin and regularity has been imposed.  In the particular case of a SPS, the vev oscillates harmonically  in phase with the source. This particular case yields an exactly vanishing value for the right hand side. However, as soon as the SPS is perturbed, $\dot m$ will start to fluctuate around zero. The average of this fluctuation will signal whether the mass starts to build up and the system eventually collapses or, else, if the net balance is zero, the perturbed solution stays regular in the future.  

An important remark is that there is a chance to find a regular solution only if the quantum system we drive is in a pure state. 
If there was a horizon, no matter how small, then part of the injected energy would fall behind it, and never reach back to the boundary. 
Unavoidably, the mass would then grow monotonically and the black hole horizon  end up reaching the boundary, i.e. reaching the dual geometry to an infinite temperature state. 

In summary, a thorough study of the stability properties of the SPS is compulsory. We will do it first in a linearized approximation and then in a fully nonlinear setup. 

\subsection*{Linear stability}

\begin{figure}[h!]
\begin{center}
\includegraphics[scale=0.5]{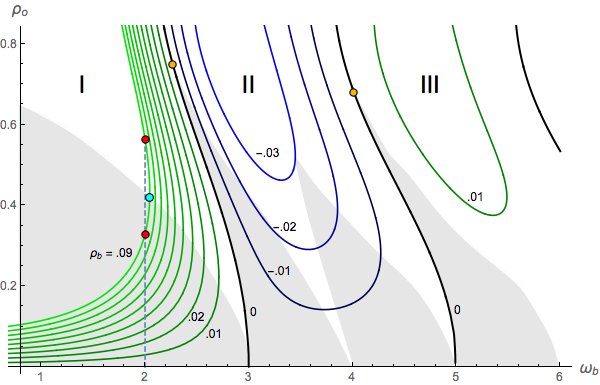}
\caption{\small Level curves of the SPS surface plotted in Figure \ref{fig:fullplot}. 
Each curve corresponds to a constant value $\rho_b$ (sometimes reported explicitly by a small number near the curve). 
The black curves denote the boson stars with $\rho_b=0$. The two red points both correspond to $\omega_b=2$ and $\rho_b=0.09$. 
The cyan dot represents the point where $\omega_b$ reaches its maximal value along the level curve.
The dark yellow dots mark the stability thresholds along the boson star lines. }
\label{fig:phibsections}
\end{center}
\end{figure}

In this section we shall establish and chart the regions of  stability within the set of SPS's given in Fig.\,\ref{fig:fullplot} by performing
a linearized fluctuation analysis. We find it convenient to move along level curves with constant source amplitude $\rho_b$. These curves are depicted in 
Fig.\,\ref{fig:phibsections}. 

Notice that from the point of view of an observer at the boundary, namely for each pair $(\rho_b,\omega_b)$, there can be more than one SPS.
They have different bulk profiles and, in particular, they reach the origin at different values of $\rho_o$. 
For instance, in Fig.\,\ref{fig:phibsections} a duplicity has been highlighted in the case $(\rho_b=0.09,\omega_b=2)$ by the  red dots on the left. 
In the case of multiple solutions corresponding to the same boundary data $(\rho_b,\omega_b)$, it is natural to suspect that one of them is stable since 
it represents the ``ground state" of the sector. 

\begin{figure}[h!]
\begin{center}
\includegraphics[scale=.4]{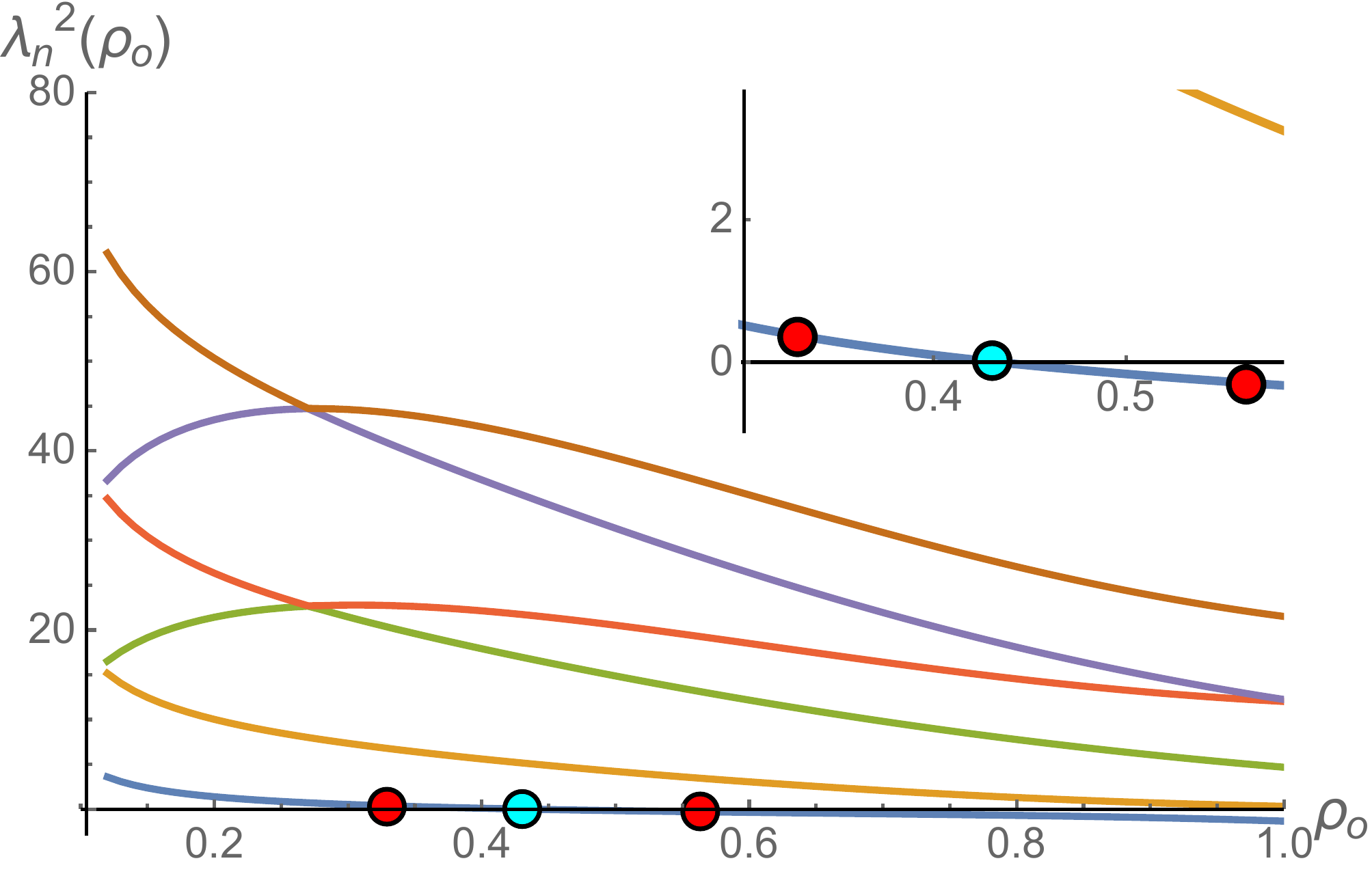} 
\caption{\small Evolution of the linearized eigenmodes along the line $\rho_b=0.09$. 
The red and cyan dots corresponds to the same solutions as in Figure \ref{fig:phibsections}.
The stability threshold occurs at $\rho_o=0.425$ and corresponds to the cyan dot where $\omega_b$ reaches its maximal value along the $\rho_b=0.09$ curve.} 
\label{fig:normalmodes} 
\end{center}
\end{figure}

Along the lines of constant $\rho_b$, one finds extremal values of the frequency where two solutions corresponding to the same $\omega_b$ become degenerate. 
At these points the spectrum of linearized fluctuations contains a zero mode that connects the two degenerate solutions.
One such example is shown in Figs.\,\ref{fig:phibsections},\ref{fig:normalmodes} by means of a cyan dot. 

Turning points of physical quantities are natural locii for the onset of  linear instability. 
The paradigmatic example is the Chandrasekhar mass of a white dwarf. 
In our analysis we also find some maxima of the mass along the level curves which are accompanied by a squared eigenfrequency 
transiting from positive to negative values, see for instance Fig.\,\ref{BS}. 
Nevertheless, looking at the extrema of the mass (or $\omega_b$) does not yield exhaustive information about the stability. 
There are cases where complex eigenfrequencies arise because two real eigenmodes merge. Further information about the mode 
structure and behavior can be found in Appendix \ref{normal}.

Charting the complete stability region involves a numerical scan of the spectrum of linearized perturbations of SPS's across the space of solutions. 
The shaded region in Fig.\,\ref{fig:phibsections} is our best approximation as to where the region of linearly stable solutions extends. 
Notice that parts of the edge of the stability region are given by the boson star solutions.  Here the passage from gray (stable) to white (unstable) 
occurs precisely because the lowest eigenmode squared turns negative. 
This observation implies that the spectrum of linearized perturbations around a boson star will always contain a zero mode. 
In Appendix \ref{normal} we prove that the zero mode generates the boson star line and we provide more information on the building and features of the stability region.

Also notice that the white wedges emerge from integer values of $\omega_b$.  Such values correspond to special eigenfrequencies of the linearized scalar field problem. 
Indeed, over global AdS$_4$, the scalar wave equation has a general spectrum of regular solutions given by $\phi(t,x)= e(x) e^{i \omega_b t}$ with  
$$
e(x) \propto \cos(x)^3 \,_2 F_1\left(\frac{3-\omega_b}{2},\frac{3+\omega_b}{2},\frac{5}{2}, \cos(x)^2\right) + \frac{3 \cot\left(\frac{\pi \omega_b}{2}\right)}{\omega_b (\omega_b^2 - 1)} \,_2 F_1\left(-\frac{\omega_b}{2}, \frac{\omega_b}{2}, -\frac{1}{2}, \cos(x)^2\right).
$$
Normal modes come in two families. Solutions with $\omega_b = 3 + 2 k$, $k = 0,1,2...$ are normalizable and have vanishing source, while solutions with $\omega_b =  2 k$, $k=1,2,3...$ are non-normalizable and have nonzero source, but vanishing vev.\footnote{This is more easy to see by rewriting the normal modes for AdS$_{d+1}$ in terms of generalised Legendre polynomials
$$
e_{k}(x) =\cos^{\Delta}x P_{k}^{(\frac{d}{2}-1, \Delta - \frac{d}{2})}(\cos(2x)), ~~~~~~~w_{k} = 2k+\Delta\,,
$$
with $\Delta = \Delta_\pm$ the two roots of $m^{2} = \Delta(\Delta-d)$. Solutions with $\Delta = \Delta_+$ are normalizable and have vanishing source. Solutions with $\Delta = \Delta_- = d-\Delta_+$ are non-normalizable and have vanishing vev.
When $m=0$, $\Delta_{+} = d$ and $\Delta_{-} = 0$. } The fact that white wedges descend to the vicinity of these linearized solutions supports the picture of linear instability as a resonance phenomenon. With this remark in mind we would expect also an instability wedge to come down to $\omega_b =2$ but actually we do not find it. 

\subsection*{Nonlinear stability}

In a nonlinear theory the results of a linear stability analysis are of limited range. A perturbed unstable solution will soon start departing largely from its original, unperturbed state. Therefore, we need to follow it to see what the end result of its evolution can possibly be. In finite dimensional nonlinear systems there are two possibilities: another stable equilibrium point or a limiting cycle. 
We have built a numerical evolution code which is a minimal adaptation of the ones employed in the study of gravitational collapse in global AdS implementing a standard
 RK4 finite difference scheme. We have checked that, when initialized at $t=0$ with a linearly stable SPS, the output yields consistently  the expected periodic geometry over as long
 as we have let the computer go. Notice that, since the profile never gets particularly spiky, rather low resolutions of $2^{10}+1$ points are sufficient. This allows for a considerable increase in computational speed.

The next numerical experiment is to initialize the code with a linearly unstable SPS slightly perturbed with the single unstable normal mode. In other words, 
if $\phi_0(x)$ is an SPS, and $\chi_1(x)$ its unstable mode with a purely imaginary eigenfrecuency $\lambda_1^2<0$, then at $t=0$ we 
will insert $\phi_0(x) + \epsilon_1 \chi_1(x)$ with $\epsilon_1 \sim {\cal O}(10^{-4})$.
In general terms, one could reasonably  expect collapse to a thermal phase as the end result of time evolutions starting from unstable initial conditions.  
Nevertheless, as we shall see now,  even in this case, the system provides regular counterexamples.

\begin{figure}[htbp]
\begin{center}
\includegraphics[scale=0.4]{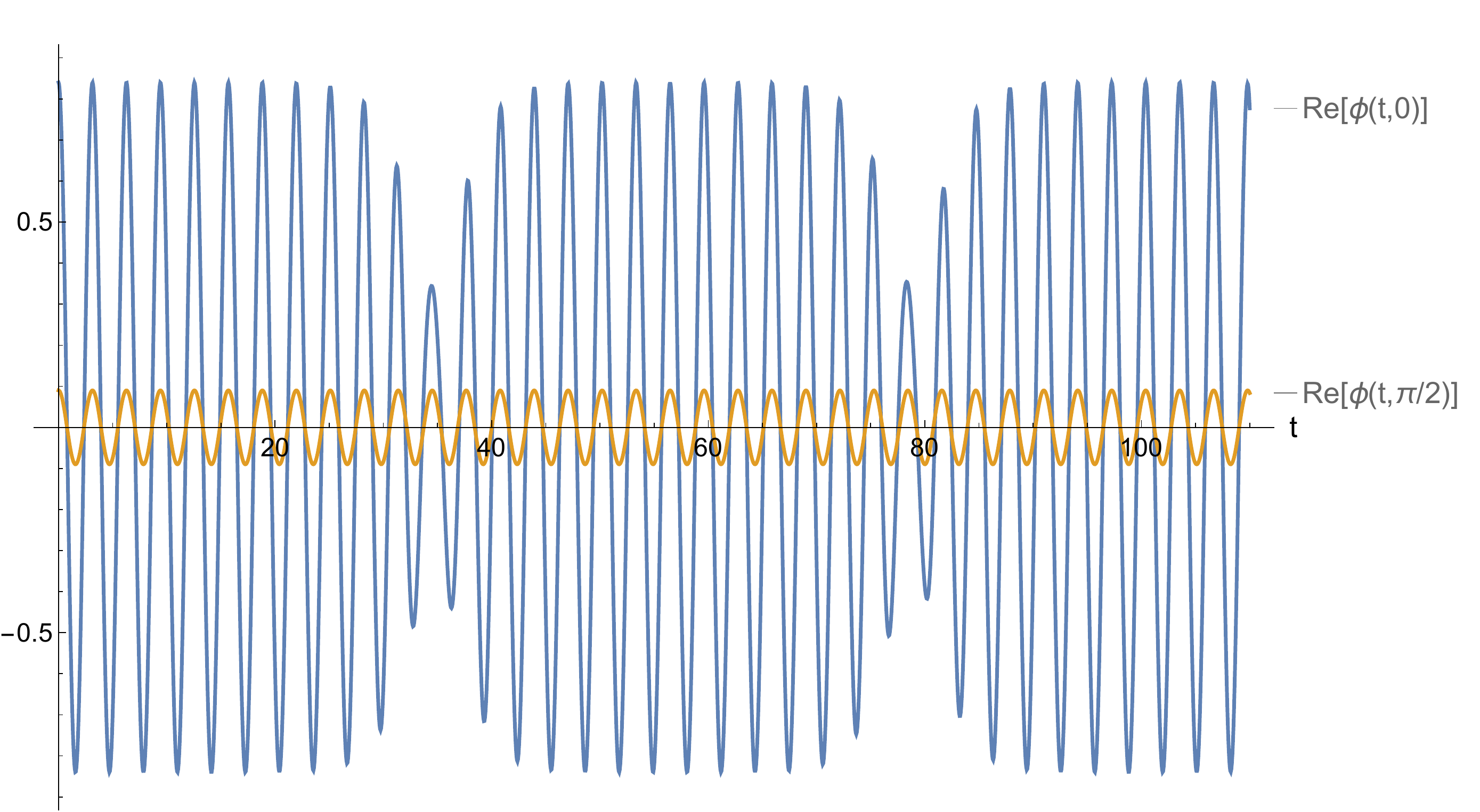} 
\caption{\small  In this plot we appreciate that the driven scalar field keeps in phase with the driving all along its profile. 
The tip at the origin and at the boundary rotate at the same pace. 
There is, a priori, no reason why this should be so and is not an artifact or truncation of the simulation.  
In fact one can find (unstable) solutions where this is not the case. }
\label{fig:pulsatingplot}
\end{center}
\end{figure}

In order to keep the analysis as systematic as possible, we continue working with the two solutions marked in red  on 
the left hand side of Fig.\, \ref{fig:phibsections}. They both represent SPS's with $\omega_b=2$ and $\rho_b=0.09$.  
The lower one lies in the region of linear stability, and we have checked that it also supports fairly ``strong kicks", $\epsilon_1\sim {\cal O}(1)$,
leading to oscillations about the initial solution.

\begin{figure}[htbp]
 \begin{center}
\includegraphics[scale=0.36]{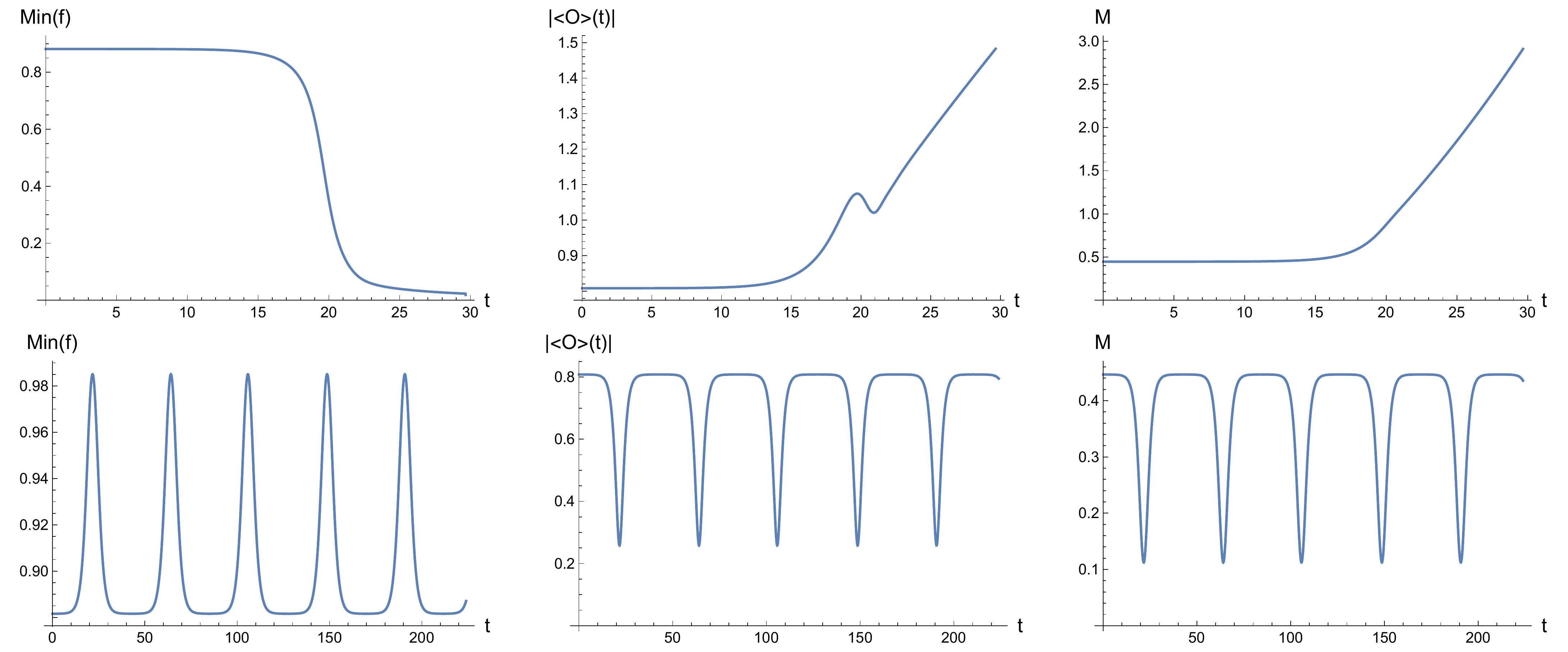} 
\caption{\small Evolution of three magnitudes for an unstable SPS with $\rho_b = 0.09$, $\rho_o=0.56$ and $\omega_b = 2$ after being perturbed 
with the first unstable mode $\phi(0,x) + \epsilon_1 \chi_1(x)$. The upper (lower) three plots correspond to  $\epsilon_1=-10^{-4}$ ($+10^{-4}$).
The magnitudes shown are the minimum of the metric function min$_{x\in [0,\pi/2) } (f(t,x))$, the energy density $m$, and the vev $|\langle \mathcal O\rangle|$. 
A relaxation oscillation with a pronounced peak and a long lived plateau is apparent on the lower plots.}
\label{fig:BHandSMS}
\end{center}
\end{figure}

In contrast, the upper one contains in its spectrum of linearized normal modes one unstable eigenmode, $\chi_1(x)$, with a 
purely imaginary frequency  $\lambda_1^2 = -0.2861478$.  This fundamental eigenmode has no nodes  and its shape  resembles the SPS itself. 
With $\epsilon_1 \sim 10^{-4}$,  the time evolution when  $t\to \infty$ differs dramatically depending upon the sign chosen for $\epsilon_1$. 
With positive sign (so that $\epsilon_1\chi_1(0)$ has the same sign as $\phi_0(0)$), the evolution collapses promptly to a black hole geometry.  
This can be seen in the top line of plots in Fig.\,\ref{fig:BHandSMS}. 
The continuous driving reflects itself in the rising of the total mass, while the horizon radius increases monotonically as does the absolute value of the vev itself. 
The system approaches the expected infinite temperature state. 

With the opposite sign, $\epsilon_1 \sim -10^{-4}$, the evolution reaches a radically different endpoint. 
Instead of a collapse, the dynamics stays regular for all times.  
In Fig.\,\ref{fig:pulsatingplot} we plot the oscillations of the real part of the scalar field at the origin  and at the boundary.
Both of them proceed in phase with one another, signalling that the profile of the scalar field stays in a plane while rotating. 
Moreover, and this is new, the oscillation of the scalar at the origin acquires a violent and sudden modulation in amplitude. 
 
Since we are working with a fully backreacted solution, the modulation of the driven oscillation affects many other physical quantities.   
Fig.\,\ref{fig:BHandSMS} highlights the impact of the modulation on three observables: the minimum of the metric function min$_x[f(t,x)]$,  
the absolute value of the vev $\langle{\cal O}\rangle (t)$ and, finally, the energy density $M$. 
Note that min$_x[f(t,x)]$ reveals the formation of an apparent horizon whenever it drops towards zero. 
 
The periodicity of the modulation should not obscure the fact that it is highly anharmonic.  
The oscillations in modulation stay on  long lived plateaux followed by sudden pulsations or beats.  
We will refer to these regular solutions as {\em sourced modulated solutions}, SMS. 
The period  $T$ is orders of magnitude away from the one due to the driving frequency ($2\pi/\omega_b$) and, actually, 
it is (weakly) dependent on the strength of the perturbation. 
Hence, in the strict sense, the solution is not a limiting cycle like the one found in nonlinear systems such as the van der Pol oscillator. 
In such systems the asymptotic dynamics eventually loses memory of the initial state whereas, in our case, it bears resemblance 
to relaxation oscillations \cite{Strogatz2001}. 
The closest mechanical analogy for a SMS would be that of an inverted pendulum slightly kicked out of its vertical unstable position, 
where it will return and stay for a long time until the next sudden turn. 

In fact, and very remarkably,  these regular solutions also occur with vanishing source. 
Indeed, unstable boson stars on the black curve slightly above the yellow dots in Fig.\,\ref{fig:phibsections} also 
exhibit this modulated  dynamics for one sign of the most relevant perturbation, 
and collapse to a black hole for the other.\footnote{As explained with care in Appendix D, the spectrum of fluctuations of a 
boson star contains a zero mode which generates the boson star branch (black lines in Fig. \ref{fig:phibsections}). 
The next mode is the one that becomes purely imaginary at the Chandraserkar-like mass. 
It is perturbing with respect to the unstable mode that the solution behaves as mentioned above in a highly correlated 
way with the sign of the coefficient of the perturbation.} Such behavior has been observed earlier, both in 
asymptotically flat \cite{Lai:2004fw} and AdS spacetime\cite{Maliborski:2016zlh,Choptuik:2017cyd}.
Also, SMS profiles like the one in Fig.\,\ref{fig:pulsatingplot}  bear some resemblance with the long time modulation of oscillating solutions that appear in confining theories 
after a global quench that injects energy below the mass gap threshold \cite{Craps:2015upq,Myers:2017sxr}. At first sight, the modulations we find
look substantially more anharmonic and strongly peaked. A  Fourier analysis should be carried out in order to reach a definite conclusion. Another interesting proposal could relate the SMS's to nonlinearly dressed multi-oscillator solutions \cite{Choptuik:2018ptp}. These still have to be constructed in the driven situation, but most likely this is feasible. 

 \begin{figure}[h!]
\begin{center}
\includegraphics[scale=0.3]{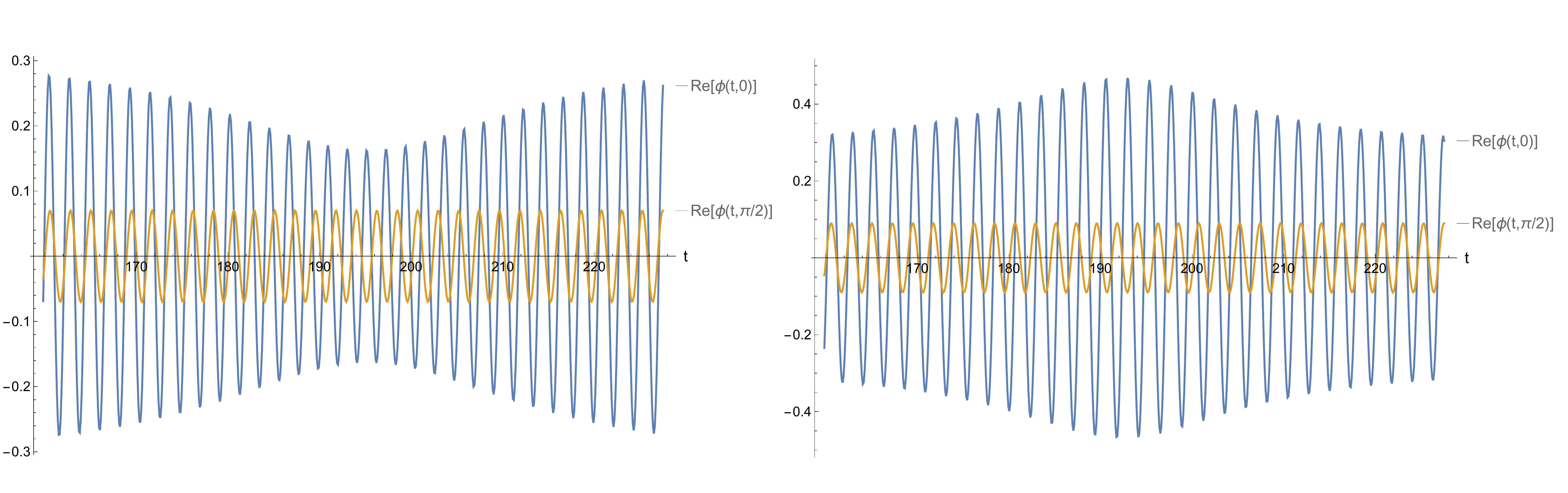} 
\caption{\small For small values of the driving amplitude and the scalar field modulus  at origin, here $\rho_b = 10^{-3}$ and $\rho_o=0.3$, 
perturbations along the unstable mode $\epsilon_1 \chi_1$ with both signs $\epsilon_1 = \pm 10^{-7}$ 
destabilize the initial unstable SPS reaching pulsating solutions with regular behavior.  }
\label{fig:pulsatingplotupdown} 
\end{center}
\end{figure}

The unstable SPS's with low enough amplitude (indicatively of the order of $\rho_o< 0.3$) undergo an exotic evolution. 
In fact, in these cases the pulsating modulation appears for {\em both} signs of the relevant perturbation $\epsilon_1$. 
With the ``potentially lethal" sign (the one that usually would drive the SPS towards a black hole) we now observe a pulsating 
modulation where the value of $\rho_o$ increases instead of decreasing, yet reaching a maximum value and bouncing back again. 
The mass and the vev also increase and decrease back and the solution stays regular forever.  

Figure \ref{fig:pulsatingplotupdown} shows a case of those just mentioned where we have plotted the real parts of $\phi$ at the origin and the boundary.
This highlights an intriguing feature: the modulation pulse does not only perturb the absolute value of the scalar field in the bulk, but also its phase. 
More precisely, during the pulsation, the two values $\phi_o(t)$ at the origin and $\phi_b(t)$ at the boundary start de-phasing in one sense or the other 
depending upon the sign of $\epsilon_1$, and the scalar profile becomes increasingly helical. 
After the pulsation has ended they again re-phase and the profile recovers its planar shape.  

The study of nonlinear stability is always a long and very much resource dependent task. 
What we have done so far is to characterize the endpoint of linearly unstable SPS's and we have found two possibilities, a dynamical black hole and a SMS. Concerning this
last one, we have indeed checked for their robustness by adding a perturbation to them. Perturbed SMS's develop wiggling plateaux but remain regular all along the 
simulation, suggesting that their normal mode spectrum contains only real eigenfrequencies. The very late time behaviour of the solution is another delicate point of the nonlinear stability analysis. As far as our codes have run, up to $t= {\cal O}(10^4)$, we have not found the slightest  evidence of
a nonlinear instability setting in at late times. After the discovery of the AdS instability \cite{Bizon:2011gg} this is an issue one should be concerned about. However, the essential ingredient found in that context, namely full resonance, 
is very unlikely to be present in the spectrum of linearized perturbations of a SMS. In the absence of any potential mechanism for destabilization at long times, any runtime is disputable in what concerns any claim for stability.  

Finally, let us stress again that all the analysis has been performed  in region I of the phase space shown in Fig.\,\ref{fig:phibsections}.
It would be interesting to do an accurate scan to see if further exotic evolutions can be obtained along the wedge of unstable solutions.
In particular, we have not analyzed the fate of unstable solutions in regions II, III... This remains for a later investigation.

\section{Quenches with periodic driving }
\label{quench}

The sourced periodic solutions we have found are eternal, extending from $t=-\infty$ to $t = +\infty$. 
Therefore, it makes sense to try to study if they can be constructed by means of a slowly growing source starting from AdS.
In this section we perform this investigation by considering quenches in a generalized sense.  
The word quench usually refers to a change in some coupling constant, which can be either sudden or slow. 
In the present context, it will refer to a certain process that interpolates between two different periodically driven Hamiltonians. 
In short, we shall explore how the system responds when the boundary data that determine the periodic driving of the dual 
QFT become, themselves, functions of time $(\rho_b(t), \omega_b(t))$. 

\subsection{Quasistatic quenches}

We analyze the system response to very slow changes of the driving parameters, whose variation occur on a typical time scale $\beta \gg \omega_b^{-1}$. 
In the context of static quantum systems the adiabatic theorem states that, for sufficiently slow quenches, the ground state of the system follows
the change in the Hamiltonian. Even if a non-vanishing transition amplitude to an excited state is generated, there is a well defined way to 
bring it to arbitrarily small values (by making $\beta$ large enough). Here we find that a similar phenomenon occurs, albeit the underlying 
unquenched Hamiltonian is time-periodic. The system follows the slow modulation of the parameters of the periodic driving, moving from one ground state to another. 

\begin{figure}[h!]
\begin{center}
\includegraphics[width=8cm]{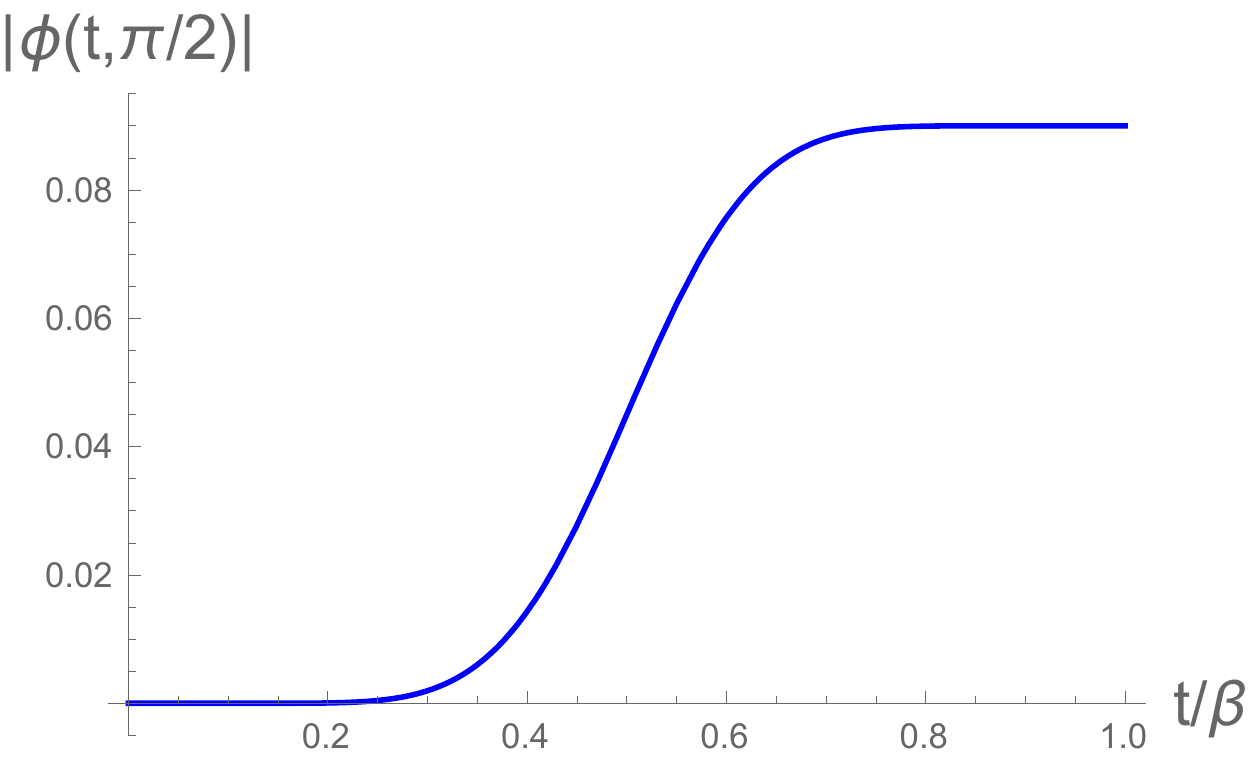}
\end{center}
\caption{\label{quench_profile} \small $\rho_b(t)$ for the quench profile \eqref{source_profile} at $\epsilon = 0.09$.}
\end{figure}

As we already know, SPS exist on a codimension-one submanifold in $(\omega_b, \rho_o, \rho_b)$ space. 
In this subsection, we demonstrate that linearly stable SPS's can be reached from the global AdS$_4$ vacuum by means of a sufficiently slow quench. 
Conversely, we also show that these linearly stable SPS determine which quench processes cannot be regarded as adiabatic.
We employ the following ansatz for the scalar field source  
\beqa
\phi(t,\pi/2) &=& \frac{1}{2}\epsilon\left(1-\tanh\left(\frac{\beta}{t} + \frac{\beta}{t-\beta} \right) \right) e^{i \omega_b t},~~~~~0\leq t<\beta, \nonumber \\
\phi(t,\pi/2) &=& \epsilon e^{i \omega_b t},~~~~~~~~~~~~~~~~~~~~~~~~~~~~~~~~~~~~~~~~~~~t \geq \beta  \label{source_profile}
\eeqa
i.e., the source starts being zero at $t = 0$, and reaches its final amplitude $\epsilon$ after a time span $\beta$.  
Afterward, it oscillates harmonically. During the whole process, the frequency $\omega_b$ is kept constant. 
We refer to the regime taking place at $t < \beta$ as the build-up phase, while the driving phase corresponds to $t \geq \beta$. 
The quench profile is plotted in Fig.\,\ref{quench_profile}, where $\rho_b(t) \equiv |\phi(t,\pi/2)|$. 

\subsubsection*{Quasistatic quench to a SPS}

Imagine that the final values of $\epsilon$ and $\omega_b$ correspond to a SPS in the region of stability, as given in Fig.\,\ref{fig:phibsections}. 
For concreteness, let us focus on region I, and consider the lower red dot depicted at $\rho_b=0.09$ and $\omega_b = 2$. 
In this case, if $\beta$ is sufficiently large, the state obtained after the build-up process is, to an excellent approximation, the SPS corresponding 
to the final harmonic driving.
Furthermore, during the driving phase the numerical solution also remains stationary (up to the largest times we have simulated and with high accuracy). 

\begin{figure}[h!]
\begin{center}
\includegraphics[width=16cm]{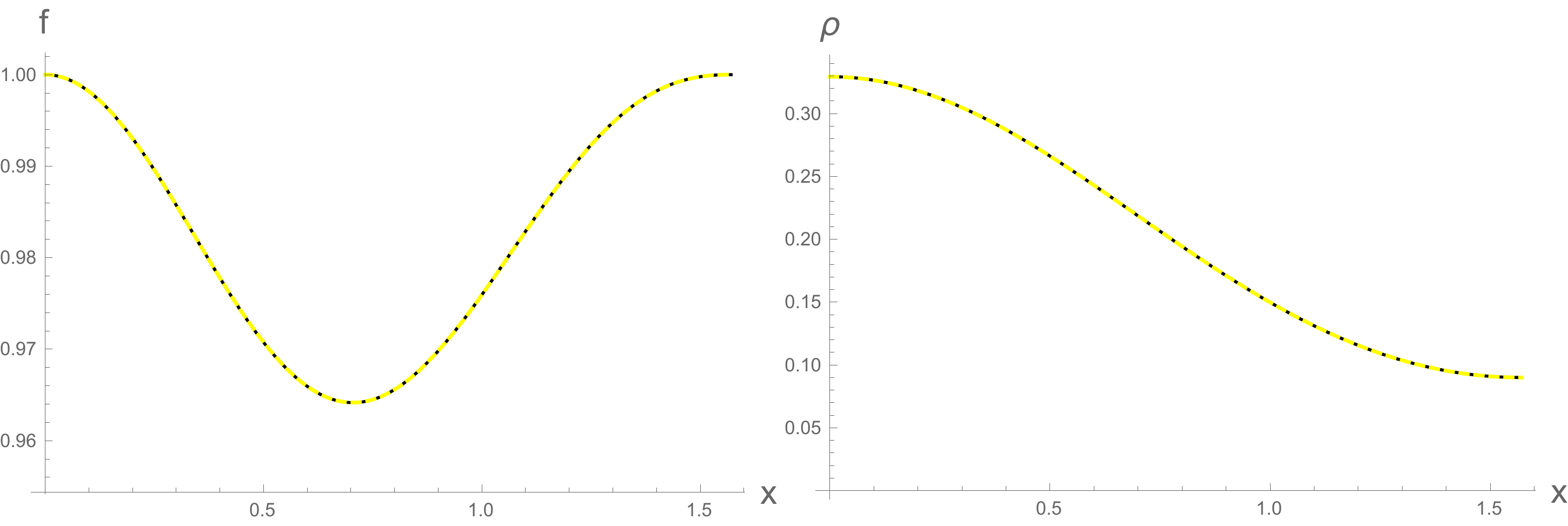}
\end{center}
\caption{\label{adiabaticity_fields} \small  In dotted black $f(x)$ and $\rho(x)$ at the end of the driving phase for a quench process of the form \eqref{source_profile} 
with time span $\beta = 2500$ and final harmonic driving data $(\omega_b = 2, \epsilon = 0.09)$. In solid yellow, $f(x)$ and $\rho(x)$ of the SPS corresponding 
to the same  data. The curves lie with high accuracy on top of each other.
}
\end{figure}

Let us choose $\beta = 2500$. 
We have determined numerically that, after the quench (i.e., for $t > \beta$), the energy density of the geometry corresponds to $m = 0.2121$, which agrees with the energy density of the SPS we expect to land on up to the fifth significant digit. 
A more detailed check is provided in Fig.\,\ref{adiabaticity_fields}, where we compare the fields $\rho$ and $f$ during the driving phase 
(solid curves) and their profiles in the corresponding SPS (dashed curves). 

Having demonstrated that the linearly stable SPS can be reached from the vacuum by a sufficiently slow quench process, 
it remains to analyze the adiabaticity of the whole procedure. 
Specifically, if the system's response to the time-dependent source is perfectly adiabatic, we expect that 
\beq
\Psi(t,x) = \Psi_{SPS}(\rho_b(t), \omega_b; x), \label{adiabatic_definition}
\eeq
where, with no loss of generality, $\Psi$ denotes any field of the geometry. 
$\Psi_{SPS}$ is the value of this field in the SPS at the given instantaneous $\rho_b$ and $\omega_b$. 
The equality \eqref{adiabatic_definition} thus entails that the dynamics does not depend explicitly on time, 
but only implicitly through the instantaneous value of the source. 
In other words, in the quasistatic large $\beta$ limit, one can draw the evolution as a path on the surface 
of SPS's (in this case, the vertical dashed line displayed in Fig.\,\ref{fig:phibsections}). 

Note that these observations should also hold for one-point functions such as $m$ and $\langle \mathcal O \rangle$: 
if the response of the system to the quasistatic quench process is perfectly adiabatic, it must be the case that  
\beq
m(t) = m_{SPS}(\rho_b(t), \omega_b), \label{adiabatic_m}
\eeq
and similarly for the scalar vev. 

\begin{figure}[h!]
\begin{center}
\includegraphics[width=16cm]{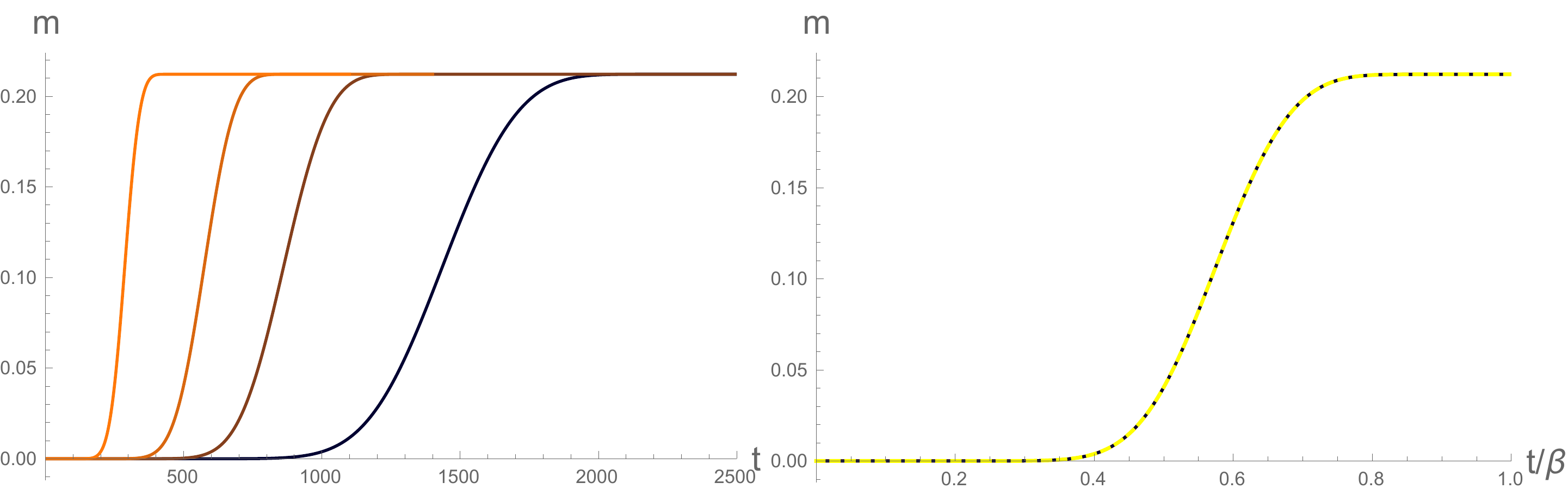}
\end{center}
\caption{\label{adiabaticity_m} \small Left: $m(t)$ for quenches of the form \eqref{source_profile} to the harmonic 
driving $\epsilon=0.09,\omega_b=2$. From left to right, $\beta = 500, 1000, 1500, 2500$. Right: $m(t/\beta)$ for the quenches of the right figure. 
The curves clearly collapse into a universal profile (dotted black). We also plot $m_{SPS}(\rho_b(t/\beta), \omega_b)$ in (solid) yellow. 
The agreement between both curves implies that the system responds adiabatically.}
\end{figure}

As we illustrate in Fig.\,\ref{adiabaticity_m}, these expectations are fulfilled. 
Let us look first at Fig.\,\ref{adiabaticity_m}a. There, we compare the time evolution of $m$ for four quench processes, with $\beta = 500, 1000, 1500, 2500$. 
Since for the quench profile \eqref{source_profile} $\rho_b(t)$ only depends on the dimensionless ratio $t/\beta$, 
relation \eqref{adiabatic_m} implies that for adiabatic response the energy density can only be a function of $t/\beta$, but 
not of $t$ and $\beta$ separately. Therefore, when plotted in terms of $t/\beta$, the four instances of the time evolution 
of $m$ shown in Fig.\,\ref{adiabaticity_m}a must collapse to a universal curve. 
This is indeed what happens, as we illustrate in Fig.\,\ref{adiabaticity_m}b (dotted black). 
Finally, in Fig.\,\ref{adiabaticity_m}b we also depict $m_{SPS}(\rho_b(t/\beta), \omega_b)$ (solid yellow).  
The evident agreement between both curves shows that relation \eqref{adiabatic_m} is satisfied.

\subsubsection*{Quasistatic quench to a black hole}

Even in the $\beta \to \infty$ limit, there exists a bound to the amplitude or the frequency of the driving that one can reach.  
Consider a quench at constant $\omega_b$ from global AdS$_4$ to a final source amplitude, $\epsilon$,  such that there is no SPS associated to $\omega_b$ and $\epsilon$. 
As the source amplitude builds up, and in the quasistatic limit, we expect that the system evolves through a succession of SPS's, at most up to the time 
$t^*$ when the SPS associated to $\rho_b^* \equiv \rho_b(t^*)$ ceases to exist.\footnote{This point corresponds to the tip of the $\rho_b = \rho_b^*$ curve 
in the $(\omega_b, \rho_o)$ plane}$^,$\footnote{Note that, strictly speaking, adiabatic evolution can break down for $t_c \leq t^*$, $t^* - t_c \ll 1$; 
for instance, $t_c$ could be the moment when the intrinsic response time of the system is above the rate of change 
of the source amplitude. In this sense, the time $t^*$ just sets an upper bound on $t_c$, and should be understood in this sense.  
On the other hand, in the quasistatic $\beta \to \infty$ limit, we must have that $t_c \to t^*$.} 
Past this point, the system exits the surface of SPS's and adiabatic evolution cannot proceed further, no matter how slowly the source amplitude increases afterwards. 

Let us analyze these questions in a specific example. We consider the quench described by equation \eqref{source_profile}, again at driving 
frequency $\omega_b = 2$, but increasing now the final source amplitude to $\epsilon = 0.1$. 
There does not exist a SPS associated to this particular driving: the highest driving amplitude compatible with the existence of a SPS 
with $\omega_b = 2$ is $\rho_b^* \in [0.099200,0.099225]$. 
According to the quench profile \eqref{source_profile}, this value is reached at a time $t^* \in [0.73482,0.73581] \beta \equiv \eta \beta$.
Here $\eta$ is critical value of the scaling variable $t/\beta$ after which we expect adiabatic evolution to break down. 

\begin{figure}[h!]
\begin{center}
\includegraphics[width=16cm]{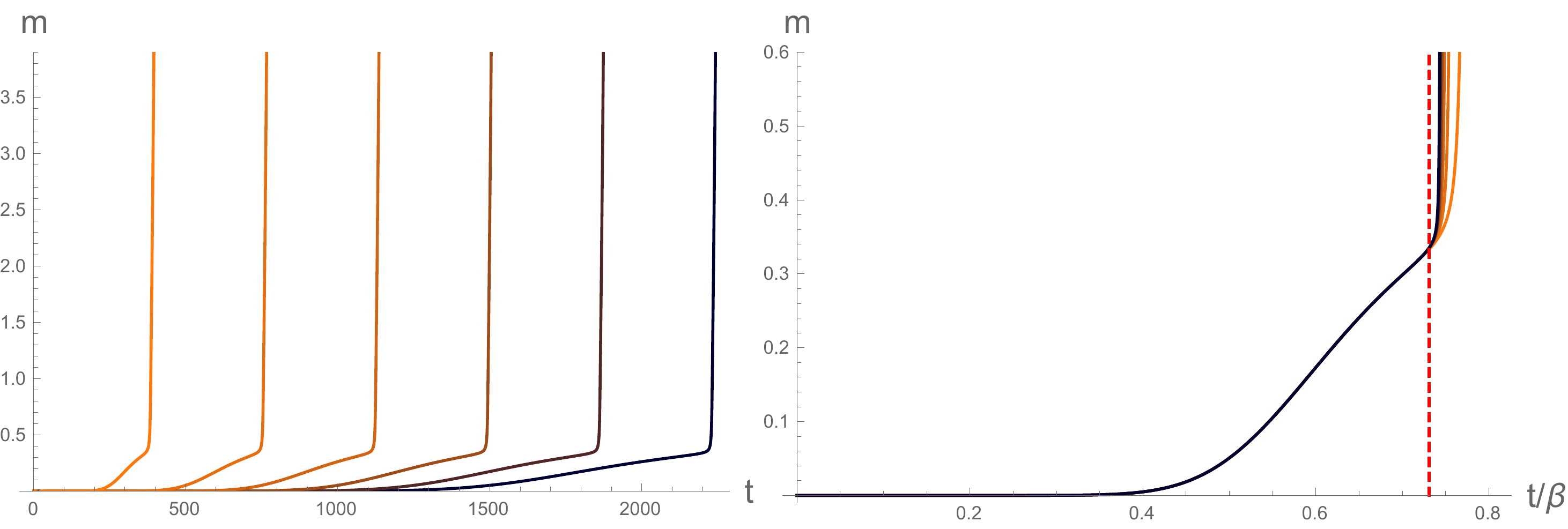}
\end{center}
\caption{\label{adiabaticity_breaking_m} \small Left: time evolution of the energy density for the quench processes described in the main text. 
From left to right, the quench time span corresponds to $\beta = 500, 1000, 1500, 2000, 2500, 3000$. Right: time evolution of the energy density, 
now plotted against the scaling variable $t/\beta$. We clearly observe that, prior to the adiabaticity breaking, $m(t/\beta)$ approaches a limiting 
curve as $\beta$ increases. The vertical line marks the time $t^*/\beta = \eta = 0.73482$.}
\end{figure}

In Fig.\,\ref{adiabaticity_breaking_m}a, we plot the time evolution of $m$ for quench profiles with time spans $\beta = 500, 1000, 1500, 2000, 2500, 3000$. 
We clearly see that gravitational collapse takes place at times $t < \beta$. In order to understand adiabaticity and its breakdown, 
in Fig.\,\ref{adiabaticity_breaking_m}b we plot $m(t/\beta)$. Two facts are manifest. 

The first one is that, for $t /\beta< \eta$ (signaled by the vertical dashed red line), all the energy density curves merge into a universal profile. 
This observation immediately leads to the conclusion that $m(t)$ is set solely by the instantaneous value of $\rho_b(t/\beta)$ and, as a consequence, 
any nontrivial dynamics that depends explicitly on $t$ is highly suppressed in the quasistatic limit. The system evolves adiabatically, as expected. 

The second one is that, for $t/\beta > \eta$, the system undergoes gravitational collapse (as can be seen in the unbounded increase of its energy density). 
Given that, for $t/\beta > \eta$, we have that $\rho_b(t) \geq \rho_b^*$ (i.e., there is no SPS associated to the driving), we conclude that adiabatic
evolution breaks immediately after  crossing the $\rho_b(t) = \rho_b^*$ threshold.

\subsection{Non-quasistatic quenches} 
\label{typeItransition}

Let us address now the non-quasistatic regime,  namely that of quenches with finite time span $\beta <\infty$. From the QFT side the standard lore is, 
according to the adiabatic theorem,  that the system will not keep up with the variation of the source, and will generically transit to an excited state. 
If this state is highly excited, it is natural to expect a decohering evolution towards an (effective) thermal mixed state.

\begin{figure}[h!]
\begin{center}
\includegraphics[width=16cm]{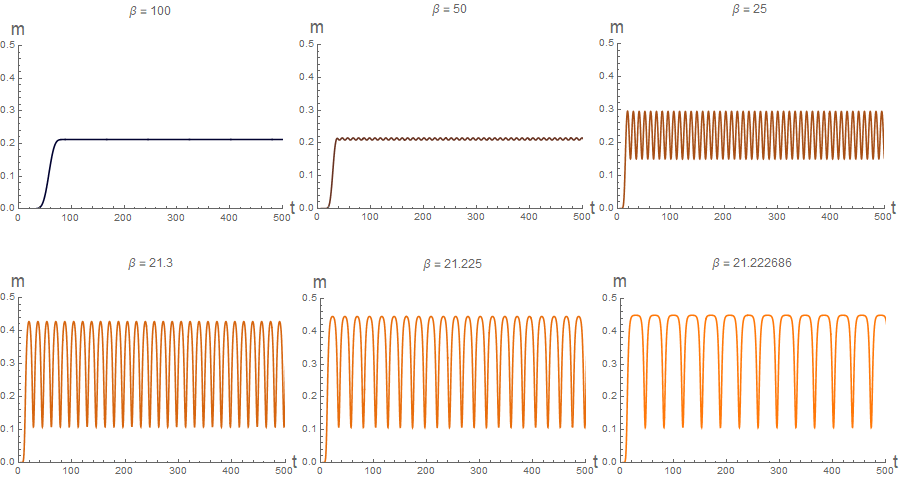}
\end{center}
\caption{\label{non_adiabatic2} \small Evolution of $m$ for different build-up processes for a driving characterized by $\omega_b = 2$, $\epsilon = 0.09$. 
From top-left to right bottom, the build-up time $\beta$ decreases as shown in the plots.}
\end{figure}

In order to inspect the nature of this excited state in the dual gravitational theory we consider a one-parameter family of 
quench profiles \eqref{source_profile} with varying $\beta\in (0,\infty)$. For definiteness, we focus on the same $\omega_b = 2, \epsilon = 0.09$ case as before.  
Recall that our findings from the previous section established that, in the $\beta\to \infty$ limit, the constructed state was the stable 
ground state SPS at $\rho_o=0.33$ given by the lowest red dot in Fig.\,\ref{fig:phibsections}. Instead,  for $\beta <\infty$, the results of our simulations vindicate the existence of a critical value $\beta_c$ separating two radically different regimes. 

\begin{itemize}

\item For $\beta \geq \beta_c$, the system always remains regular in the driving regime, and settles down to a time-periodic geometry.  
For finite $\beta$, in addition to the harmonic response given by $e^{i \omega_b t}$, it develops an additional periodic modulation. 
This can be seen in several one-point functions such as  $m$ and $\left|\left< \mathcal O \right>\right|$, as illustrated in Fig.\,\ref{non_adiabatic2}. 
This additional modulation is tantamount to the fact that our system is not in the ground state anymore, in agreement with expectations. 
As $\beta$ is lowered two phenomena can be seen from the plots. On one side, both the amplitude and the periodicity, $T$, of the modulation grow. At the same time,  the injected mass also increases and ends up  oscillating around larger mean values. 
This reflects the fact that the quench has injected more energy and, consequently, the state is more excited. On the other, periodic modulations become less and less harmonic, 
and start developing plateaux where the system stays for progressively longer times as  $\beta \to \beta_c$. 
This can be observed in Fig.\,\ref{non_adiabatic2} as we move to plots on the lower right part. 
Interestingly, the mass curve in the last plot matches very well with the corresponding one in Fig.\,\ref{fig:BHandSMS}. 
As we will show in the next section, this is more than a coincidence, and indeed, in the limit $\beta\to\beta_c$ the end 
result of the non-quasistatic quench is the SMS to which the unstable SPS with the prescribed values of $(\omega_b,\epsilon)$ 
will decay when perturbed with the appropriate sign of the most relevant fluctuation. 
To be most transparent, and referring to Fig.\,\ref{fig:phibsections}, the implied unstable SPS is the top red dot  
sitting at $\rho_o=0.56$ along the vertical dashed line. In summary, a quench to the same final boundary data $(\epsilon, \omega_b)$ yields 
the stable solution (lower red dot) if performed in a quasistatic way $\beta\to\infty$, and  the SMS associated to the 
unstable SPS (higher red dot) in the critical limit  $\beta\to\beta_c$. 
However, the dashed line should not cause confusion in the later case. 
The process, not being quasistatic, has not proceeded through a sequence of SPS's. 
It cannot be drawn as a path in the solution space. 

\begin{figure}[h!]
\begin{center}
\includegraphics[width=16cm]{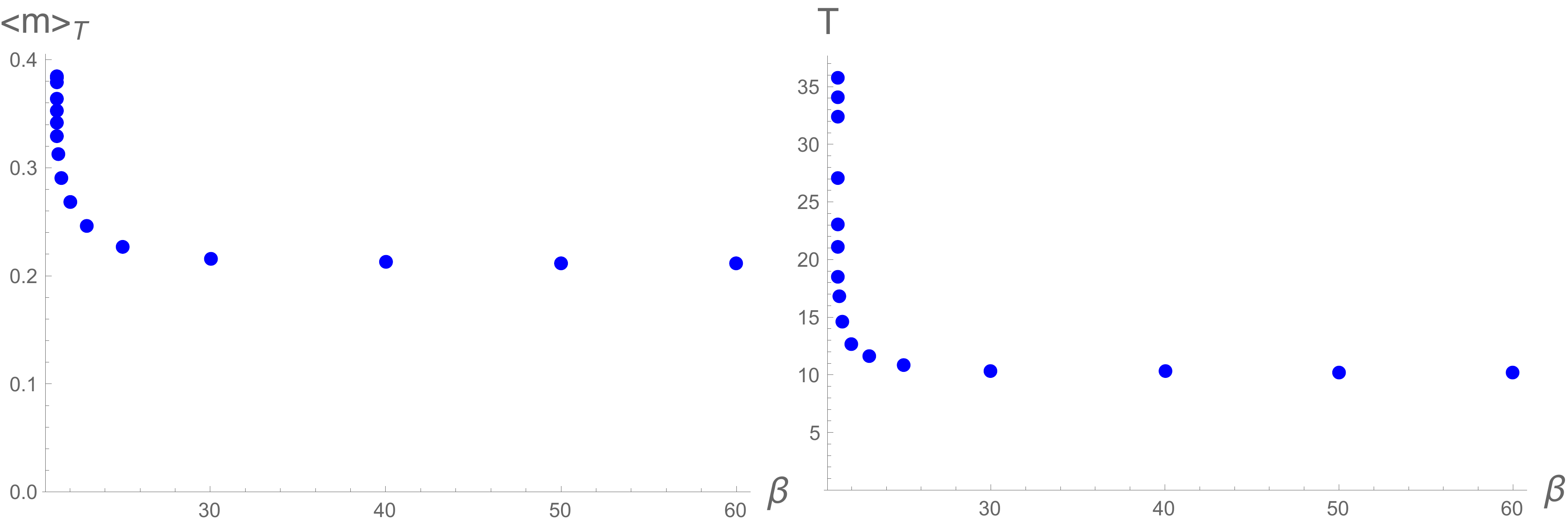}
\end{center}
\caption{\label{mT} \small Average mass $\left<m\right>_T$ (left) and period $T$ (right) of the time-periodic solutions 
generated by the build-up phase when $\beta>\beta_c$.}
\end{figure}

In Fig.\,\ref{mT}, we plot the period $T$ of the periodic modulation of the final solutions as well as the average mass
\beq
\left<m\right>_T \equiv \frac{1}{T}\int_t^{t+T} dt \, m(T)  
\eeq
as a function of $\beta$.
Both functions increase monotonically with decreasing $\beta$. For $\beta \to \infty$, $\left<m\right>_T$ tends to its value 
on the stable SPS, while $T$ is compatible with the period of the fundamental normal mode of this SPS.

\item For $\beta < \beta_c$, the quench results in an energy injection process  strong enough to trigger  gravitational collapse.  
After a transitory regime  both $m$ and $\left|\left< \mathcal O \right>\right|$ increase without bound, while the  
value $\textrm{min}_x[f(t,x)]$ drops to zero, signaling the approach to an apparent horizon.  
The duration of this transitory regime  grows with smaller $|\beta-\beta_c|$.  
After the collapse,  the harmonic driving keeps injecting energy continuously into the system, which increases its mass monotonically. 
Representative examples of the behavior just described can be found in Fig.\,\ref{non_adiabatic}. 

A pertinent question that cannot be definitely answered with our numerical methods is whether the system reaches an infinite mass state 
in finite or infinite field theory time.\footnote{We cannot discard that the growing black hole finally equilibrates into a steady state of finite entropy, 
although it seems intuitively unlikely, giving that we are considering a system with an infinite number of d.o.f. 
It would be nice to have some sharp physical arguments regarding this point.}
 A careful analysis of this post-collapse regime should connect with the findings in \cite{rangamani2015driven}. 
We have indeed pinned down different regimes for the growth of the one-point functions and refer the interested reader 
to Section \ref{postcol}.
\begin{figure}[h!]
\begin{center}
\includegraphics[width=16cm]{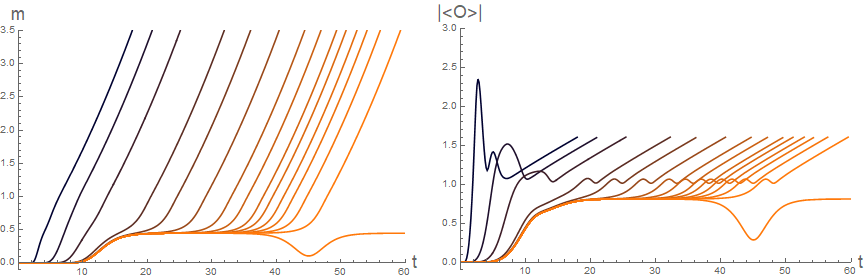}
\end{center}
\caption{\label{non_adiabatic} \small Energy density (left) and absolute value of the vev (right) for a family of build-up processes of 
the harmonic driving with $\omega_b = 2$, $\epsilon =0.09$. $\beta$ increases from $\beta = 5$ to $\beta = 21.222683$ from left to right. 
The non-collapsing simulation corresponds to $\beta = 21.222690 > \beta_c$.}
\end{figure}  

\end{itemize}

\subsubsection*{The unstable attractor}

We have clearly demonstrated that, for non-quasistatic build-up processes, the major property is the existence of the time scale $\beta_c$, 
which separates two different phases during the driving regime. These phases are distinguished by the final fate of the solution. 
The new time scale seems to be related to the existence of an intermediate time attractor, and we mentioned that this attractor was nothing 
but the unstable SPS associated to $\omega_b$ and $\epsilon$. In the following discussion we provide evidence in favor of this statement. 

\begin{figure}[h!]
\begin{center}
\includegraphics[width=16cm]{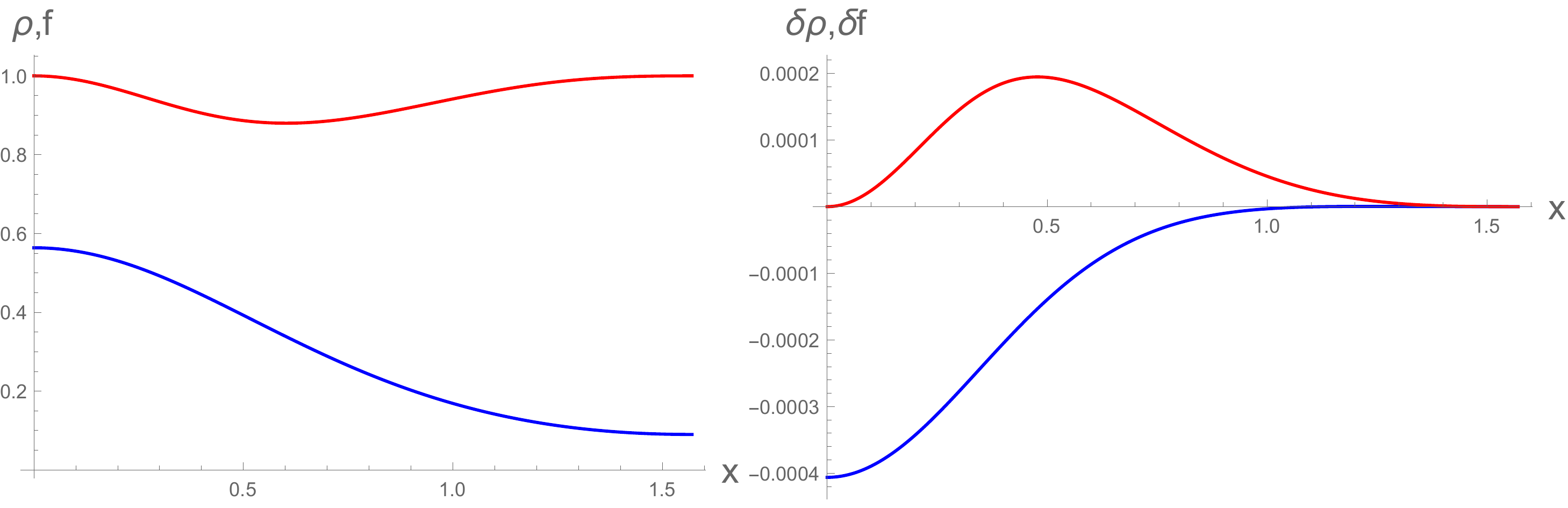}
\end{center}
\caption{\label{attractor1} \small Left: $\rho(t_0, x)$ (blue) and $f(t_0, x)$ (red), as obtained from a numerical simulation 
with $\omega=2$, $\epsilon = 0.09$, $\beta = 21.222683$ at $t_0 = 26.075$. Right: difference of the functions on the left plot 
with respect to their values on the unstable SPS, $(\delta \rho)(t_0,x)$ (blue) and $(\delta f)(t_0, x)$ (red).}
\end{figure}

For concreteness, let us consider the $\beta = 21.222683 < \beta_c$ case. In Fig.\,\ref{attractor1}a, we plot $\rho(t_0, x)$, $f(t_0, x)$, as obtained from the numerical simulation at $t_0 = 26.075$. In Fig.\,\ref{attractor1}b, these functions are compared with their values on the unstable SPS, $\rho_{SPS}(x)$, $f_{SPS}(x)$, where we plot the differences
\beqa
(\delta \rho)(t,x) = \rho(t,x) - \rho_{SPS}(x),~~~~~~(\delta f)(t,x) = f(t,x) - f_{SPS}(x), \label{df}
\eeqa
at $t_0$. Clearly, these differences are small.

\begin{figure}[h!]
\begin{center}
\includegraphics[width=10cm]{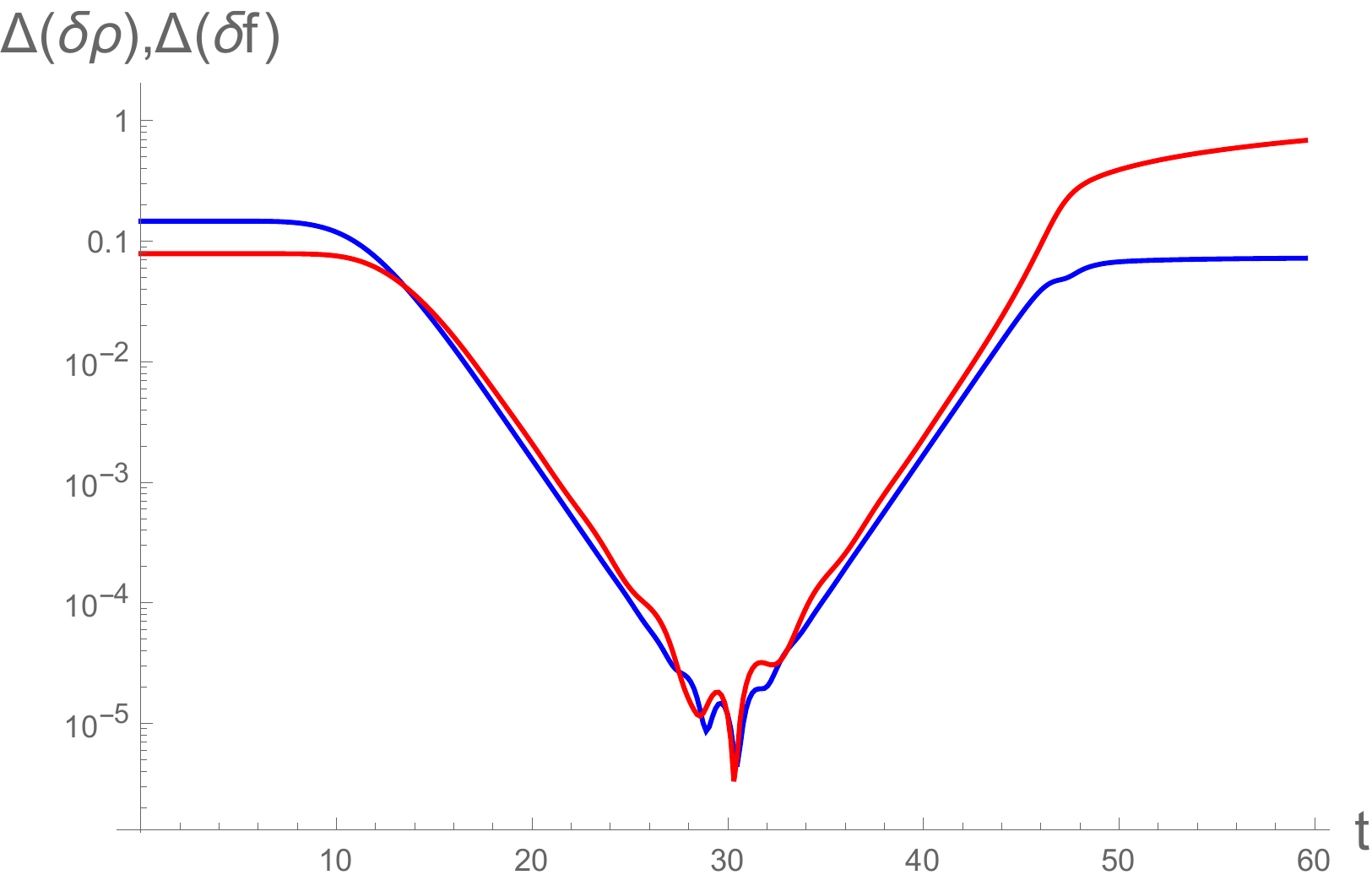}
\end{center}
\caption{\label{attractor2} \small Time evolution of $\Delta (\delta \rho)$ (blue) and $\Delta (\delta f)$ (red) for $\omega=2$, $\epsilon = 0.09$ and $\beta = 21.222683$.}
\end{figure}

In order to gain further understanding, let us compute the time evolution of the norms of $\delta \rho$ and $\delta f$, 
defined by\footnote{$\rho(t,x)$ is not normalizable when the source is nonzero; we define its norm as the norm of its normalizable part.}
\beqa
&&\Delta (\delta \rho) = \left(\int_0^{\frac{\pi}{2}}\tan^2(x) (\delta \rho - \delta \rho|_{x=\pi/2})^2 \right)^{\frac{1}{2}}, \label{drho}\\
&&\Delta (\delta f) = \left(\int_0^{\frac{\pi}{2}} \tan^2(x) (\delta f)^2 \right)^{\frac{1}{2}}. \label{df}
\eeqa
We plot these quantities in Fig.\,\ref{attractor2}. It is clearly seen that there is a time window in which the distance 
between the fields in the driving phase and the unstable SPS is negligible. 

In Section \ref{type1}, we employ the identification of the intermediate attractor and the unstable SPS to argue that the results presented 
in this section can be naturally understood as a {\it dynamical} type I phase transition.

\section{Type I phase transitions in driven AdS/CFT}
\label{type1}

Our results bear some resemblance to critical phenomena in asymptotically flat gravitational collapse. 
In both cases, there exists an intermediate attractor with one unstable mode that separates two qualitatively different dynamical regimes. 
In the flat case, they are gravitational collapse versus dispersion to asymptotic infinity; in our case, they are the flow to the infinite energy 
phase or the establishment of a regime of persistent, exactly periodic oscillations in the system.

There are also apparent differences between our findings and the type II phase transition uncovered by Choptuik for a massless scalar 
field \cite{Choptuik:1992jv}; actually, our results are akin to the type I phase transitions studied in asymptotically flat space  \cite{Choptuik:1996yg,Bizon:1998qd,Hawley:2000dt}. 
In the first of these works, the authors considered a $SU(2)$ non-Abelian gauge field minimally coupled to gravity. 
There, for certain one-parameter families of initial data, a static soliton - the Bartnik-Mckinnon soliton-\footnote{There is actually an 
infinite countable family of them, we are referring to the one of lowest mass.} played the role of an intermediate time unstable attractor. 
The existence of this attractor entailed that, for supercritical data and as the critical surface was approached, the mass of the black hole 
formed did not display Choptuik scaling: there was a mass gap, set precisely by the soliton mass. 

As summarized in \cite{Bizon:1998qd}, there are two model-independent assumptions that must be satisfied in order to have a first-order phase 
transition in spherically symmetric gravitational collapse, which we state here for completeness: 
\begin{itemize}
\item Assumption {\it a}. {\it There exists a static, regular solution with only one unstable eigenmode}. 
Let $\Psi^{u}(x)$ denote a generic field of this unstable geometry. Having only one unstable eigenmode, the 
evolution of any small spherically symmetric fluctuation around $\Psi^u(x)$  can be decomposed as 
\beq
\delta \Psi^u(t,x) = \Psi(t,x) - \Psi^u(x) = \alpha e^{\lambda t} \chi_\lambda(x) + \sum_{n>1} \alpha_n e^{\lambda_n t} \chi_n(x),   \label{linearization}
\eeq 
where $\lambda \in \mathbb R^+$ is the eigenvalue associated to the unstable eigenmode. Modes with $ n > 1$ have $\textrm{Re}\, \lambda_n \leq 0$. 

\item Assumption {\it b}. {\it The final fate of the perturbation $\delta \Psi^u(t,x)$ depends solely on the sign of $\alpha$}. 
For one sign, a black hole forms; for the other, gravitational collapse does not take place.\footnote{The final state 
depends on the details of the Einstein-matter action.} 
\end{itemize}
As pointed out in \cite{Bizon:1998qd}, assumption $a$ means that the stable manifold $W_S$ of the solution $\Psi^u$ has codimension one. 
Assumption $b$ entails that the stable manifold is a critical surface that divides the phase space into collapsing and non-collapsing initial data. 

Consider now a one-parameter family of initial data $\Psi_0(x;p)$ that crosses the critical surface $W_S$ at $p=p^*$. 
$\Psi_0(x;p^*)$ flows to $\Psi^u(x)$ along $W_S$. 
By continuity, initial data $\Psi_0(x;p)$ with sufficiently small $\left|p-p^*\right|$ have a time development $\Psi(t,x;p)$ that remains sufficiently 
close to $W_S$ so as that, after some time $t_1$, the linearization \eqref{linearization} holds. 
The precise numerical values of the coefficients $\alpha, \alpha_n$ are set by the initial data: $\alpha = \alpha(p), \alpha_n = \alpha_n(p)$. 

Given that by definition $\alpha(p^*) = 0$, we must have that 
\beq
\alpha(p) = a (p - p^*) + \ldots 
\eeq
and, as a consequence 
\beq
\Psi(t,x;p) = \Psi^u(x) + a (p - p^*) e^{\lambda (t - t_1)} \chi_\lambda(x) + \ldots
\eeq
The above linearization is valid up to some time $t_2 > t_1$ at which the unstable eigenmode becomes dominant and the solution is repelled 
away from the critical surface. Therefore, at $t = t_2$ we have that
\beq
(p-p^*)e^{\lambda \Delta t} = O(1), 
\eeq
where we have defined $\Delta t = t_2 - t_1$. 
This last condition forces $\Delta t$ to scale as 
\beq
\Delta t \sim -\frac{1}{\lambda}\log\left|p - p^*\right|. \label{dt_scaling}
\eeq
In Section \ref{quench}, we have analyzed a one-parameter family of build-up processes with different time spans $\beta$, 
but the same final harmonic driving, characterized by $\rho_b = 0.09$ and $\omega_b = 2$. 
We found out that there exists a critical $\beta_c$ that divides the post-quench response of the system into two different regimes,
which correspond to gravitational collapse ($\beta < \beta_c$) or the establishment of persistent and exactly periodic oscillations in 
the system ($\beta > \beta_c$).  Furthermore, we demonstrated that, as $\beta \to \beta_c$, at intermediate times, the system is 
attracted to a SPS with just one unstable eigenmode. 
This observation, together with the fact that the nonlinear evolution of a perturbed, unstable SPS falls generically into two different dynamical 
regimes depending solely on the sign of the perturbation (as shown in Section \ref{comp}), implies that assumptions $a$ and $b$ apply 
naturally to the problem we are considering. As a consequence, we expect the scaling \eqref{dt_scaling} to hold, where now $p$, $p^*$ 
correspond to $\beta, \beta_c$. In particular, the lowest eigenfrequency of the unstable SPS, $\lambda$, has to be extractable from our numerical simulations. 

\begin{figure}[h!]
\begin{center}
\includegraphics[width=10cm]{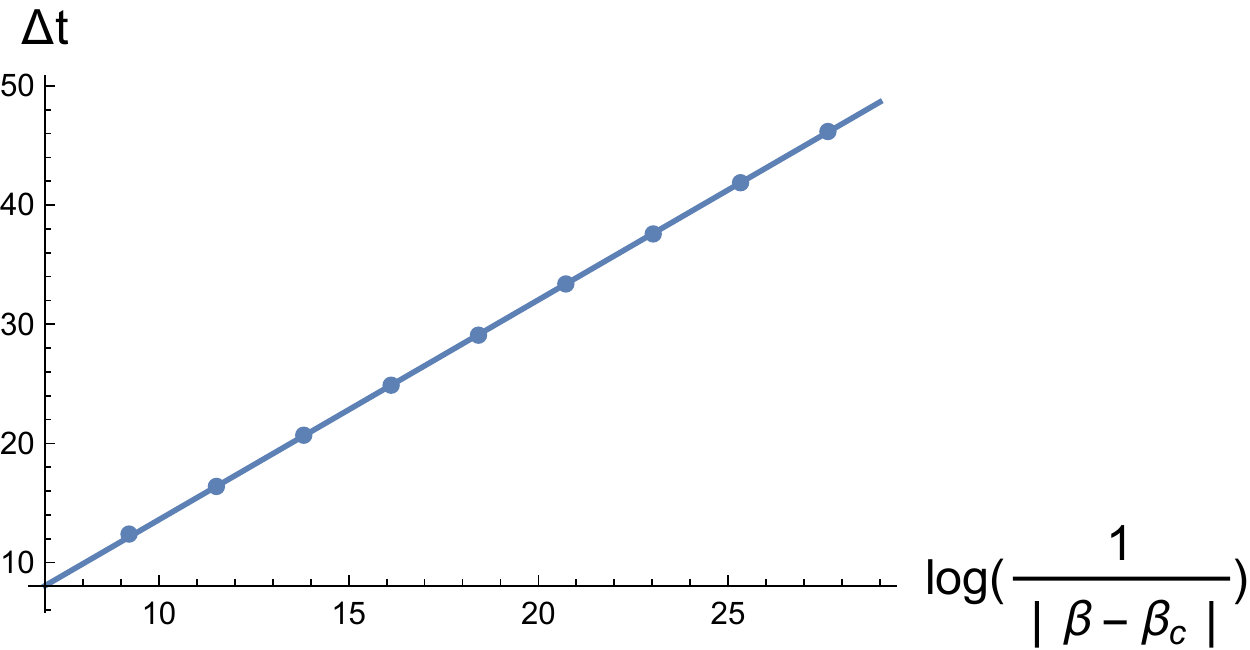}
\end{center}
\caption{\label{fig:scaling} \small Scaling of the permanence time $\Delta t$ along the unstable SPS with $\omega_b = 2$. }
\end{figure}

In Fig.\,\ref{fig:scaling} we plot $\Delta t$ versus $|\log(\beta - \beta_c)|$ for a series of quenches with different 
values of $\beta >\beta_c$. The linear fit yields a value of $\lambda = 0.542$,  matching the value of $\lambda$ we 
have obtained independently by means of the numerical solution to the eigenmode equations. 

\section{Dynamical construction of a boson star}
\label{BSlego}

We are starting to get operational control over the phase diagram of periodic solutions. Armed with a quench protocol like \eqref{source_profile}
we are able to target both at SPS's as well as at SMS's by appropriately tuning the parameters $\omega_b, \epsilon$ and $\beta$. 
In the cases we have analyzed so far, the end result is some sourced periodic solution, whether modulated or not. 
An intriguing legitimate question is if we can engineer a protocol such that the end solution is unsourced, yet not AdS but, instead, a boson star. 

Boson stars are dual to excited QFT states that, despite being unsourced, exhibit a periodic one-point 
function.\footnote{The boson star can be envisaged as circular polarized solution for a pair of real scalar 
fields rather than a complex field. The one-point function of each of these real scalar fields varies harmonically in time. 
A natural extension of the model would include a gauge field with which one could gauge away the phase rotation by  
a gauge transformation linear in time. This is a different scenario which also deserves a thorough study. 
However, the single real scalar field will also exhibit time periodic unsourced solutions. 
In this case no gauge transformation is available to gauge the periodic motion away.}
It is more than tempting to try to establish a link between this type of solutions to the so called time crystals.
Strictly speaking, these last systems are postulated to be true ground states, yet to break 
time translation invariance spontaneously \cite{Wilczek:2012jt}. However, their existence as ground 
states has been ruled out in \cite{Watanabe:2014hea}.\footnote{The no-go theorem \cite{Wilczek:2012jt} refers to an infinite-volume thermodynamic limit. 
As we comment later, global AdS holography relies on a different large $N$ thermodynamic limit}
The no-go theorem leaves open the possibility of finding spontaneous 
time shift symmetry breaking in excited states (see \cite{Sacha:2017fqe} for a review with references). 

In parallel to what just described, there is another possible connection to time crystals.
Due to interactions, upon a small periodic driving certain systems can respond sub-harmonically; 
they are dubbed discrete time crystals or Floquet time crystal. Such proposal has been realized experimentally in spin systems in which the response beats 
at twice the driving periodicity \cite{else2016floquet,khemani2016phase,von2016phase}.



We only take these as inspiring considerations trying to highlight similarities and differences. 
As a major target, we try to devise a quench protocol that connects continuously vacuum-AdS$_4$  with a boson star solution. 
The most naive guess immediately faces a seemingly unavoidable obstruction. 
Imagine, for concreteness, drawing a straight vertical line at $\omega_b = 2.5$ that connects AdS with the BS curve 
(i.e., lying within region I in Fig.\,\ref{fig:phibsections}).
Physically, this corresponds to cranking up, from AdS, a periodic driving source with rising 
module, $\dot \rho_b(t)>0 $, at a constant frequency $\omega_b$.  
If performed quasistatically, $\beta \to \infty$, the quench protocol of the previous section moves 
the state along the vertical segment drawn in the figure. As explained before, this can only proceed up 
to the point where the line exits the gray (stability) region. 
If once there one continues increasing the source amplitude, unavoidably will exit the surface and face collapse and thermalization. 
If one  starts decreasing quasistatically  the source amplitude,  the system will proceed backward through the same states 
until AdS is again recovered.  
It is impossible to  slide back along the  vertical line in the white (unstable) sector down to zero $\rho_b$.  
Bosons stars look unreachable from this side of the phase diagram by means of a quasistatic process. 
We will provide two ways to get around this caveat. 
Needless to say, one of the assumptions made in the previous argument will have to be relaxed. 
For example, if one insists in using a quasistatic quench, one may still reach the BS curve from region II  
without crossing any instability curve. 
A remarkable possibility arises by employing, instead, non-quasistatic processes. 
This takes us out of the surface of SPS's, but an unexpected attractor will bring us back to it.

\subsection{Quasistatic method}
\label{sec:Quasistatic_method}
In this section we explicitly construct a boson star, starting from AdS, via a quasistatic quench.  
As advertised above, the clue to this procedure comes from drawing a quasistatic path on the space of 
SPS's across region II, always staying within the region of linear stability. 
It is fortunate that boson stars lie at the boundary of such region. 
The advantage of this method is that we control the frequency of the final boson star. 
We must start by extend the quenches considered in the previous sections, and allow now for time dependence both on the modulus and phase of the driving
\beq
\phi(t,\pi/2) = \phi_b(t) =  \epsilon(t) e^{i \omega_b(t) t}, 
\eeq
where 
\beqa
\epsilon(t) = \frac{1}{2}\epsilon_m\left(1-\tanh\left(\frac{\beta}{t} + \frac{\beta}{t-\beta} \right) \right),~~~~~0\leq t<\beta, \nonumber \\
\epsilon(t) = \frac{1}{2}\epsilon_m\left(1 + \tanh\left(\frac{\beta}{t-\beta} + \frac{\beta}{t-2\beta} \right) \right),~~~~~\beta\leq t<2 \beta 
\label{source_profile_epsilon} \\
\omega_b(t) = \omega_i + \frac{1}{2} (\omega_f - \omega_i) \left(1-\tanh\left(\frac{2\beta}{t} + \frac{2\beta}{t-2\beta} \right) \right), ~~~~~0\leq t<2\beta  \, .\label{source_profile_omega}
\eeqa
Note that, although  $\omega_b(t)$ has the quench shape of Fig.\,\ref{quench_profile} 
(interpolating between $\omega_i$ and $\omega_f$),  the actual instantaneous frequency of 
the source is given rather by the time derivative of its 
phase, $\omega_{ins}(t) = d/dt (\omega_b(t) t) = \omega_b'(t) t + \omega_b(t)$, which has the same asymptotic values as $\omega_b(t)$. 
This quench process starts from the global AdS$_4$ vacuum at $t = 0$; 
we expect it to land on the stable boson star with $\omega_b = \omega_f$ at $t = 2 \beta$ when $\beta$ is sufficiently 
large (modulo small corrections that vanish in the $\beta \to \infty$ limit). 

\begin{figure}[h!]
\begin{center}
\includegraphics[width=16cm]{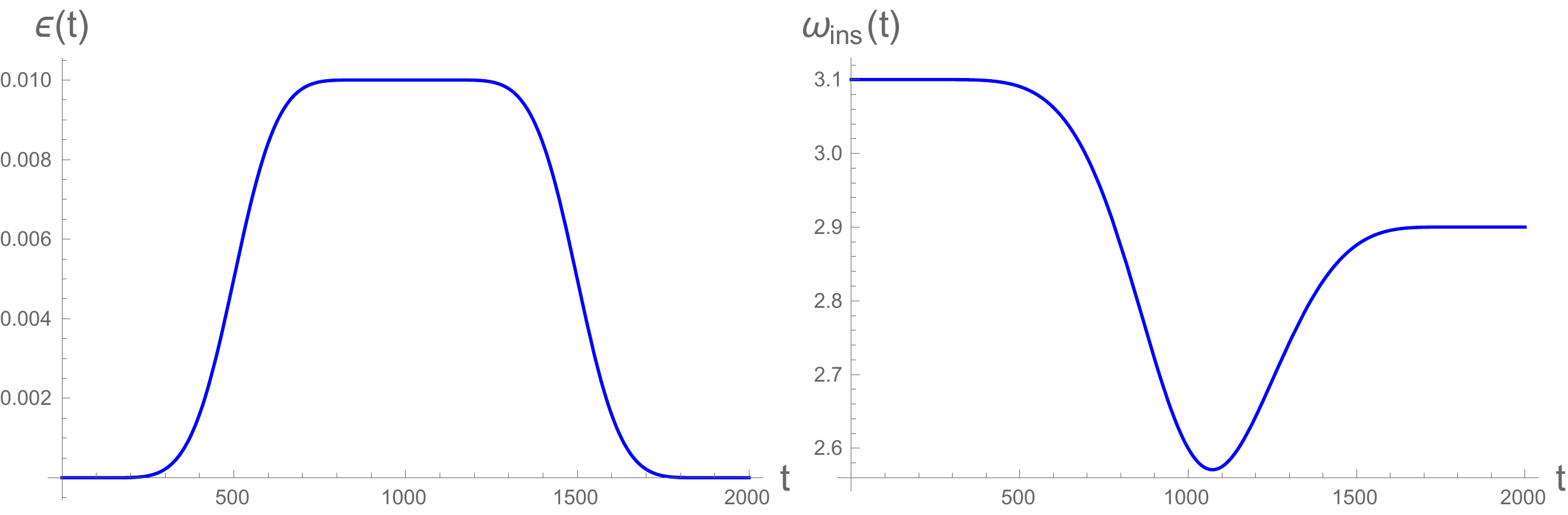}
\end{center}
\caption{\label{source_plot} \small The functions $\epsilon(t)$ and $\omega_{ins}(t)$ employed to build a boson star starting from AdS.}
\end{figure}

As an example, we consider \eqref{source_profile_epsilon}, \eqref{source_profile_omega} 
with $\omega_i = 3.1$, $\omega_f = 2.9$, $\epsilon_m = 0.01$ and $\beta = 1000$. 
See Fig.\,\ref{source_plot} for a plot of $\epsilon(t)$ and $\omega_{ins}(t)$ in this case. 

\begin{figure}[h!]
\begin{center}
\includegraphics[width=16cm]{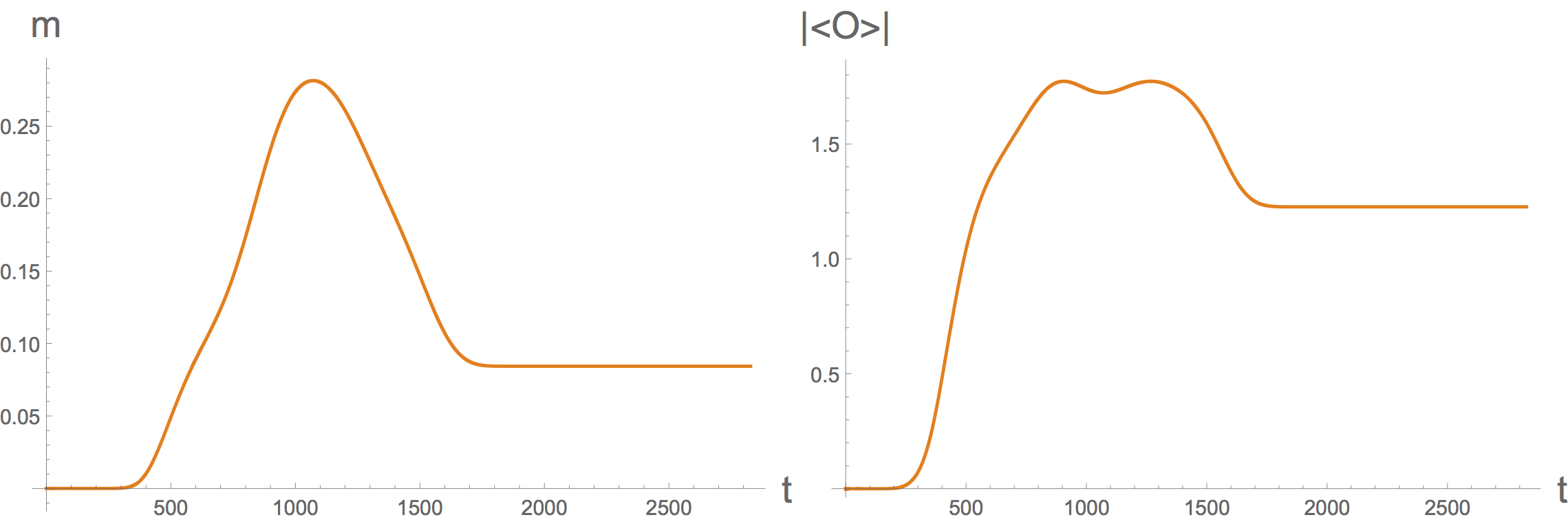}
\end{center}
\caption{\label{bs_made} \small $m$ and $\left| \left<\mathcal O \right> \right|$ for the quench process described in the main text. 
After the quench, both functions (approximately) relax to their values in the boson star solution corresponding to $\omega_f = 2.9$.}
\end{figure}

\begin{figure}[h!]
\begin{center}
\includegraphics[width=16cm]{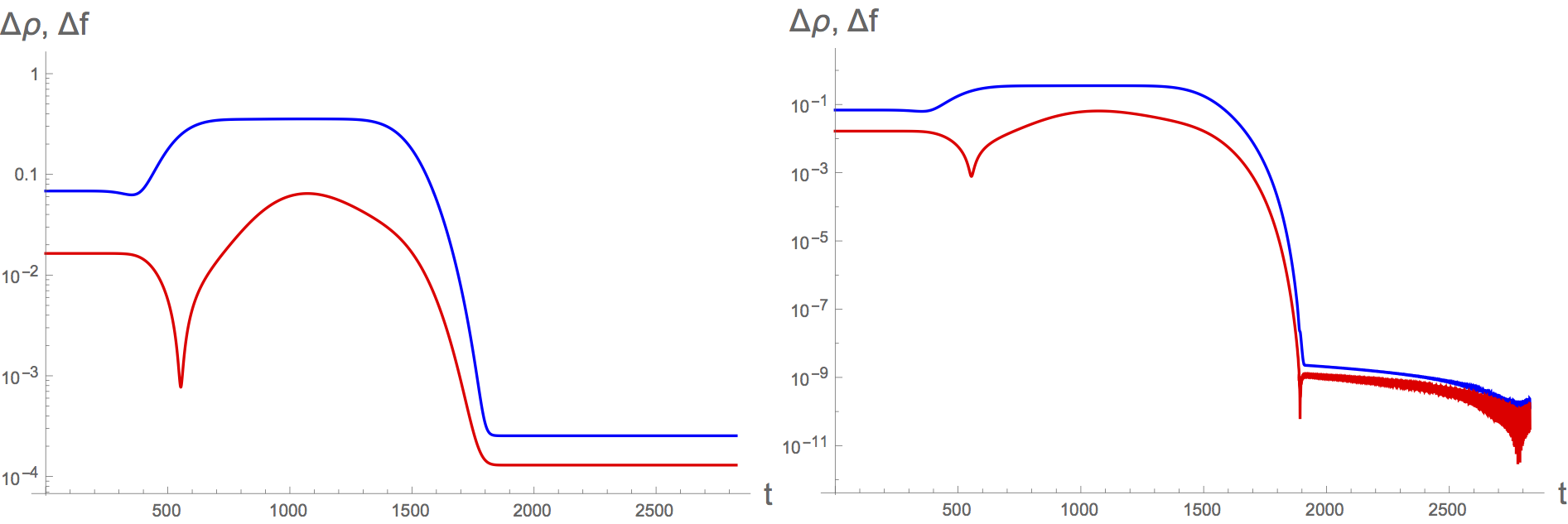}
\end{center}
\caption{\label{bs_made_check}  \small Time evolution of $\Delta \rho$ (blue) and $\Delta f$ (red) for the quench process described in the main text. Left: we compare the simulation results to the boson star at the target frequency $\omega_f = 2.9$. Right:  we compare the simulation results to the boson star at the actual frequency  $\omega_{BS} = 2.8992$.}
\end{figure}

In Fig.\,\ref{bs_made}, we plot the time evolution of $m$ and $\left| \left< \mathcal O\right> \right|$. 
It is clearly seen that they relax to (almost) time-independent values after the quench. 
In particular, from the numerical simulation, we have that 
\beq
m(t > 2 \beta) = 0.0843513,~~~~~~\left| \left< \mathcal O(t>2\beta) \right> \right| = 1.22637. 
\eeq
On the other hand, by solving the ODE system that determines the boson star solution with $\omega_f = 2.9$, we obtain
\beq
m_{BS} = 0.0837265,~~~~~~\left| \left< \mathcal O \right> \right|_{BS} = 1.22247. 
\eeq
The agreement between them is rather good, in particular the relative error between the values 
obtained from the numerical simulation and the solution of the ODE system is below $1 \%$.
Another useful quantities to monitor the difference between the end product of the quench and the 
stable boson star are $\Delta \rho$, $\Delta f$ -cf. equations \eqref{drho},\eqref{df}-. We plot them in Fig.\,\ref{bs_made_check}a. 

By looking at Fig.\,\ref{bs_made_check}a, we clearly see that, although small, 
the discrepancy between the end state of the quasistatic quench process and the target boson star is not negligible. 
It is natural to wonder about the origin of this difference. Is the boson star oscillating at a frequency different than 
the target frequency? Does this difference vanish when $\beta \to \infty$?

Let us address the first issue. In our example we have that, for $t > 2 \beta$, $\rho(t,0) = 0.230379$. 
The boson star associated to this number has $m_{BS} = 0.0843514$ and $\left| \left< \mathcal O \right> \right|_{BS} = 1.22635$. 
As we see, these quantities agree with the results of the numerical simulation in five significant digits. 
On the other hand, the frequency of this boson star is $\omega_{BS} = 2.8992 < \omega_f = 2.9$. 
The simulated solution and the actual boson star the system equilibrates to are compared in Fig.\,\ref{bs_made_check}b. 
In this case, after the quench both $\Delta \rho$ and $\Delta  f$ are compatible with the numerical noise. 

The discussion in the previous paragraph shows that the frequency of the boson star obtained after the slow quench does 
not need to agree with the target frequency. We expect that the difference between both frequencies vanishes in the 
strict $\beta = \infty$ limit, i.e., for a  quasistatic quench. 
Equivalently, we expect that, when we employ the target boson star to perform the comparison, $\Delta \rho, \Delta f \to 0$ as $\beta \to \infty$. 

\begin{figure}[h!]
\begin{center}
\includegraphics[width=8cm]{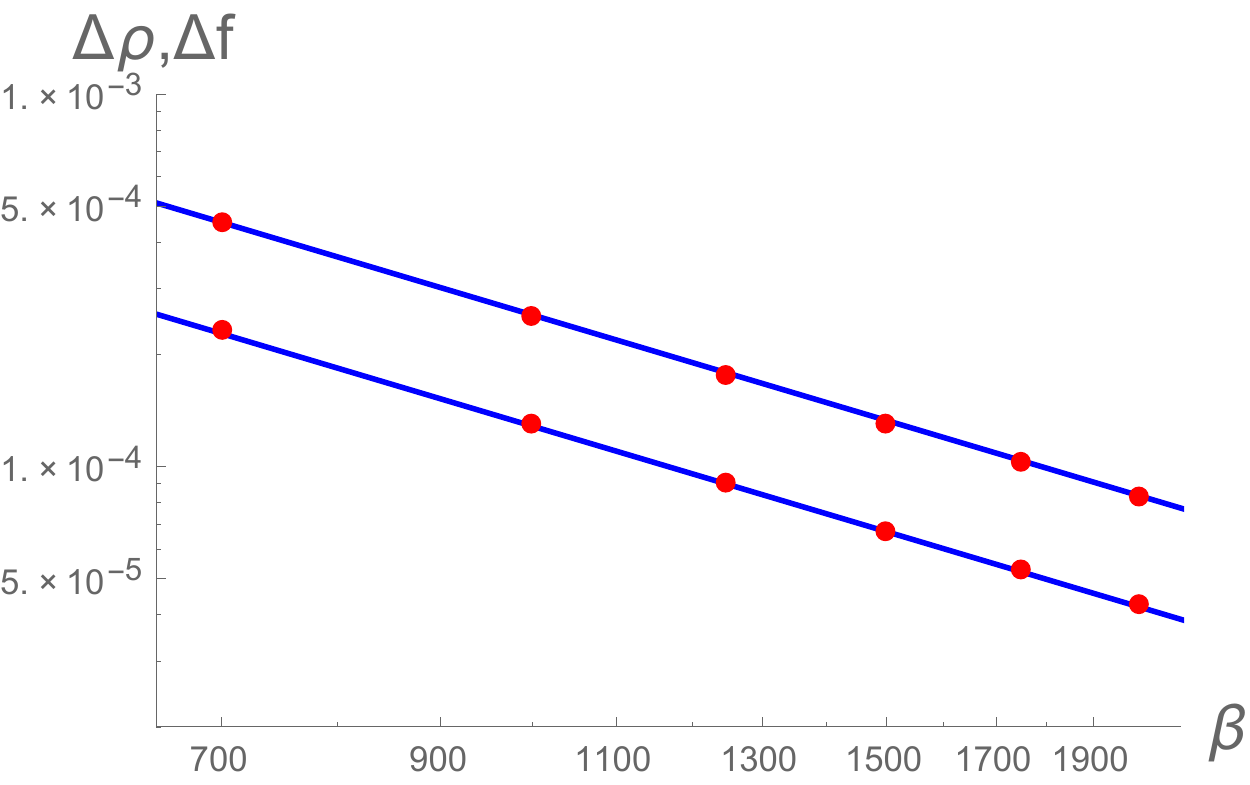}
\end{center}
\caption{\label{bs_beta_scaling} \small $\Delta \rho$ (upper curve) and $\Delta f$ (lower curve) vs $\beta$ for the 
boson star building processes described in the main text. The blue lines correspond to the fits $\Delta \rho$, $\Delta f \sim \beta^{-1.6079}$.}
\end{figure}

In Fig.\,\ref{bs_beta_scaling}, we plot $\Delta \rho$ and $\Delta f$ for the quench 
profile \eqref{source_profile_epsilon},\eqref{source_profile_omega} with $\omega_i = 3.1$, $\omega_f = 2.9$, $\epsilon_m = 0.01$ and 
varying $\beta$. We consider $\beta = 700, 1000, 1250, 1500, 1750$ and $2000$. 
It is clearly seen that $\Delta \rho$, $\Delta f$ decrease with increasing $\beta$. In particular, a log-log fit of these data gives
\beq
\Delta \rho, \Delta f \sim \beta^{-1.6079}. 
\eeq
This behavior makes plausible to assume that in the strict quasistatic limit we will recover the target boson star. 

\subsection{Non-quasistatic method}

As advertised before, we can also reach BS's with a quench that starts from region I but, in turn, we must  give up quasistaticity. 
The crucial observation is that, if instead of using a large value of $\beta$, we let $\beta\to \beta_c$, we reach a SMS associated to an unstable SPS. 
Once in the SMS, which is stable by definition, we may slowly turn off the source amplitude. 
If the end result is a stationary solution that corresponds neither to a BH nor to AdS, the only remaining option is a BS. 
We shall provide numerical evidence that this is indeed the case.

\begin{figure}[h!]
\begin{center}
\includegraphics[width=17cm]{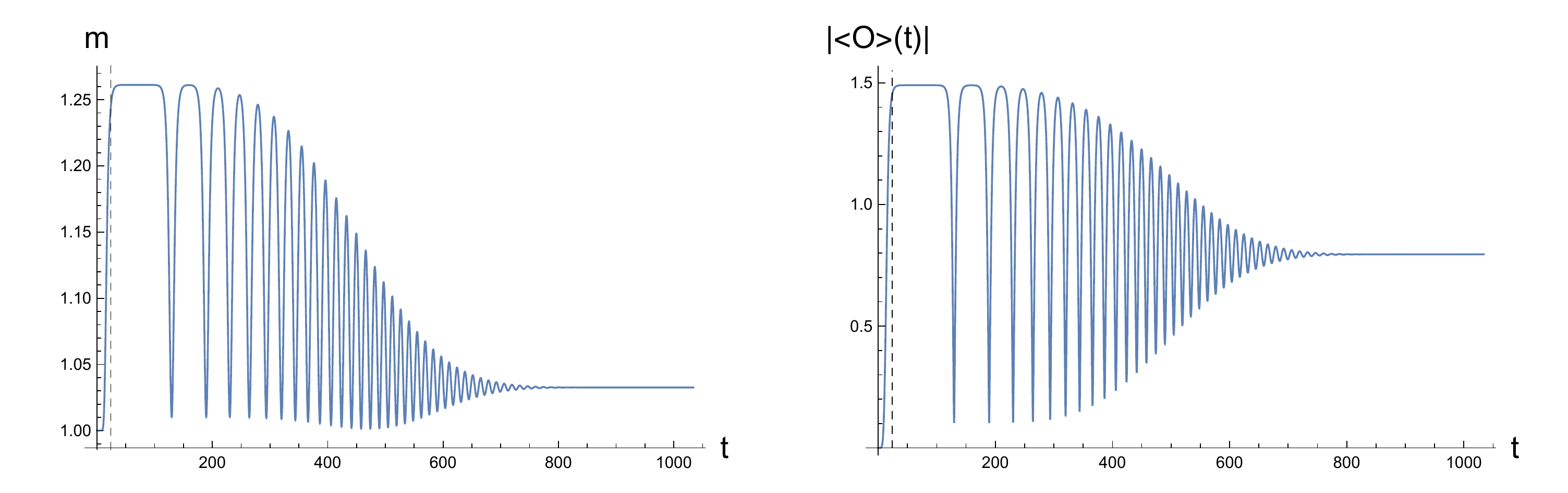}
\end{center}
\caption{\label{bs2_nonquasi} \small A close-to-critical quench with build up time 
scale $\beta_{in} =23.811857278$ (dashed line) brings AdS close to the unstable attractor, where is stays for a long time before decaying to the modulated SMS. The initial time 
given by the dashed vertical line. 
A quasistatic quench then turns the driving amplitude very slowly off $\beta_{out} = 1000$ and leaves a very slightly perturbed BS at the end.}
\end{figure}

We construct the SMS by the procedure described in section \ref{typeItransition}. We work with the source profile  \eqref{source_profile} with $\beta = \beta_{in}\gtrsim \beta_c $,  and wait until a time $t_{out} \geq \beta_{in}$ after build up. Once the SMS has formed, we start to turn off the source amplitude, reaching zero at $t = t_{out} + \beta_{out}$. The extinction quench follows the time reversed profile with, now, a larger time scale $\beta_{out}$
\beqa
\phi(t,\pi/2) &=& \frac{1}{2}\epsilon\left(1+\tanh\left(\frac{\beta_{out}}{t-t_{out}} + \frac{\beta_{out}}{t-t_{out}-\beta_{out}} \right) \right) e^{i \omega_b t},~~~~~t \geq t_{out}, t < t_{out}+\beta_{out}, \nonumber \\
\phi(t,\pi/2) &=& 0.~~~~~~~~~~~~~~~~~~~~~~~~~~~~~~~~~~~~~~~~~~~~~~~~~~~~~~~~~~~~~~~~~~t \geq t_{out} + \beta_{out}  \, .\label{source_profile_out}
\eeqa
As an example, in Fig.\,\ref{bs2_nonquasi} we have considered a build-up phase characterized 
by $\omega_b = 2.5$, $\epsilon = 0.09$ and a quench-in time $\beta_{in} = 23.811857278$, which is the 
closest to  $\beta_c$ that we have managed to approach from above. 
As shown in Fig.\,\ref{bs2_nonquasi}, the system first ``jumps" on top of the unstable SPS attractor, 
where is stays for more than $100$ seconds, until it falls onto the side of the modulated SMS, as expected. 
After the first  beating, we begin to turn off the source amplitude slowly, with a time scale $\beta_{out} = 1000$. 
The final, sourceless solution has a non-vanishing mass and frequency $\omega$ consistent with some BS. 

A very important difference between the setup at hand and the quasistatic method described in the previous subsection is that, 
in the present case, even for large $\beta_{out}$, the frequency of the reached BS, $\omega_{BS}$, has no obvious relation  
with the driving frequency,  $\omega_b$,  that has been used throughout the quench. 

A second very important point is that, presumably due to the violence of the build-up phase 
(i.e. the smallness of $\beta_{in}$), the system does not  return to a strict BS state after the source has been turned off:
the final geometry can be understood as the linear superposition of a BS and its first nontrivial eigenmode. 
Our numerical experiments point to the fact that this conclusion holds irrespectively of the value of $\beta_{out}$.
In fact, a stronger statement seems to be true: the final BS is actually independent of $\beta_{out}$ and there is a univocal relation between this BS and the 
SMS generating it; $\beta_{out}$ only controls the amplitude of a final oscillation about the final BS, which occurs 
with the frequency of its first nontrivial normal mode. An interesting open question is determining which property of the original SMS sets $\omega_{BS}$. 

In order to strengthen these statements above, we have considered a set of non-quasistatic quenches leading to a BS, 
now with $\omega_b=2$, $\beta_{in}=22$, and quench-out time scales $\beta_{out} = 1000,2000,3000, 4000,5000$. 
In each one of them, the post quench value of $\rho_o(t) = \rho(t,0) = |\phi(t,0)|$ oscillates around $\rho_{o,BS} = 0.092828$, 
which corresponds to a BS frequency $\omega_{BS} = 2.983133$, very different from the driving frequency, $\omega_b = 2$. 
For normal mode fluctuations around this BS, the first nontrivial eigenfrequency is $\lambda_1 = 1.975822$. 
According to our discussion so far, we expect that, after the quench out, the time evolution of the $\rho(t,x)$ drawn from the simulation is given by  
\beq
\rho(t,x) = \rho_{BS}(x) + \alpha \chi_1(x)\cos(\lambda_1 t + \delta) + e(t, x), \label{rho_analytic}
\eeq
where $\chi_1(x)$ is the spatial profile of the BS first nontrivial eigenmode, and $e(t,x)$ an error term that takes 
into account both linear contributions from higher eigenmodes as well as possible nonlinear corrections. 
The parameters $\alpha$ and $\delta$ are fixed by demanding that \eqref{rho_analytic} and its first time 
derivative (both without the error term) hold exactly at the origin at $t = t_{out}+\beta_{out}$.\footnote{Of course, this choice is not unique. 
Alternative ones do not modify the conclusions of the main text.}

\begin{figure}[h!]
\begin{center}
\includegraphics[width=17cm]{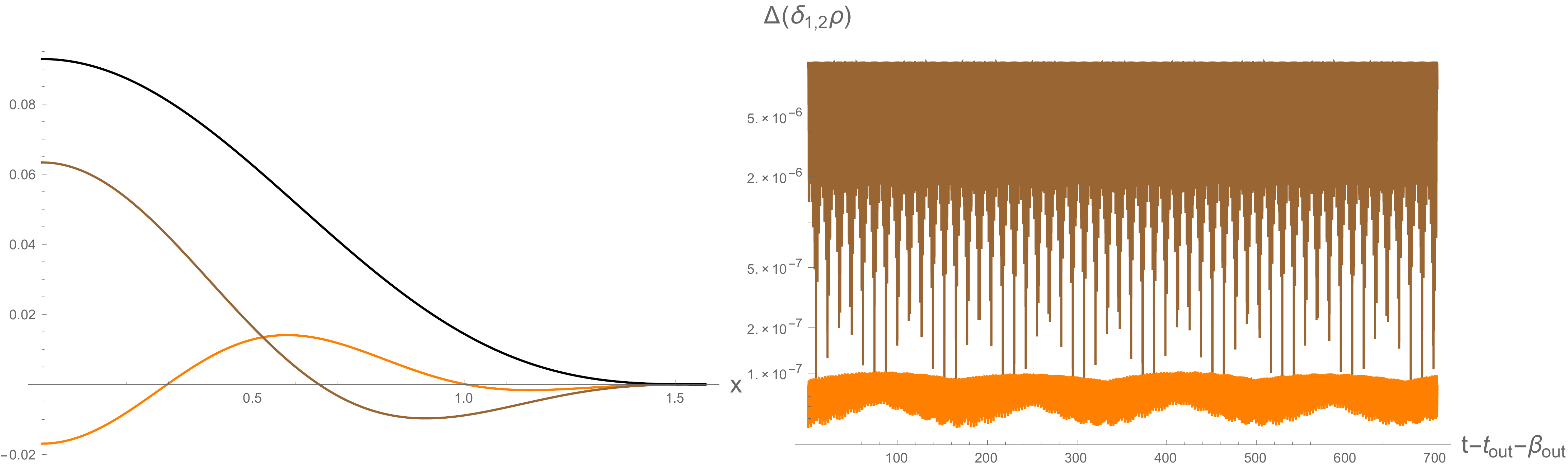}
\end{center}
\caption{\label{bs2_comparison} \small Left: $\rho$ (black), $10^3 \delta_1 \rho$ (brown) and $10^5 \delta_2 \rho$ (orange) 
for $\beta_{out} = 5000$ at a time $t = 600$ after the quench-out. The three magnitudes clearly display the hierarchy 
described in the main text. Right: $\Delta(\delta_1 \rho)$ (brown) and $\Delta(\delta_2 \rho)$ (orange) during the 
whole post-quench-out regime for $\beta_{out} = 5000$. We clearly see that the inequality $\Delta(\delta_2 \rho) \ll \Delta(\delta_1 \rho)$ holds.}
\end{figure}

We define 
\beqa
&&(\delta_1 \rho)(t,x) \equiv \rho(t,x) - \rho_{BS}(x), \\
&&(\delta_2 \rho)(t,x) \equiv \rho(t,x) - \rho_{BS}(x) - \alpha \chi_1(x) \cos(\lambda_1 t + \delta) = e(t,x). 
\eeqa
After the source has been turned off, we expect that $\Delta(\delta_1 \rho) \ll 1$ (i.e., we are close to the BS solution) and, 
furthermore, that $\Delta(\delta_2 \rho) \ll \Delta(\delta_1 \rho)$ (i.e., the leading order deviation with respect to the BS 
solution is controlled by its first nontrivial eigenmode). Both of these expectations hold, as the reader can convince 
herself by looking at Fig.\,\ref{bs2_comparison}. 

\section{The post-collapse regime}
\label{postcol}

Our numerical techniques allow us to follow the system after gravitational collapse has taken place, at least during some time. 
In the examples we have analyzed, we have not found traces of equilibration to a stationary regime: the harmonic driving always 
led to a monotonic increase of the energy density. This increase is responsible for a steady growth of the apparent horizon, 
which gets progressively closer to the boundary.\footnote{Our results indicate, but not demonstrate, that the boundary is not hit in finite time.} 
The way in which the energy density increases is not universal, and depends on the specific form of the harmonic driving. 

\begin{figure}[h!]
\begin{center}
\includegraphics[width=14cm]{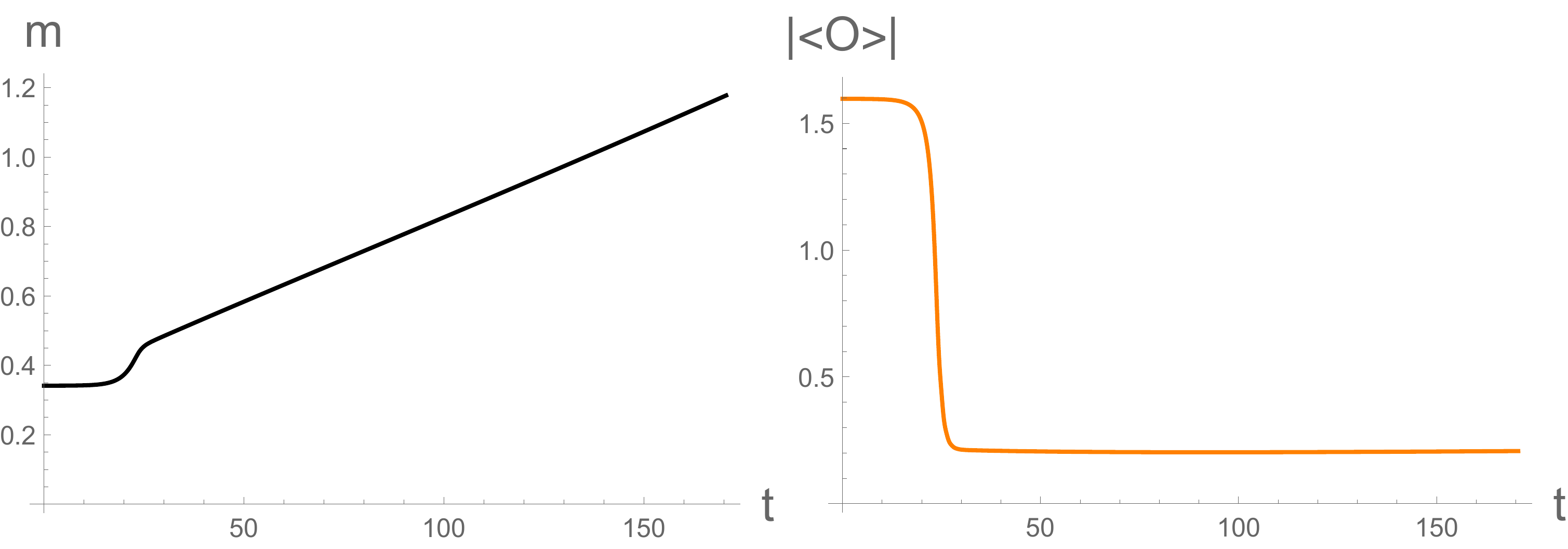}
\end{center}
\caption{\label{regimeI_mvev} \small Time evolution of the energy density (left) and vev modulus (right) of 
the SPS with $\rho_b = 0.01, \omega_b = 2.375$ when perturbed by its fundamental, unstable eigenmode with 
amplitude $\epsilon = 0.001$.} 
\end{figure}

\begin{figure}[h!]
\begin{center}
\includegraphics[width=14cm]{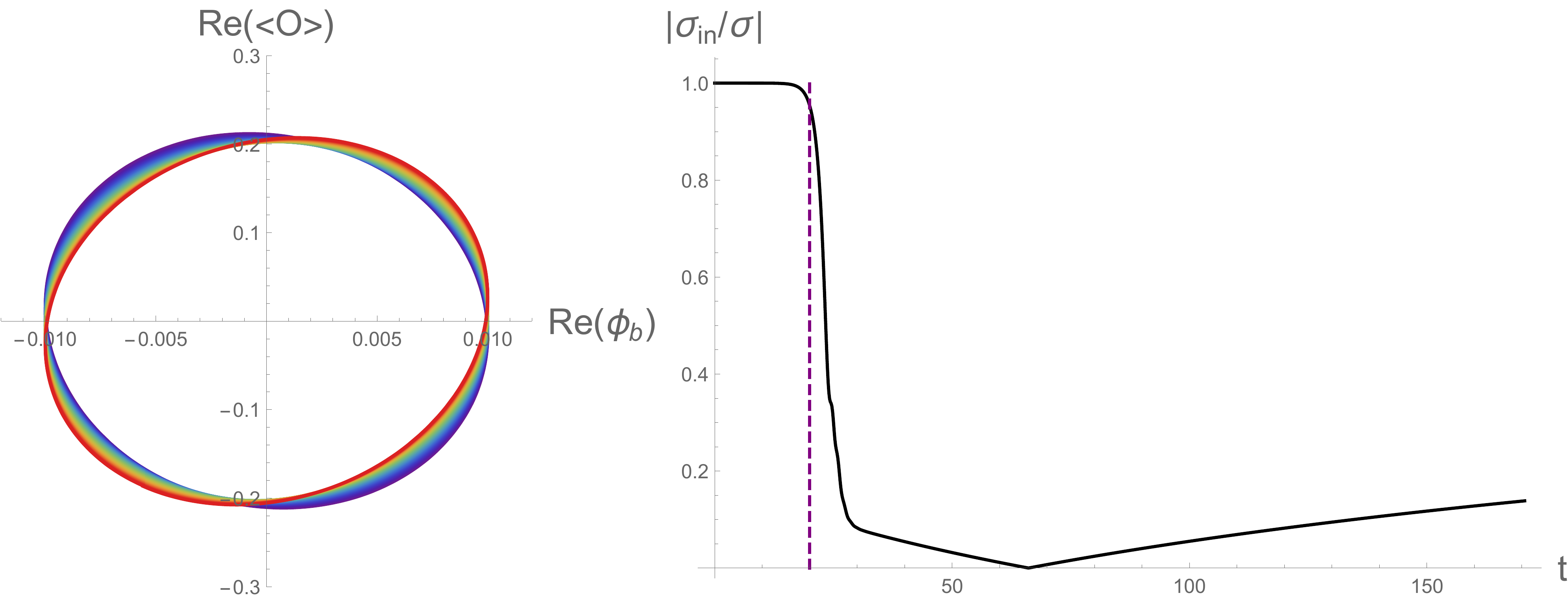}
\end{center}
\caption{\label{regimeII_phase} \small Left: Phase portrait of the time evolution of the SPS 
with $\rho_b = 0.01, \omega_b = 2.375$ when perturbed by its fundamental, unstable eigenmode 
with amplitude $\epsilon = 0.001$. We start at $t = 30$. Each period of the source has a different color, 
whose wavelength increases the later it starts. Right: Relative value of the imaginary part of nonlinear 
conductivity $\sigma$. A nonzero value indicates that the system response is partially in-phase with the driving.} 
\end{figure}

Let us discuss some examples in detail. The first one involves a perturbation of the unstable SPS with $\rho_b = 0.01, \omega_b = 2.375$ by its first, unstable eigenmode. In Fig.\,\ref{fig:phibsections} this
point is to be located to the right of the upper red dot in region I, on the isocurve with $\rho_b = 0.01$.
As we have illustrated in Section \ref{comp}, choosing the right sign of the perturbation leads right away to gravitational collapse. 
Following the post-collapse evolution of the system, we find the results depicted in Fig.\,\ref{regimeI_mvev}: 
$\left| \langle \mathcal O \rangle \right|$ stays approximately constant,\footnote{Our results show a 
small oscillation around a nonzero mean.} while the energy density increases in an approximately linear fashion, as prescribed by the differomorphism Ward identity given in eq. \ref{WardId}.

As explained in \cite{rangamani2015driven}, it is useful to portrait the system trajectory in the $(\phi_b, \left<\mathcal O\right>)$ 
plane in order to understand its response to the harmonic driving. We plot the real section of this complex curve in Fig.\,\ref{regimeII_phase}a. 
It is clearly observed that the trajectory remains bounded. In order to understand the phase alignment of the source and the response, 
it is convenient to introduce the nonlinear conductivity 
\beq
\sigma(t) \equiv \frac{1}{i \omega_b} \frac{\langle \mathcal O(t) \rangle}{\phi_b(t)} \equiv \sigma_{out}(t) + i \sigma_{in}(t). 
\eeq
We plot the relative contribution of $\sigma_{in}$ to the total conductivity in Fig.\,\ref{regimeII_phase}b. 
It is clearly observed that, while small, it typically has a non-negligible value. The boundedness of the response, 
together with the fact that it is not in complete phase opposition with respect to the source, leads to identify 
the post-collapse evolution of the system as belonging to the {\it dynamical crossover tilted regime} obtained in  \cite{rangamani2015driven}. 

\begin{figure}[h!]
\begin{center}
\includegraphics[width=14cm]{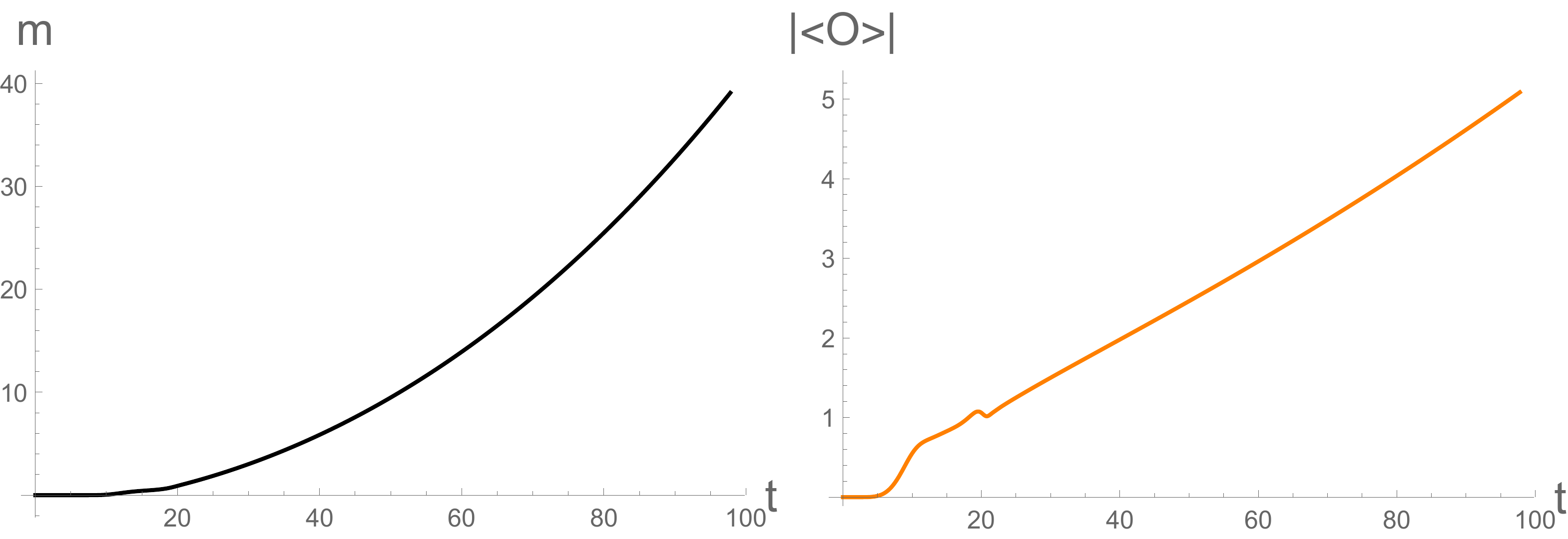}
\end{center}
\caption{\label{regimeIII_mvev} \small Time evolution of the energy density (left) and vev modulus (right) for a quench 
process of the form \eqref{source_profile} with $\epsilon = 0.09, \omega_b = 2$ and $\beta = 20$.}  
\end{figure}

\begin{figure}[h!]
\begin{center}
\includegraphics[width=14cm]{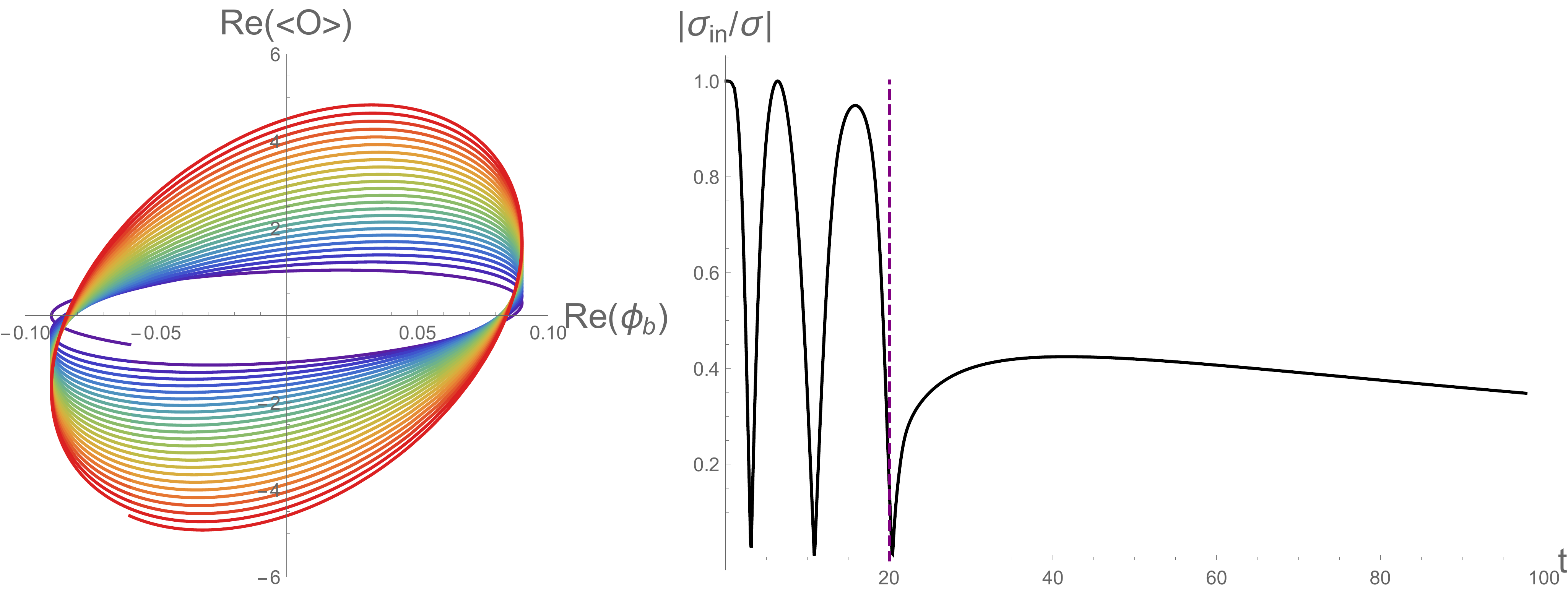}
\end{center}
\caption{\label{regimeIII_phase} \small Left: Phase portrait of the late-time dynamics of a quench process of 
the form \eqref{source_profile} with $\epsilon = 0.09, \omega_b = 2$ and $\beta = 20$. We start at $t = \beta$. 
Each period of the source has a different color, whose wavelength increases the later it starts. Right: Relative value 
of the imaginary part of the nonlinear conductivity $\sigma$. A nonzero value indicates that the system response is partially in-phase with the driving.} 
\end{figure}

The second example of post-collapse evolution we would like to discuss involves a SPS
with $\rho_b = 0.09, \omega_b = 2$, perturbed by its fundamental, unstable eigenmode. In Fig.\,\ref{fig:phibsections} this is precisely  the upper red dot  in sector I. Since this solution controls the late-time 
dynamics of the non-quasistatic quench processes we discussed in depth in Section \ref{quench}, we can instead focus on those. 
For definiteness, we take $\beta = 20$. In Fig.\,\ref{regimeIII_mvev}, we plot the time evolution of $m$ and $|\left<\mathcal O \right>|$, 
while in Fig.\,\ref{regimeIII_phase}a we depict the phase diagram (along the lines of Fig.\,\ref{regimeII_phase}a). 

It is clear that this late-time regime is qualitatively different from the previous one: $|\left<\mathcal O\right>|$ 
does not remain bounded and, as a consequence, $m$ increases faster than linearly. On the phase portrait, we observe a clear precession 
of the system's trajectory, which is accompanied by an increase in its amplitude. This precession is due to the fact that the response 
is not entirely out-of-phase with the source, as illustrated by Fig.\,\ref{regimeII_phase}b. This new late-time phase is best 
identified with the {\it unbounded amplification regime} found in \cite{rangamani2015driven},\footnote{In particular, compare 
our Fig.\,\ref{regimeII_phase} with with Fig.10a in \cite{rangamani2015driven}.} 
as the trademark of this regime is the unbounded and partially in-phase response to the harmonic source. 

\begin{figure}[h!]
\begin{center}
\includegraphics[width=14cm]{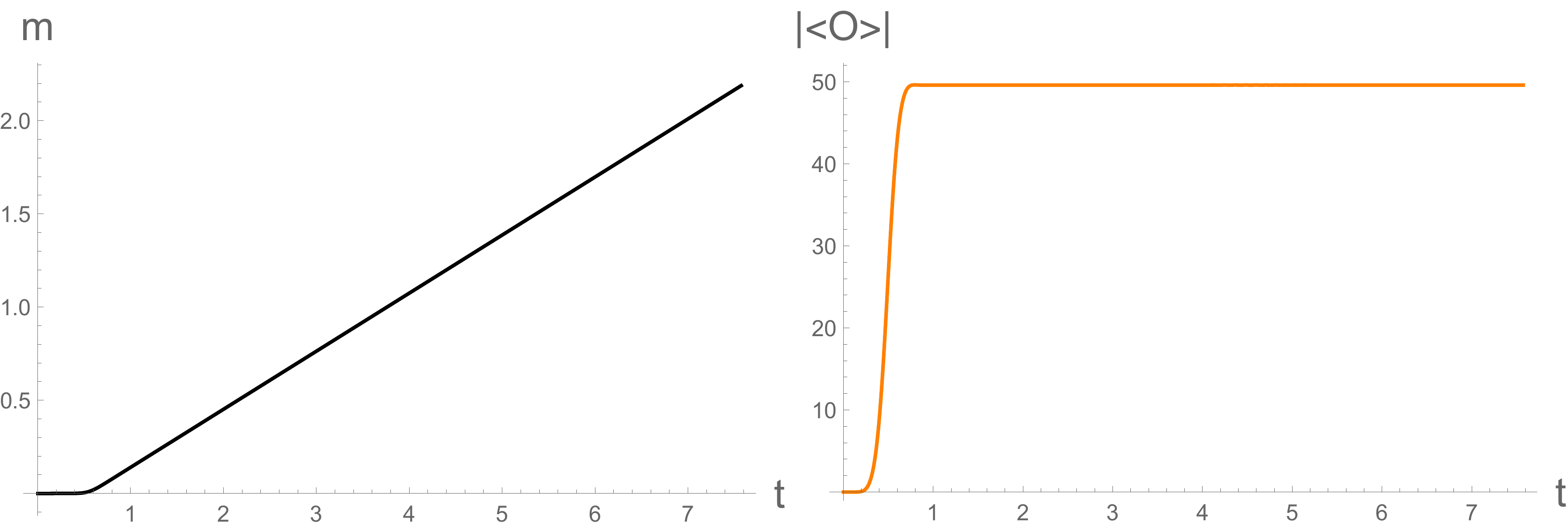}
\end{center}
\caption{\label{regimeI_mvev} \small Time evolution of the energy density (left) and vev modulus (right) for a quench 
process of the form \eqref{source_profile} with $\epsilon = 10^{-4}, \omega_b = 20 \pi$ and $\beta = 1$.}  
\end{figure}

\begin{figure}[h!]
\begin{center}
\includegraphics[width=14cm]{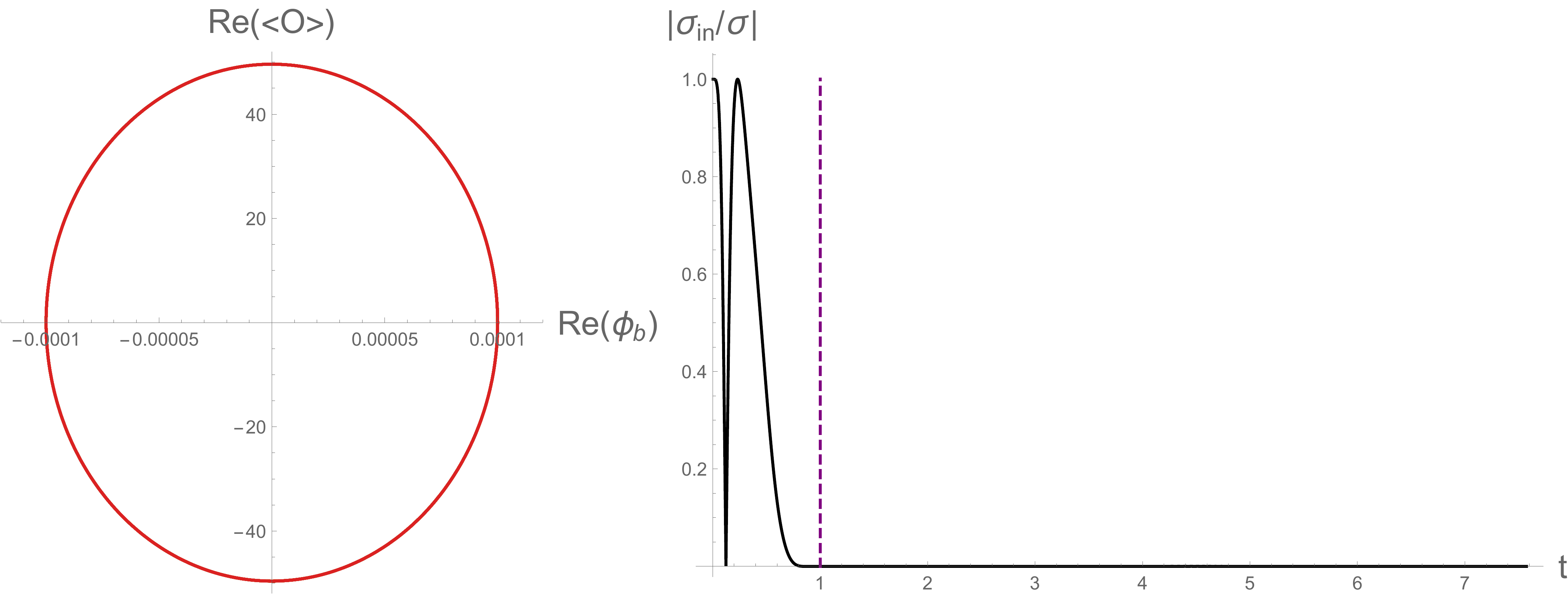}
\end{center}
\caption{\label{regimeI_phase} \small Left: Phase portrait of the late-time dynamics of a quench process of 
the form \eqref{source_profile} with $\epsilon = 10^{-4}, \omega_b = 20\pi$ and $\beta = 1$. We start at $t = \beta$. The different periods of time evolution collapse into a single curve. Right: Relative value of the imaginary part of the nonlinear conductivity $\sigma$. Its vanishing after the quench indicates that the response is in perfect phase opposition with the source.} 
\end{figure}

We close this section by providing one last example. It corresponds again to a quench profile of the form \eqref{source_profile}, this time with frequency $\omega_b = 20 \pi$, amplitude $\epsilon = 10^{-4}$ and time span $\beta = 1$. As illustrated in Fig.\,\ref{regimeI_mvev}, the energy density grows linearly after the quench, while the absolute value of the scalar vev remains constant. In accordance with these facts, the phase diagram of the system corresponds to a closed, non-precessing trajectory of fixed amplitude (see Fig.\,\ref{regimeI_phase}a). Consistently, the imaginary part of the nonlinear conductivity vanishes, i.e., the system response is in complete phase opposition with respect to the harmonic source (see Fig.\,\ref{regimeI_phase}b). The features discussed so far imply right away that the system finds itself in the linear response regime.

\section{Periodically driven real scalar field}
\label{real}

We now turn the attention to the case of a real scalar field. 
Now, unlike the complex case,   the metric of a time-periodic real solution will not be static. 
This implies a dramatic change in the formalism and the methodology. 
Naturally, what makes the problem tractable is  time-periodicity itself which, for instance, allows us to replace the 
linear analysis of normal modes by the equivalent Floquet analysis. 
Qualitatively, the final results we obtain are surprisingly close to those of the complex case. 

\begin{figure}[h!]
	\begin{center}
		\includegraphics[scale=0.6]{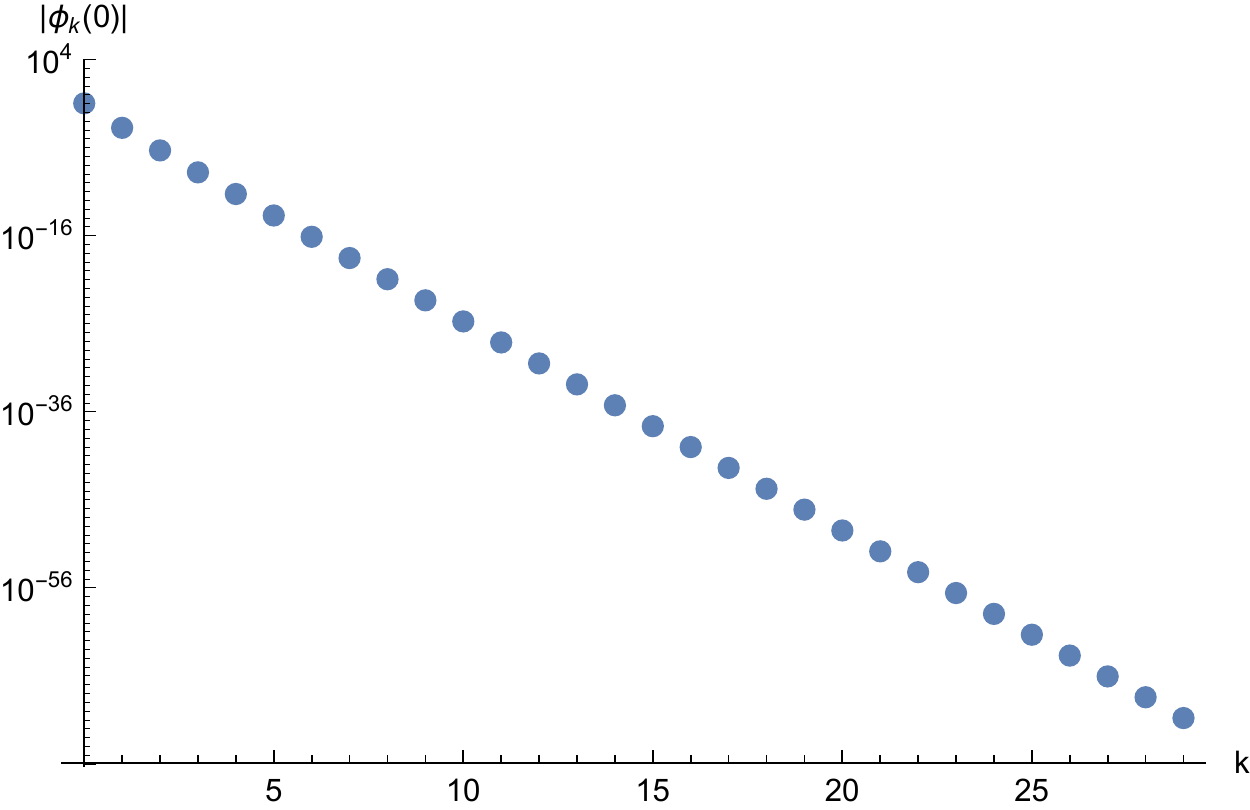}
		\caption{Fourier spectrum at the origin of the periodic solution given by $\omega_b = 2$, $\rho_b = 0.01$ and $\rho_0 = 0.1$.}
		\label{fig:real_spectrum}
	\end{center}
\end{figure}

Another important  difference with respect to the complex case is the fact that the solution, albeit  being periodic, is not harmonic. 
For the complex scalar the time dependence can be factorized as a complex exponential -cf. eqn. \eqref{bsansatz}-, but for the real one the time 
periodicity requires an infinite spectrum of Fourier modes given by the following ansatz	
\be
\phi(t,x) = \sum_{k=0}^{\infty}\phi_{k}(x)\cos((2k+1)\omega_b t). \label{periodicansatz}
\ee
Only the boundary source remains harmonic, $\phi(t,\pi/2) = \rho_b \cos \omega_b t$. 
As before, $\omega_b = 2\pi/T$ is the driving frequency in  the boundary gauge $\delta(\pi/2)=0$ (the 
techniques needed to obtain the fully nonlinear solutions are reviewed in appendix \ref{realmethods}). 
Although  we cannot factorize the radial dependence into a function $\rho(x)$, we will keep the same notation as 
for the complex case, i.e., $\rho_{o} \equiv \phi(0,0)$ and $\rho_b \equiv \phi(0,\pi/2)$. Fig.\ref{fig:real_spectrum} shows 
an example of the Fourier spectrum of these solutions.

\begin{figure}[h!]
	\begin{center}
		\includegraphics[scale=0.6]{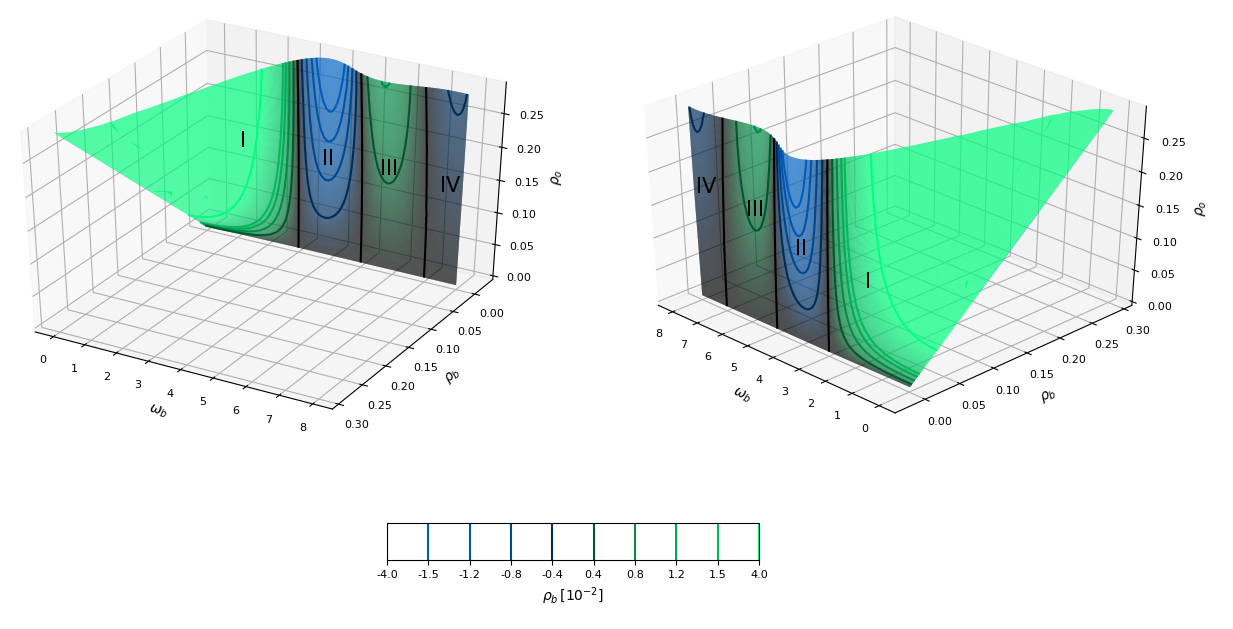}
		\caption{\small The surface of time-periodic geometries responding to a periodically driven  real scalar field. 
		The surface traverses the {\em nonlinear oscillon} plane $\rho_b=0$ at a set of curves.  
		This plot corresponds to the one on Fig.\,\ref{fig:fullplot}, to which the similarity is evident. 
		The range exhibited is considerably smaller in this plot and this is due to the higher technical difficulty in computing each solution individually.}
		\label{fig:fullplotreal}
	\end{center}
\end{figure}

As in the complex case, the space of {\em sourced periodic solutions} -SPS- is a surface embedded in the three 
dimensional space spanned by coordinates  $(\omega_b,\rho_b,\rho_o)$. In the $\rho_o\rightarrow 0$ limit the periodic 
solutions go to (sourced and unsourced) linear modes of AdS$_4$, while the limiting, unsourced cases with $\rho_b=0$ are 
sometimes termed {\em nonlinear oscillon} solutions in the literature, and were first 
constructed in  \cite{Maliborski:2013jca} (see also \cite{Maliborski:2016zlh} for an extensive account of results and methods). 
Figure \ref{fig:fullplotreal}  is the counterpart of Fig.\,\ref{fig:fullplot} in the real case. 
The similarity is remarkable. The quality of the plot is substantially smaller, as oscillatory solutions are much more demanding 
in terms of computational resources than stationary ones. This implies that going to high values of $\rho_o$ is very difficult. 
Although not being apparent from Fig.\,\ref{fig:fullplotreal}, the oscillon curves tilt and wiggle in a similar way as the BS lines do.  

\subsection*{Linear stability}

The study of stability  is  more involved now than  in the complex case. There, a set of static equations was perturbed, 
and the linearized spectrum of normal modes was easy to establish. In the present situation, the linearized fluctuations obey a time-dependent system of equations.  
The pertinent tool to employ now is provided by the Floquet analysis. In our situation we consider linearized fluctuations for the fields of the form
\begin{equation}
{\bf x}(t,x) = {\bf x}_{p}(t,x) + \tilde {\mathbf {x}}(t,x)
\end{equation}
where the subindex $p$ stands for ``periodic solution" and $\tilde {\mathbf {x}}$ denotes the perturbations. 
At first order in the perturbations amplitude we obtain a system of the form
\begin{equation}
\dot{\tilde {\mathbf {x}}}= {\bf L}(t) \tilde {\mathbf {x}}
\label{eqpertper}
\end{equation}
where ${\bf L}(t) = {\bf L}(t+T)$ is a time periodic linear integro-differential operator (in $x$) built out of the solution ${\bf x}_{p}$. 
The Floquet theorem establishes the existence of a solution of the form
\begin{equation}
{\bf x} (t) = e^{\lambda t} {\bf P}(t)~~~~\hbox{with} ~~~~{\bf P}(t) = {\bf P}(t+T)\, .
\end{equation}
The information about the stability resides in $\lambda_i$, with $i=1,2,\ldots\,$ In the general case, both $\lambda$ and ${\bf P}$ will be complex. 
From equation \eqref{eqpertper}, we immediately see that $\lambda^*$ and ${\bf P}^*$ are also solutions. 
The real solution thus involves the appropriate linear combination of both. 
Also, though not evident, one can show that solutions come in pairs, with eigenvalues of opposite signs, $\pm\lambda$. 
This is ultimately a reflection of the Hamiltonian character of the dynamical system \eqref{eqpertper}. 
Hence the stability analysis collapses to one of two possible cases, depending on whether Re($\lambda)$ is zero or not. 
The first case yields a stable (cycle) solution, and the second one an unstable (saddle point) one.

\begin{figure}[h!]
	\begin{center}
		\includegraphics[scale=0.5]{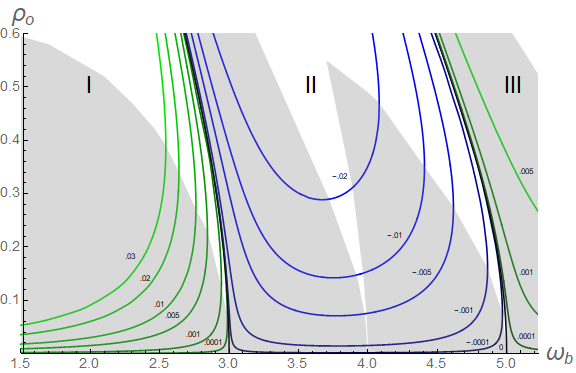}
		\caption{\small Level curves of the SPS surface plotted in Figure \ref{fig:fullplotreal}. Each curve corresponds to a constant value $\rho_b$ (reported explicitly by a small number near the curve). Solutions belonging to shaded (white) regions are linearly stable (unstable).}
		\label{fig:Level_Plot_real}
	\end{center}
\end{figure}

Computing $\lambda_{i}$ requires an exact integration of \eqref{eqpertper} over a full period $T$. 
The same techniques used to obtain the exact nonlinear SPS's can be applied here; the reader can find further details 
in appendix \ref{realmethods}. The numerical analysis gives the region of stability that can be observed in Fig.\,\ref{fig:Level_Plot_real}, whose similarity with 
Fig.\ref{fig:phibsections} is manifest.
The linear analysis says that this similarity goes beyond the shape, also at  boundary of the stability region we find the same algebraic structure, either the lowest eigenvalue $\lambda_1$ becoming purely real or two real eigenvalues fusing and developing imaginary components. 

\subsection*{Nonlinear stability}

As for the complex case, analyzing nonlinear stability requires to study the full numerical evolution of the system departing from the perturbed unstable solution.  
The main conclusions remain similar to the ones obtained in the case of a complex scalar. 
The generic end result of an initial unstable SPS is a black hole.  However, in the lower part of the white wedge in sector I 
unstable solutions  decay either to a black hole or to a {\em sourced modulated solution}, SMS, depending on the sign of the single unstable linear perturbation.  As for the complex case, the pulsation is a modulation of the driven oscillation, with sudden beats separated by  a substantially larger plateaux.  

Like the Boson Stars, nonlinear oscillons become unstable beyond the  turning point for the mass \cite{Maliborski:2016zlh}. At such high values of the field at the origin, the SMS's are far more fragile than their complex counterparts
and usually decay after a couple of modulating beats. This may come from the fact that the whole geometry now shakes. Or maybe from the fact that the harmonic ansatz for the source starts conflicting with the anharmonic response of the field in the interior and calls itself for a generalized periodic ansatz.

\begin{figure}[h!]
	\begin{center}
		\includegraphics[scale=0.09]{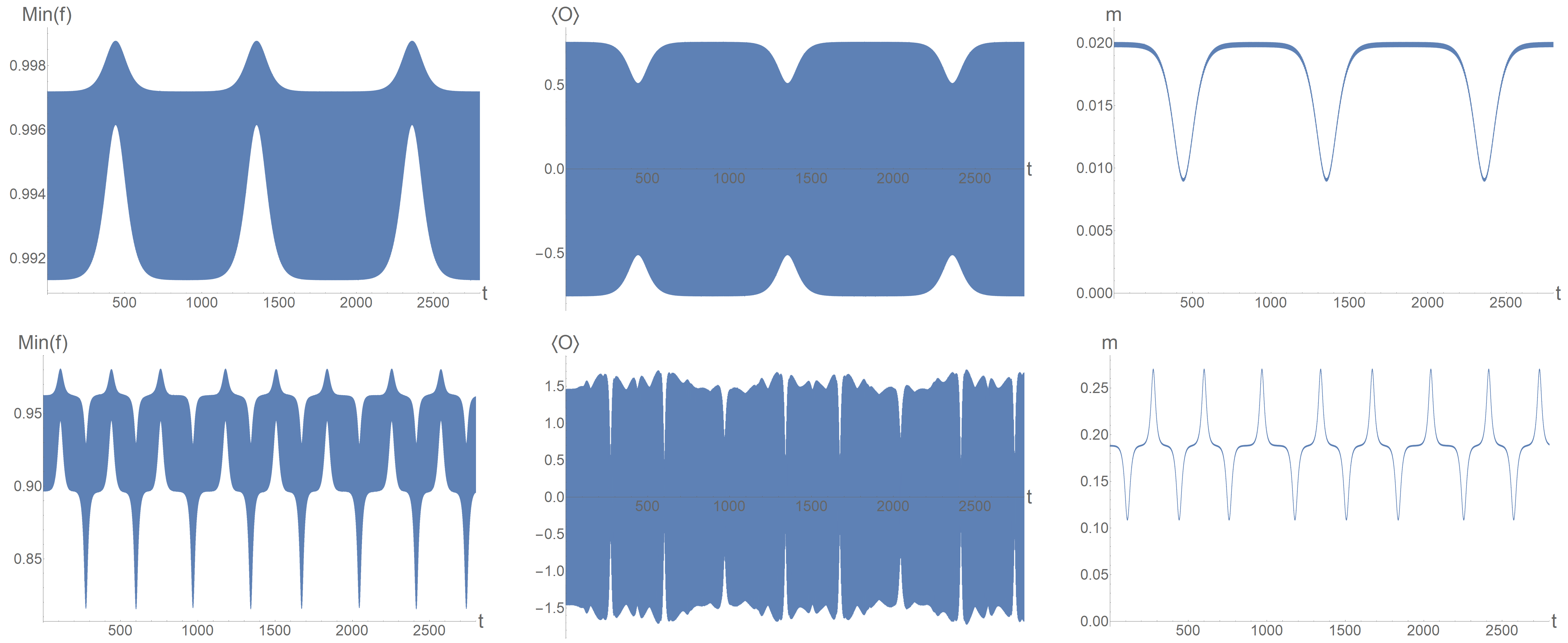}
		\caption{\small Sourced modulated solutions (SMS) with $\rho_b = 10^{-3}$. In the first line we depict a standard SMS with $\omega_b = 2.944$ and $\rho_o = 0.15$. The blue filling corresponds to the fast oscillations of the scalar field (as compared to the pulsating modulation). In second line, we provide an example of an exotic SMS, where pulsating modulations of both signs alternate. The unstable initial SPS has, in this case, $\omega_b = 2.722$ and $\rho_o = 0.54$.}
		\label{fig:pulsatingreal}
	\end{center}
\end{figure}

Similarly to the complex case, for sufficiently small values of the source, $\rho_b \sim 10^{-3}$, we have also found exotic real SMS's where the 
bounce occurs in both directions (i.e, towards lower or higher masses) when the initial perturbation is very weak. 
We plot an example in the second line of Fig.\,\ref{fig:pulsatingreal}.

\subsection*{Type I phase transition in the real case}

In the real case, the ingredients necessary for the assumptions $a$ and $b$ to hold are present (see Section \ref{type1}). 
Therefore, we expect the same type I phase transition we have uncovered in the complex case. 
In order to confirm this expectation, we consider a time-dependent source given by the real 
part of eq. \eqref{source_profile} (i.e., we replace $\exp(i\omega_b t) \to \cos(\omega_b t)$). 

\begin{figure}[h!]
\begin{center}
\includegraphics[width=16cm]{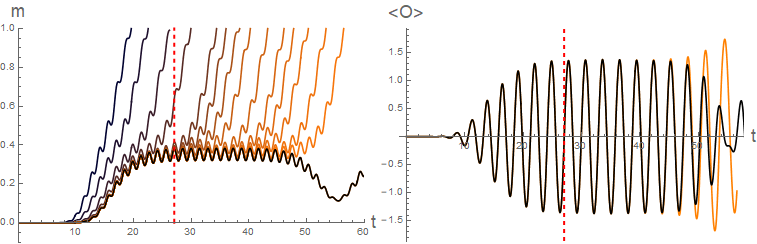}
\end{center}
\caption{\label{non_adiabatic_real} \small Left: Energy density for a family of build-up processes of 
the harmonic driving $\omega_b = 2.15$, $\epsilon =0.09$. The $\beta$ of the last build-up process ending in 
the runaway phase has been marked by the vertical dashed red line. 
Right: $\left<\mathcal O \right>$ for the last simulation ending in the runaway phase (solid orange) and the 
first one ending in the non-collapsing phase (solid black).}
\end{figure}

In Fig.\,\ref{non_adiabatic_real} we plot $m$ and $\left<\mathcal O \right>$ for the one-parameter family of build-up processes of 
the harmonic driving $\omega_b = 2.15$, $\epsilon = 0.09$. We clearly see that there exists a $\beta_c$ that separates two different 
late-time phases, as for the complex scalar field. Furthermore, as $\beta \to \beta_c$, the system seems to spend progressively larger 
time around an intermediate attractor. 

\subsection*{Dynamical construction of a  nonlinear oscillon}
	
Nonlinear undriven real SPS's ({\em nonlinear oscillons}) are the equivalent to BS's in the real case. In this section we will show how nonlinear oscillons can be dynamically constructed, in analogy to the results we presented in section \ref{sec:Quasistatic_method} for the BS case. Same as there, here we will  describe a quasistatic quench connecting the AdS$_4$ vacuum with a nonlinear oscillon across the linearly stable part of region II. The protocol requires the appropriate tuning of  the source parameters,  frequency and amplitude, as a function of time
\begin{equation}
\phi(t,\pi/2) = \phi_{b}(t) = \epsilon(t)\cos(\omega_b(t)t)\, .
\end{equation} 
The specific profiles for $\epsilon(t)$ and $\omega_b(t)$ we shall employ are again given by (\ref{source_profile_epsilon})-(\ref{source_profile_omega}), and plotted in Fig.\,\ref{source_plot}.

\begin{figure}[h!]
\begin{center}
\includegraphics[width=16cm]{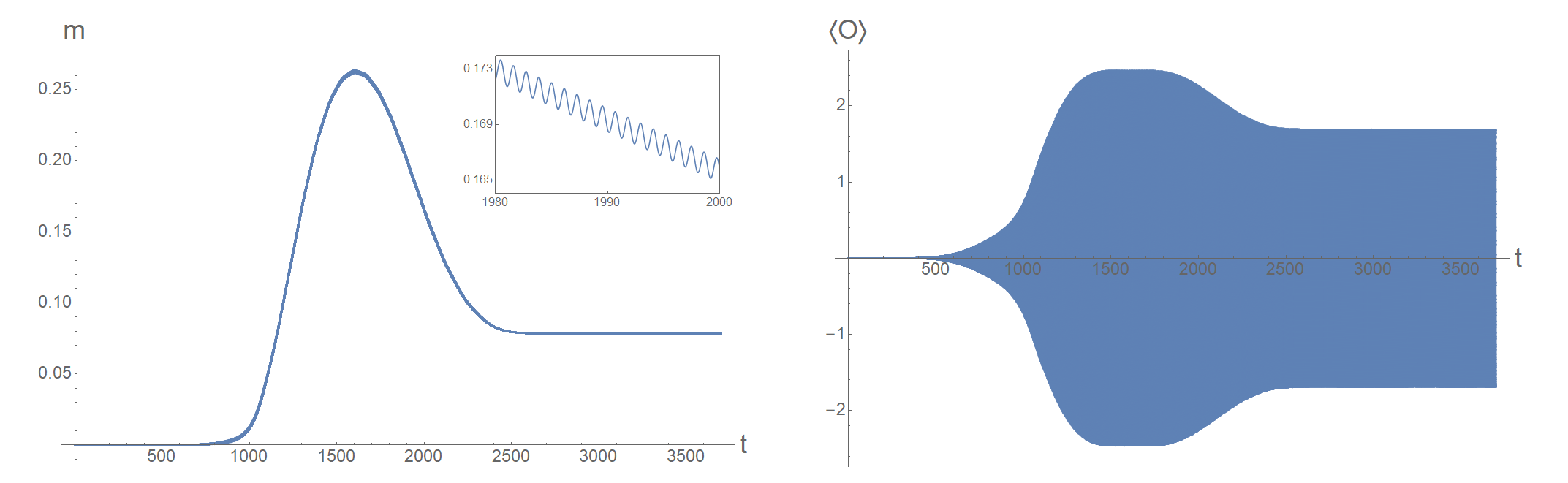}
\end{center}
\caption{\label{nonlinear_oscillon_made_check}  \small Time evolution of $m$ and $\langle \mathcal O \rangle$ along the quench described in the main text. The blue region corresponds to a highly dense number of fast oscillations compared with the duration of the process. In the first plot this region is not appreciated but as the inset shows it is also present.}
\end{figure}	
	
\begin{figure}[h!]
\begin{center}
\includegraphics[width=17cm]{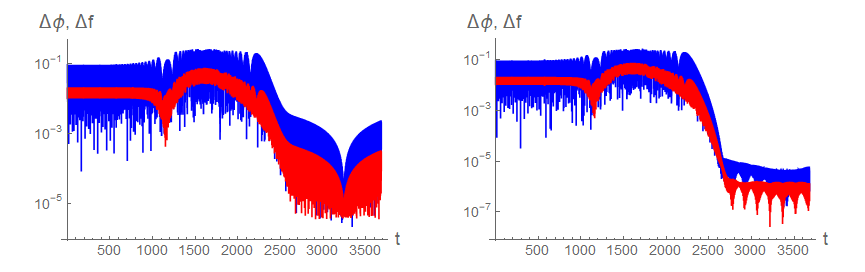}
\end{center}
\caption{\label{D_phi_f_real_quench} \small Time evolution of $\Delta\phi$ (blue) and $\Delta f$ (red). We compare the end field configurations $\phi$ and $f$ after the quench with two  nonlinear oscillons: one  with $\omega_{\text{NLO}} = 2.9$ (left), 
 and another one with $\omega_{\text{NLO}} = 2.89992$ (right). The agreement is better in the second case.}
\end{figure}

As a particular example, we consider a quench protocol given by  $\omega_i = 3.1$, $\omega_f = 2.9$, $\epsilon_m = 10^{-3}$ and $\beta = 1500$.   Figure \ref{nonlinear_oscillon_made_check} contains the time evolution of $m$ and $\langle \mathcal O \rangle$ along the process. The main difference with respect to Fig.\,\ref{bs_made} is that now we are plotting the oscillation of real quantitities. The fast oscillations pile up and, due to their high density, form the blue shaded area visible in the figure. Now, instead of  (\ref{drho})-(\ref{df}), we define two new quantities, $(\Delta\phi$,$\Delta f$), which measure the distance to the nonlinear oscillon ($\phi_{\text{NLO}},f_{\text{NLO}}$) built with the target data, 
\begin{eqnarray}
\Delta\phi(t) & = & \left(\int_{0}^{\frac{\pi}{2}}\tan^2(x)\left(\phi(t,x)-\phi_{\text{NLO}}(t,x)\right)^2\right)^{\frac{1}{2}}\,, \\
\Delta f(t) & = & \left(\int_{0}^{\frac{\pi}{2}}\tan^2(x)\left(f(t,x)-f_{\text{NLO}}(t,x)\right)^2\right)^{\frac{1}{2}} \, .
\end{eqnarray}
In Fig.\,\ref{D_phi_f_real_quench} we have plotted the time evolution of these two quantities for our particular example. On the left plot we observe that, after the quench, the geometry is very close to the oscillon with frequency $\omega_{\text{NLO}} = 2.9$ but, as also happened in the complex case, even more to the one with $\omega_{\text{NLO}} = 2.89992$ (see right plot).

\section{Summary and outlook} 
\label{futuro}

In this paper we analyzed periodically driven scalar fields on global AdS$_4$.
This framework allows one to study different aspects of holographic Floquet dynamics, 
such as dynamical phase diagrams and late-time regimes alternative to thermalization.

We constructed zero-temperature solutions subjected to a constant-amplitude periodic 
driving of the scalar and dubbed them Sourced Periodic Solutions (SPS). They are dual to so called Floquet condensates. 
We have characterized the SPS solution space in detail and studied throughly both linear and nonlinear stability properties. 
SPS extend beyond linearity known perturbative solutions about AdS.

The unstable SPS's feature a rich phenomenology upon time evolution. In particular, there exist horizonless 
stable late-time solutions which evade gravitational collapse and develop instead a pulsating behavior. We named them Sourced Modulated Solutions (SMS).
SMS's are themselves stable solutions where the modulation pulsation impacts in the vev profile and the total mass  while, remarkably,  its frequency is not imposed by that of the driving. 

We addressed various types of quench processes concerning both the amplitude and the frequency of the scalar source.
We focused on both \emph{slow} and \emph{fast} quenches. 
The first allow the study of quasistatic processes and showed that they can be used to prepare SPS's, dual to Floquet condensates, starting from the AdS vacuum.
The study of non-quasistatic quenches uncovered a nice surprise: by suitably fine-tuning the quench rate, the system stays for an arbitrary long time on  an unstable
SPS and then decays either to a SMS or a black hole. These SPS's act therefore as attractors in a Type I gravitational phase transition. 

Next, we examined the possibility of 
using such quenches to prepare  boson stars starting from AdS. From the gauge/gravity duality, this is motivated by the exciting possibility of preparing experimentally  sourceless Floquet condensates in  strongly coupled systems. In fact, we find explicit solutions to this problem using both quasistatic and non-quasistatic quench protocols.

Finally we have studied the post-collapse evolution. We have unravelled the three basic behaviours found  elsewere \cite{rangamani2015driven}.

The analysis has been performed  both for the case of complex and real scalars. We showed that the phenomenology is qualitatively similar 
despite the technical treatment is different (and considerably more involved in the real case). 
We mainly focused on massless scalars, but preliminary results for the massive situation with $m^2=-2$ point to the persistence of the same picture.

The complex field setup in the boson star configuration can be regarded as a time-like version of the spatial Q-lattice  model\cite{Donos:2013eha}.
These models have been recently shown to provide a framework where to study spontaneous symmetry breaking of space translations 
\cite{Amoretti:2016bxs,Amoretti:2017frz,Amoretti:2017axe}.
This observation gives support to  the speculation that relates boson stars to time crystals. 
The statement as such could be affected by subtleties which need to be carefully studied. 
It is however interesting to note that the holographic setup could avoid the known no-go theorems  \cite{bruno2013impossibility,Watanabe:2014hea}.
Indeed, the no-go theorems rely on standard arguments about the large-volume thermodynamic limit,
while holography in global AdS is related to an alternative thermodynamic limit at finite volume.

The precise origin of the frequency of the pulsating solutions is an interesting open problem to be addressed. 
It resembles the possible spontaneous generation of a time scale observed for similar late-time solutions in 
the pumping setup \cite{Carracedo:2016qrf}, although in the present case the pulsating frequency does depend 
on the amplitude of the initial perturbation of the unstable harmonic solution. 
There is a lower bound to this amplitude set by the numerical noise, hence we cannot establish whether 
in the limit of vanishing amplitude the period of the modulating pulsation diverges or not.

While entering the final stages of this work, paper \cite{Kinoshita:2017uch} appeared. Albeit in a totally different context, their results bear intriguing resemblance with
ours. In fact their Fig.\,3 and our Fig.\,\ref{fig:fullplot}  are very similar. Our boson stars and nonlinear oscillons are, in their language,  vector meson Floquet condensates. However, the physics behind looks very different. In their case the instability is related to a Schwinger pair dielectric breakdown. Also their process of building a 
sourceless Floquet condensate differs from ours. We intend to investigate further these similarities by going to richer models for a scalar field in AdS involving potentials which have a phase transition.

\section*{Acknowledgements}
We would like to thank Andrea Amoretti, Daniel Are\'an, Riccardo Argurio, An\'ibal Sierra-Garc\'ia, 
Blaise Gout\'eraux, Carlos Hoyos, Piotr Bizo\'n, Keiju Murata and Alfonso Ramallo for pleasant and insightful discussions.

This work of  was supported by grants FPA2014-52218-P from Ministerio de Economia y Competitividad, by  Xunta de Galicia ED431C 2017/07, by FEDER and by Grant Mar\'\i a de Maeztu Unit of Excellence MDM-2016-0692. A.S. is happy to acknowledge support from the International Centre for Theoretical Sciences (ICTS-TIFR), and wants to express his gratitude to the ICTS community, and especially to the String Theory Group, for their warm welcome. D.M. thanks the FRont Of pro-Galician Scientists (FROGS) for unconditional support. A.B. thanks the support of the Spanish program ''Ayudas para contratos predoctorales para la formaci\'on
de doctores 2015'' associated to FPA2014-52218-P. This research has benefited from the use computational resources/services provided by the Galician Supercomputing Centre (CESGA).

\begin{appendix}

\section{Complex periodic solutions}
This is by now standard material but we include it here for completeness and in order to fix the notation. 
The general gravitational action for a complex scalar field is
\be
S= \frac{1}{2\kappa^2} \int d^{d+1}x \sqrt{-g}\left( R - 2\Lambda\right)   -\int d^{d+1}x\sqrt{-g}\left(   \pd_\mu\phi  \pd^\mu\phi^*  + V(|\phi |)\right) \ ,
\ee
with $\kappa^2 = 8\pi G$,  $\Lambda = -d(d-1)/2l^2$ for AdS$_{d+1}$. 
Note that it is assumed $V(0)=0$ otherwise the constant term of the scalar potential would contribute to the cosmological constant. 
The specific action \eqref{com_act} corresponds to taking $d=3$ and $V(|\phi |)= -m^2 \phi^* \phi $.
As the scalar field is complex, the action is invariant under the global $U(1)$ transformations $\phi\to e^{i\alpha}\phi$.
The equations of motion are
 \beqa
 R_{\mu\nu} - \frac{1}{2} g_{\mu\nu} R +\Lambda g_{\mu\nu} &=& \kappa^2  \left[  \pd_\mu\phi^* \pd_\nu\phi+  \pd_\nu\phi^* \pd_\mu\phi - g_{\mu\nu} \left( | \pd \phi |^2
+ V(|\phi |)\right) \right] \label{eqeins}\ ,\\
\frac{1}{\sqrt{-g}} \pd_\mu \left( \sqrt{-g} g^{\mu\nu} \pd_\nu\phi\right) &=&  \frac{\pd V(\phi,\phi^*)}{\pd\phi^*} \ .\label{eqphi}\\
 \eeqa
The isotropic ansatz for the metric of $AdS_{d+1}$ can be expressed as follows
\be
ds^2 = \frac{l^2}{\cos^2 x}\left( - f e^{-2\delta} dt^2+ f^{-1} dx^2 + \sin^2 x \, d \Omega_{d-1}^2\right) \ , \label{line1}
\ee
with $x\in [0,\pi/2)$. A useful field redefinition is given by
\beqa
\Phi(t,x) = \phi'(t,x) \ , \qquad \Pi(t,x) = \frac{e^\delta(t,x)}{f(t,x)} \dot \phi(t,x)\, .
\label{PhiPi}
\eeqa
Upon this redefinition, the scalar field equations of motion can be cast in the form
\beqa
\dot\Phi &=& \left(f e^{-\delta}  \Pi \right)'  \ , \\
\dot\Pi &=&   \displaystyle \frac{1}{\tan^{d-1} x}\left(\tan^{d-1} x f e^{-\delta} \Phi\right)' - \frac{l^2}{\cos^2x} e^{-\delta} \partial_{\phi_c} V(|\phi |) \, . 
\eeqa

From the Einstein equations we obtain, after setting $\kappa^2 = (d-1)/2$ as well as $l=1$,
\beqa
f' &=&\displaystyle \frac{d-2 + 2 \sin^2 x}{\sin x \cos x} (1-f) -\sin x \cos x \, f\,  \left( |\Phi|^2 + |\Pi |^2   \right)
- \displaystyle \tan x \, V(|\phi |)\ ,  
\\
\rule{0mm}{10mm}
\delta' &=&-\displaystyle   \sin x \cos x \left(  |\Phi|^2 + |\Pi |^2 \right) \, .  
\eeqa

In the main text  we consider the time-periodic ansatz \eqref{bsansatz} where $\omega$ is an angular frequency and $\rho(x)$ is a real function. 
We are adhering here to the  boundary gauge $\delta(\pi/2)=0$ that makes $t$ equal to the proper time of an observer at the boundary. 
For any given $\omega$ there are static solutions for $\rho(x)$, $f(x)$ and $\delta(x)$ to be obtained from the equations of motion 
\beqa
&& \rho''  +\left( \frac{d-1}{\sin x\cos x} + \frac{f'}{f}  - \delta' \right)  \rho' +\left(\omega^2 \frac{e^{2\delta}}{f^2} - \frac{m^2 }{f \cos ^2 x}\right) \rho = 0\ ,  \nonumber \\
\rule{0mm}{7mm} 
&&f' - \displaystyle \frac{1+ 2 \sin^2 x}{\sin x \cos x} (1-f) - f \delta' + m^2 \rho^2 \tan x  =0\ , \label{drivenbosoneq}
\\
\rule{0mm}{8mm}
&& \delta' + \displaystyle   \sin x \cos x \left( \rho'^2 + \omega^2\frac{ e^{2\delta}}{f^2}  \rho^2 \right)  = 0 \, , \nonumber 
\eeqa
that result from the above-mentioned ansatz.

\section{The role of the pumping solution}

In a previous work \cite{Carracedo:2016qrf}, some of the authors analyzed a massless, real scalar field in AdS$_4$ subjected to 
a linearly rising {\it pumping} source, $\phi(t,\pi/2) = \alpha_b t$. One of the major findings was that there exist extremely 
simple {\it pumping solutions}, in which the scalar field profile is flat in the radial direction, $\phi(t,x) = \alpha_b t$, 
while the metric functions $f(x), \delta(x)$ are nontrivial, but static.  

Our objective in this appendix is to comment on the interplay between these pumping solutions and the four-dimensional SPS's we have found in this paper.
From Fig.\,\ref{fig:fullplot} it is apparent that the surface of SPS's extends to values $\omega_b=0$ on a line $\rho_b = \rho_o$. Naively these would be related
to the mentioned pumping solutions. The correct answer involves a careful scaling limit.

\subsubsection*{The scaling limit}

Consider the SPS equations of motion \eqref{drivenbosoneq} (for $d=3, m=0$) and set  
\beq
\rho(x) = \hat \rho_b/\omega_b. 
\eeq
By taking the  $\omega_b \to  0$ limit at fixed $\hat \rho_b$, these equations reduce to 
\beq
f'(x) - \frac{1 + 2 \sin^2 x}{\sin x \cos x} (1 - f(x)) - f(x) \delta'(x) = 0, 
\eeq
\beq
\delta'(x) + \cos x \sin x \frac{e^{2\delta(x)} \hat \rho_b^2}{f(x)^2} = 0, 
\eeq
\beq
\frac{e^{\delta(x)}}{f(x)} \hat \rho_b \omega_b = 0.  
\eeq
The equations are nothing but the  equations of motion for the pumping ansatz (as reported in \cite{Carracedo:2016qrf}), 
provided we identify $\hat \rho_b = \alpha_b$ and fix $\omega_b = 0$. This simple observation establishes that the pumping 
solution controls the SPS's in the limit of infinite source amplitude and zero source frequency (i.e., on the upper left part 
of the diagram displayed in Fig.\,\ref{fig:phibsections}). 

As demonstrated in \cite{Carracedo:2016qrf}, there is a critical pumping rate $\alpha_b^*$ above which no pumping solution 
exist. Numerically, $\alpha_b^* \approx 0.785187$. This fact, together with the relation between the pumping solutions and 
the SPS's in the scaling limit, leads to the prediction that, 
when $\omega_b \ll 1$, the upper boundary of the region containing the linearly stable SPS's in the $\omega_b - \rho_b$ plane must approach the curve 
\beq
\omega_b \rho_b = \alpha_b^*.  \label{w0_scaling}
\eeq 
Note in particular that, if this expectation is correct, $\rho_b$ must diverge as $\omega_b \to 0$. 
The analogous statement $\omega_b \rho_o = \alpha_b^*$ also has to hold in this limit, since the difference 
between $\rho_b$ and $\rho_o$ is also expected to vanish as $\omega_b \to 0$: the pumping solution is radially flat and has no vev. 

\begin{figure}[h!]
\begin{center}
\includegraphics[width=8cm]{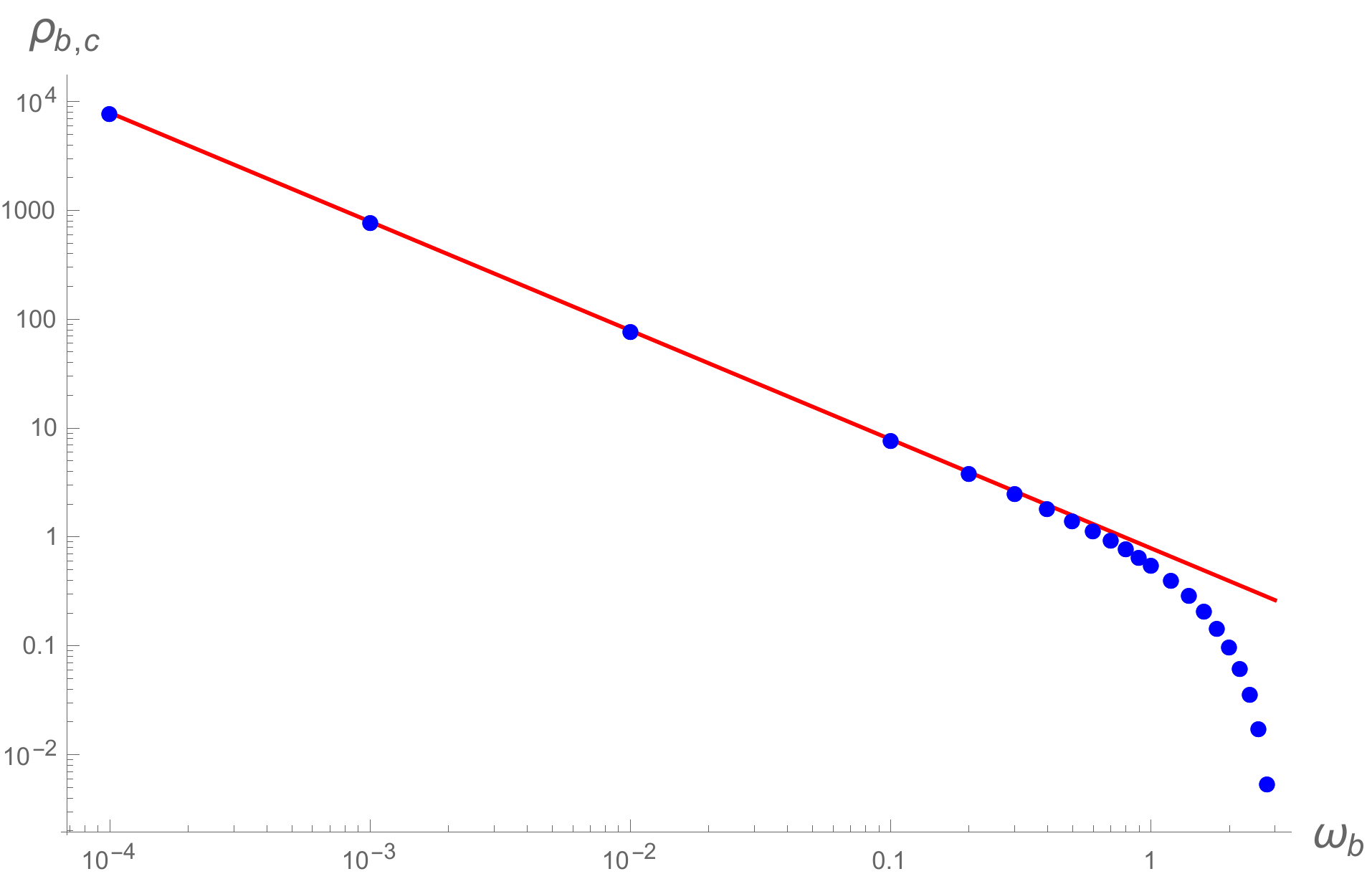} \includegraphics[width=8cm]{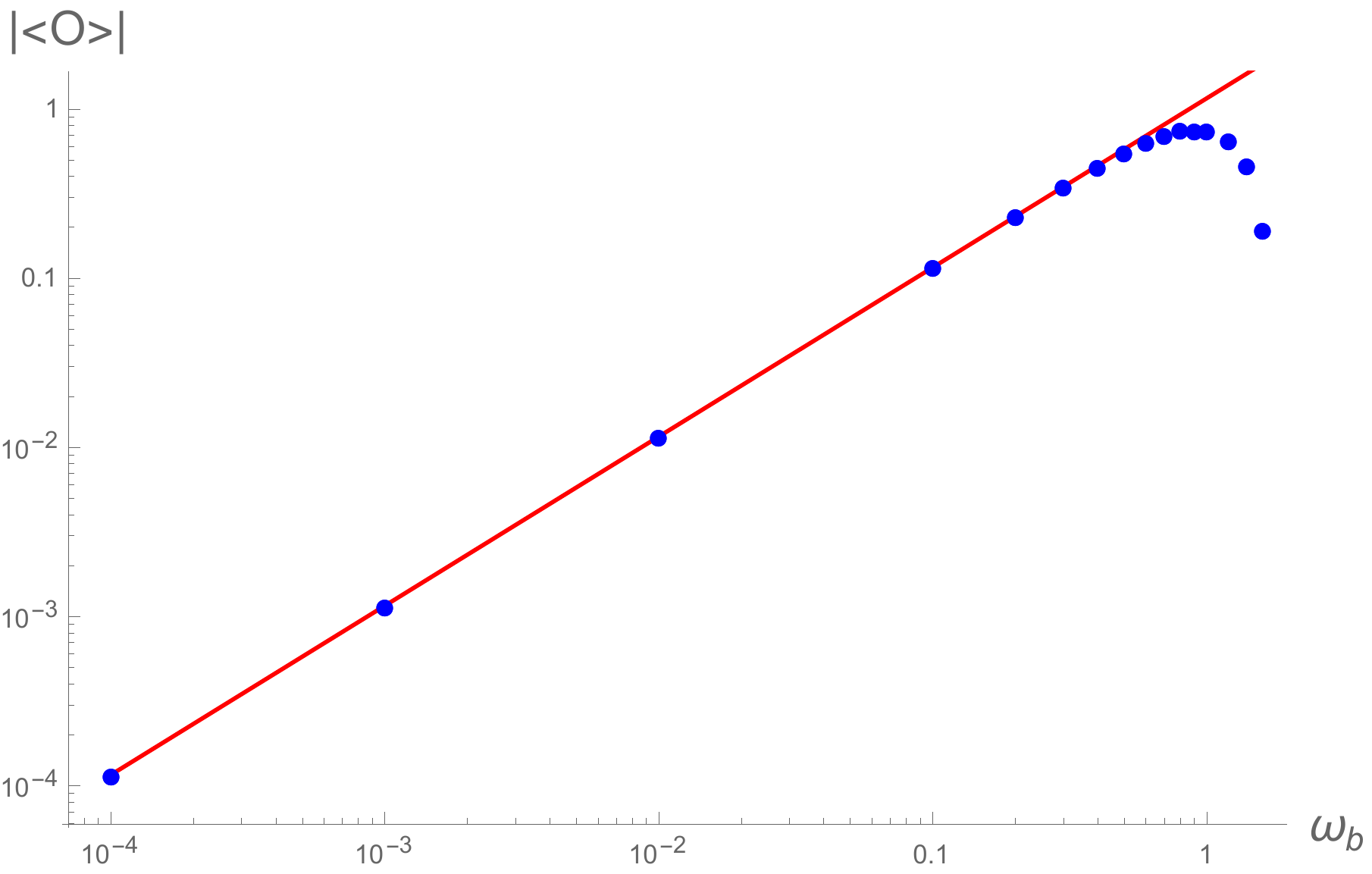}
\end{center}
\caption{\label{pumping_numerical_check} \small Left: Critical values $\rho_{b,c}$ as a function of $\omega_b$ (blue dots); the red line corresponds to the analytical prediction $\rho_{b,c} = \alpha_b^*/\omega_b$. Right: $\left| \left< \mathcal O_c \right> \right|$ of the critical SPS's (blue dots), together with the linear fit  $\left| \left< \mathcal O_c \right> \right| = 1.1543165... \omega_b$ (red line).}
\end{figure}

In order to confirm these expectations, we have determined numerically the upper boundary of the stability region in the 
$(\omega_b, \rho_b)$ phase diagram (at small frequency). The results are shown in Fig.\,\ref{pumping_numerical_check}a.  
It is clearly seen that, as $\omega_b \to 0$, this upper boundary lies progressively closer to the predicted curve $\rho_{b,c} = \alpha_b^*/\omega_b$. 

As a further consistency check, in Fig.\,\ref{pumping_numerical_check}b we establish that 
$\left| \left< \mathcal O_c \right> \right| \to 0$ as $\omega_b \to 0$ -here $\left| \left< \mathcal O_c \right> \right|$ is 
the vev of the last linearly stable SPS-. The numerical data are compatible with a linear decrease in driving frequency, 
$\left| \left< \mathcal O_c \right> \right| \sim \omega_b$.

\section{Study of the normal modes}
\label{normal}

This appendix provides details on the spectrum of linearized normal modes  of the complex SPS's and BS's presented in the main text. 

Linear stability is the first important piece of information one gets from the normal mode spectrum. 
A point-wise study of several individual solutions in the plane $(\omega_b,\phi_o)$ leads to establish the stability diagram 
reported in Fig.\,\ref{fig:phibsections} whose details are highlighted below.

\begin{figure}[h!]
\begin{center}
\includegraphics[scale=.35]{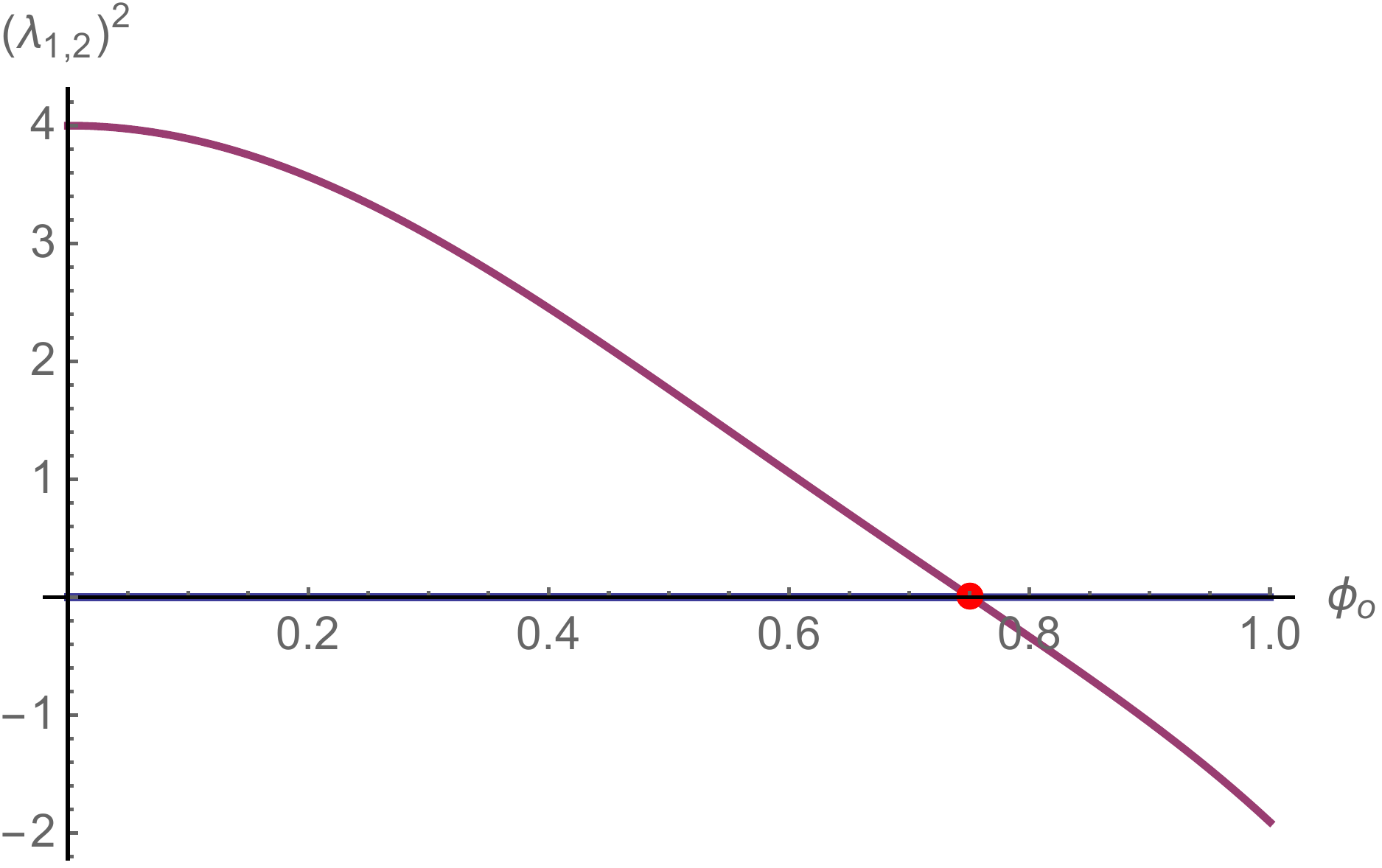}
\includegraphics[scale=.35]{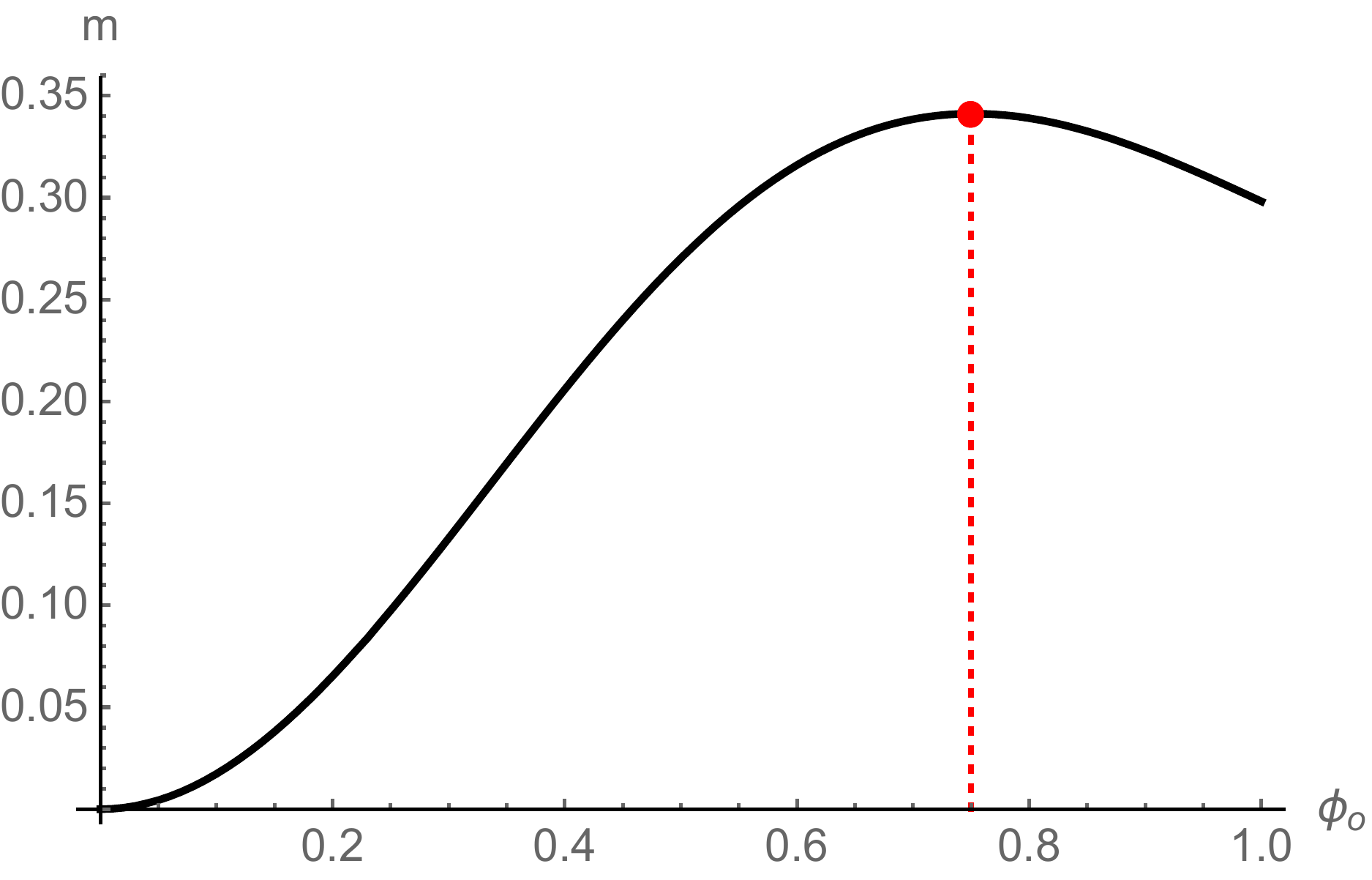}
\includegraphics[scale=.35]{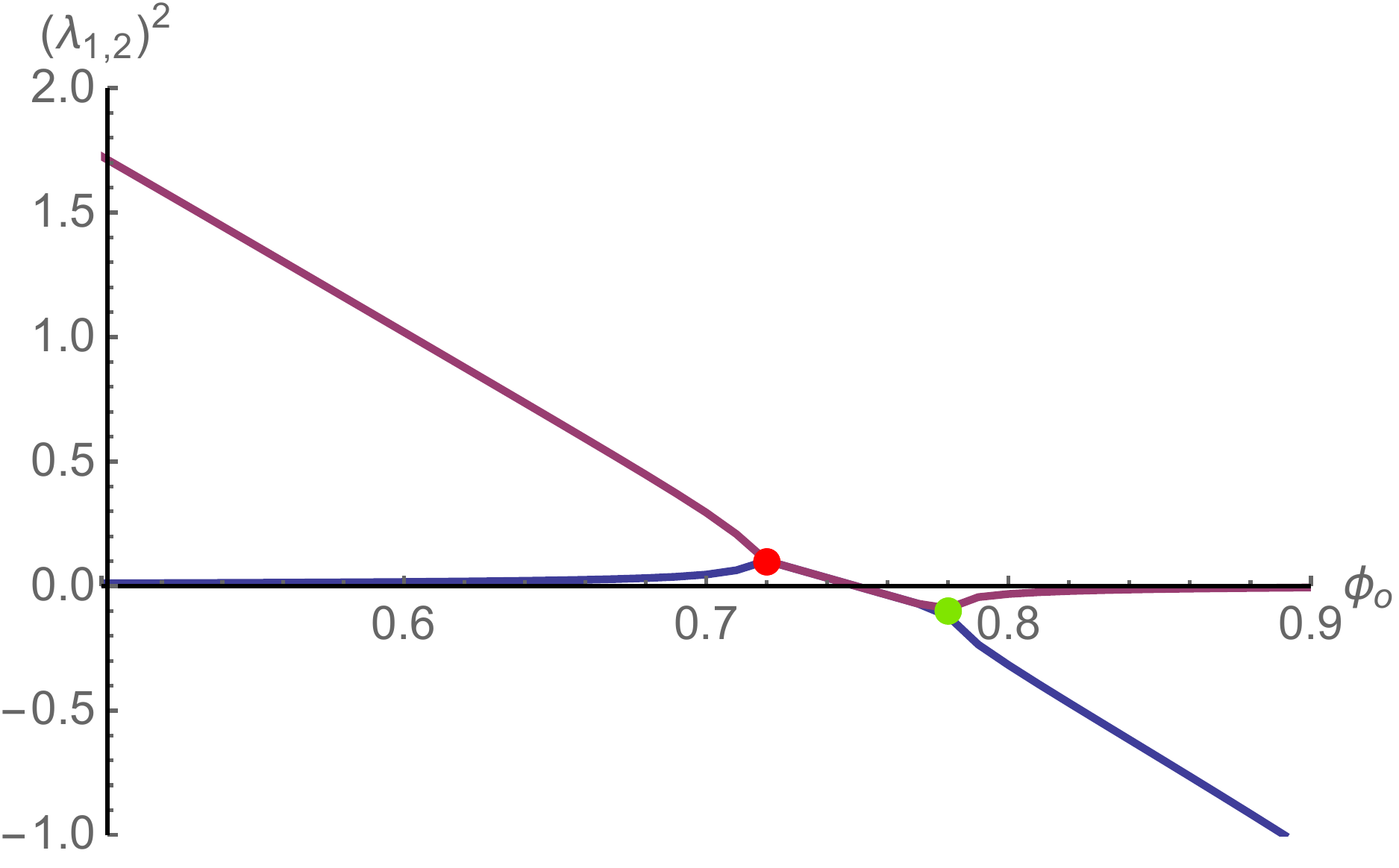}
\includegraphics[scale=.35]{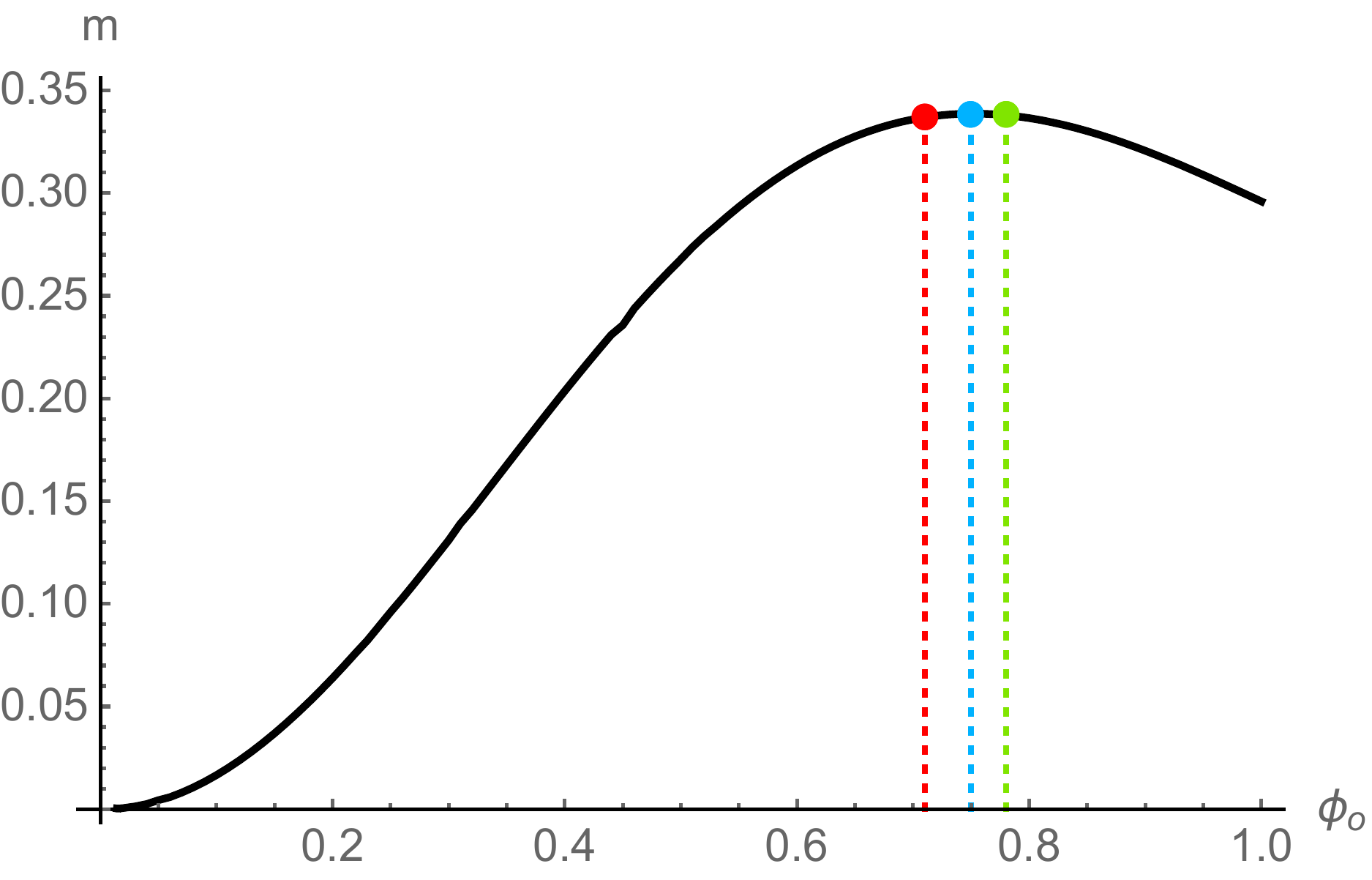}
\caption{\small Top: Real part of the first two squared eigenfrequencies of the lower Boson Star branch. The red point signals the onset of linear instability which corresponds
to the maximum value attained by the mass along the same branch. Bottom: Analogous plots for the $\phi_s=-.001$ sourced solution. The red point 
represents the onset of linear instability.
Here the two lower eigenfrequencies meet and develop opposite complex parts. The instability (red dot) occurs before the maximum 
value of the mass along the branch is attained (light-blue dot).}
\label{BS}
\end{center}
\end{figure}

In Fig.\,\ref{fig:phibsections}, the solid lines departing from the $\phi_o=0$ axis at $\omega_b=3+2n$ with $n \in N^+$ representing the BS branches 
that dress nonlinearly the AdS oscillons. Moving along the BS branches by raising $\phi_0$, one encounters BS's with increasing mass which are 
stable until the mass reaches a maximum. This points provides a Chandrasekar-like bound beyond which the BS's become linearly unstable. This is seen in 
Fig.\,\ref{BS}  were we plot the evolution of the squared frequencies of the first two normal eigenmodes $\lambda_{1,2}^2$. Precisely at the Chandrasekar mass, 
the second eigenmode $\lambda_2$ becomes purely imaginary.

The second set of plots shows the evolution of the same modes along a line of SPS's with a small source
in sector II (profiles with one node), running closely parallel to the BS curve. The situation here changes qualitatively, though continuously. 
The instability is now triggered by the first two modes fusing at positive values and developing
imaginary parts of opposite sign. This now happens before the maximum of the mass is reached. 
The continuation of this point away from the line of BS deep into sector II draws the upper limit of the stability 
region displayed on Fig.\,\ref{fig:phibsections} (in gray) on the segment $\omega_b\in (3,4)$.

\begin{figure}[h!]
\begin{center}
\includegraphics[scale=1]{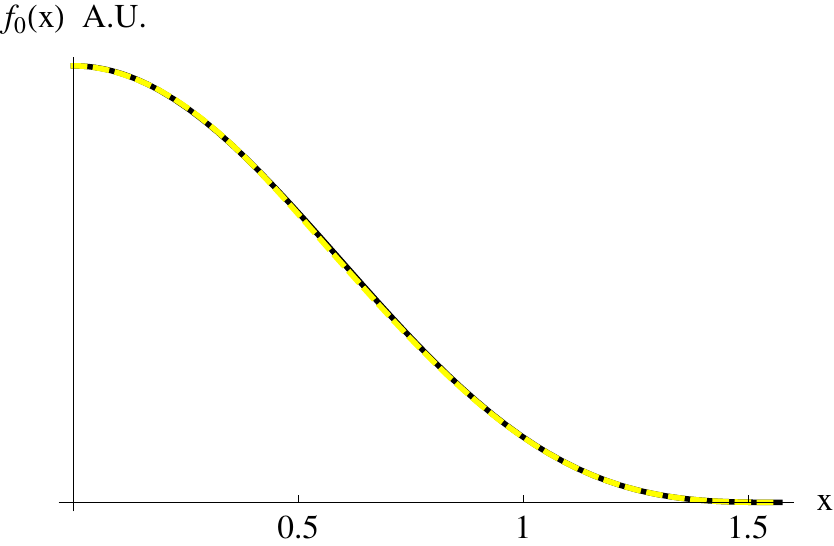}
\caption{\small Zero-mode of a Boson Star compared with the ``profile increment'' of the scalar field along the Boson Stars branch. 
The plots are rescaled in order to overlap.}
\label{BS0}
\end{center}
\end{figure}

Figure \ref{BS} shows also that the lowest normal eigenmode in the fluctuation spectrum of BS's is a zero mode (in blue). 
This  eigenmode corresponds to the zero-frequency deformation 
generating the BS branch itself. To verify this statement, one can compare the profile of the zero-frequency eigenfunction with 
the increment of the scalar field profile along the branch (see Fig.\,\ref{BS0}).

\begin{figure}[h!]
\begin{center}
\includegraphics[scale=.35]{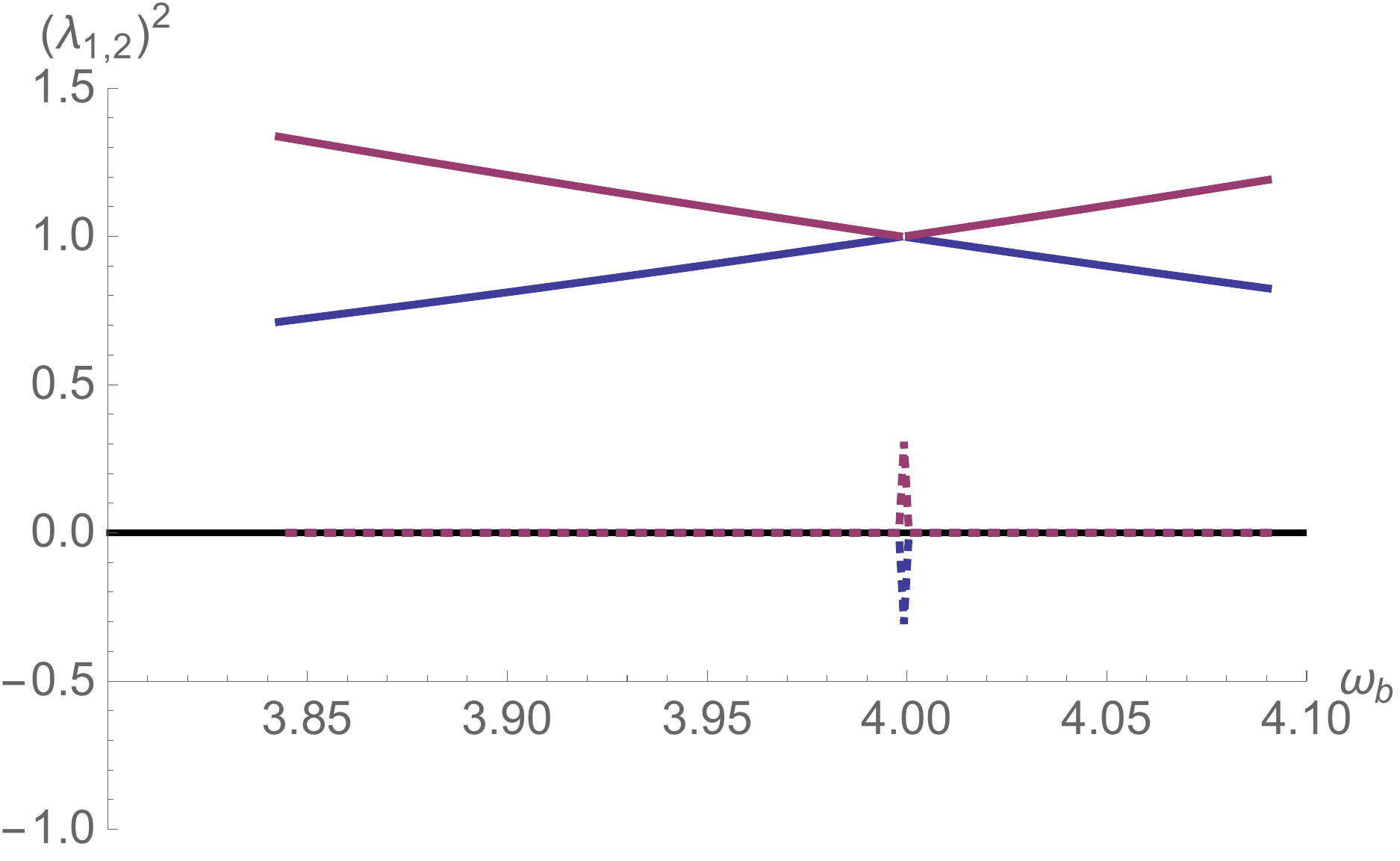}
\includegraphics[scale=.35]{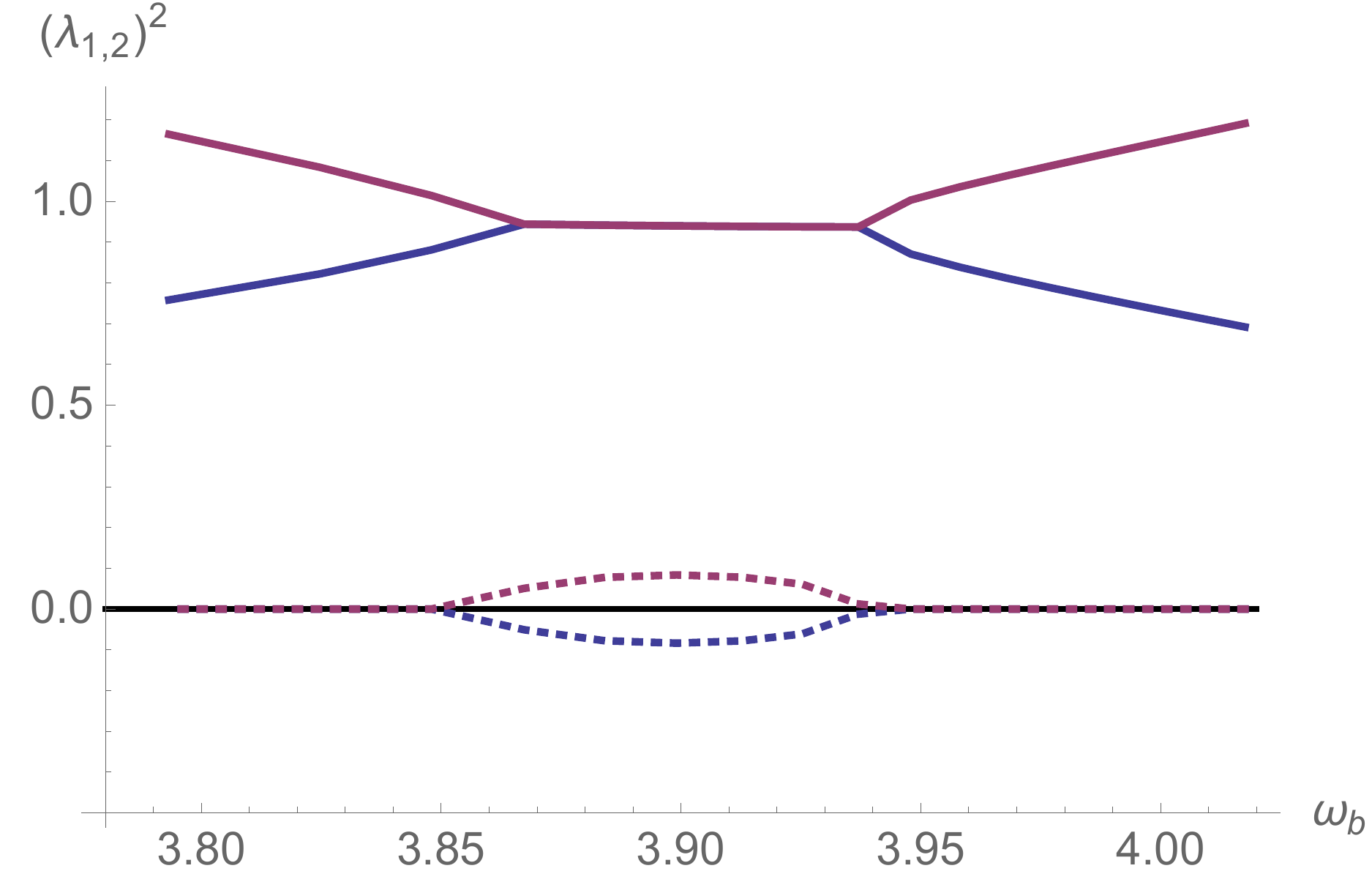}
\caption{\small First two normal modes squared developing an imaginary part (dashed) upon fusing.
The plots correspond to $\phi_b=-.001$ (left) and $\phi_b=-.01$ (right). In the left plot 
the imaginary parts have been rescaled by a factor 500 to be visible.}
\label{bubble}
\end{center}
\end{figure}

The stability diagram displayed in Fig.\,\ref{fig:phibsections} presents a peculiarity at $\omega_b=4$ and very small $\phi_o$: 
the solutions are unstable, even though surrounded by stable regions.
Such qualitative behavior is repeated for $\omega_b = 2n$ with $n\geq 2$ but not for $\omega_b=2$.
A direct look to the modes shows that the instability here is again triggered by the fusion of the first two modes
developing opposite imaginary parts (see Fig. \ref{bubble}).

It should be noted that the ``encounter'' of two normal modes does not lead always to a fusion and an instability.
For instance, at $\omega_b\sim 2$ for the SPS's with very small source the modes do not fuse
and do not spoil linear stability.

\begin{figure}[h!]
\begin{center}
\includegraphics[scale=.35]{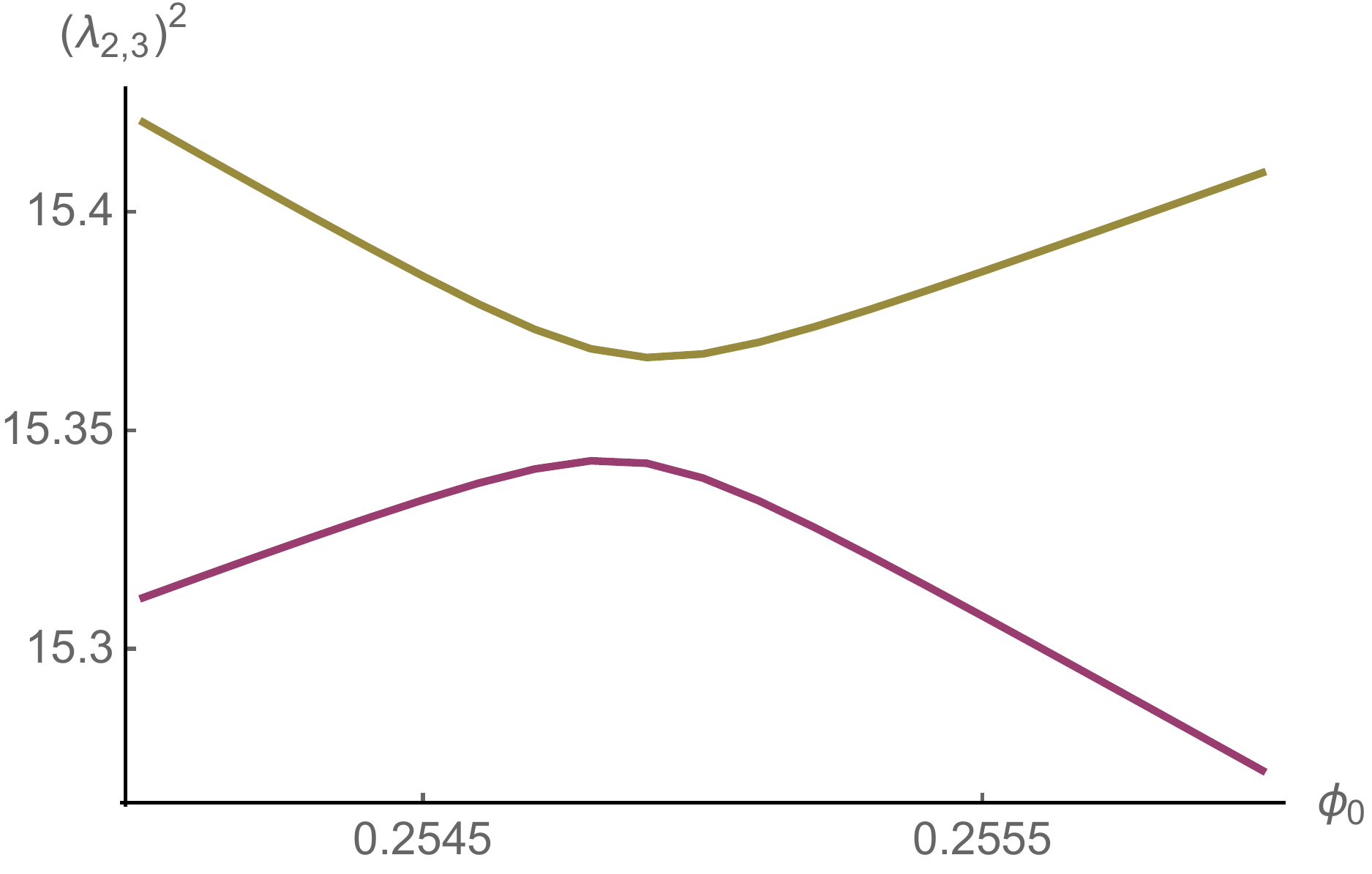}
\includegraphics[scale=.35]{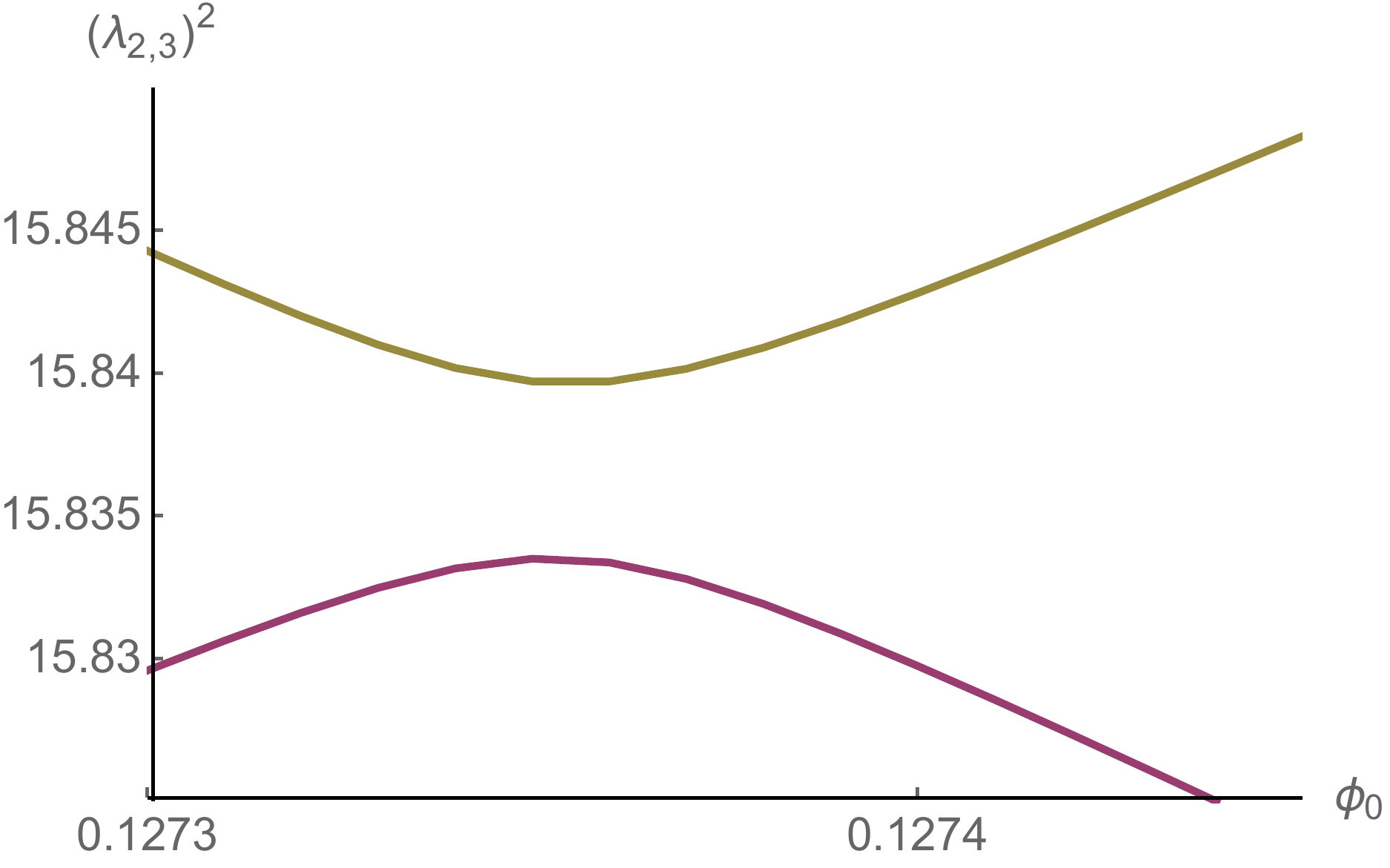}
\caption{Second and third modes repelling at $\phi_b=.2$ (left) and $\phi_b=.1$ (right).}
\label{rep}
\end{center}
\end{figure}

When two modes encounter they can either cross or repel. Deciding which is the case could be however
numerically demanding, especially at small values of the source $\phi_b$. Let us show this by means 
of an explicit example. Consider two SPS's corresponding to $\phi_b=0.2$ and $\phi_b=0.1$ respectively.
In a region where $\omega_b \sim 1$ the second and third modes approach and repel each other.
The two modes however get closer as the source is lowered. Specifically, the minimal 
distance between two modes decreases more than linearly with the source, as illustrated in Fig.\,\ref{rep}.

\begin{figure}[h!]
\begin{center}
\includegraphics[scale=.4]{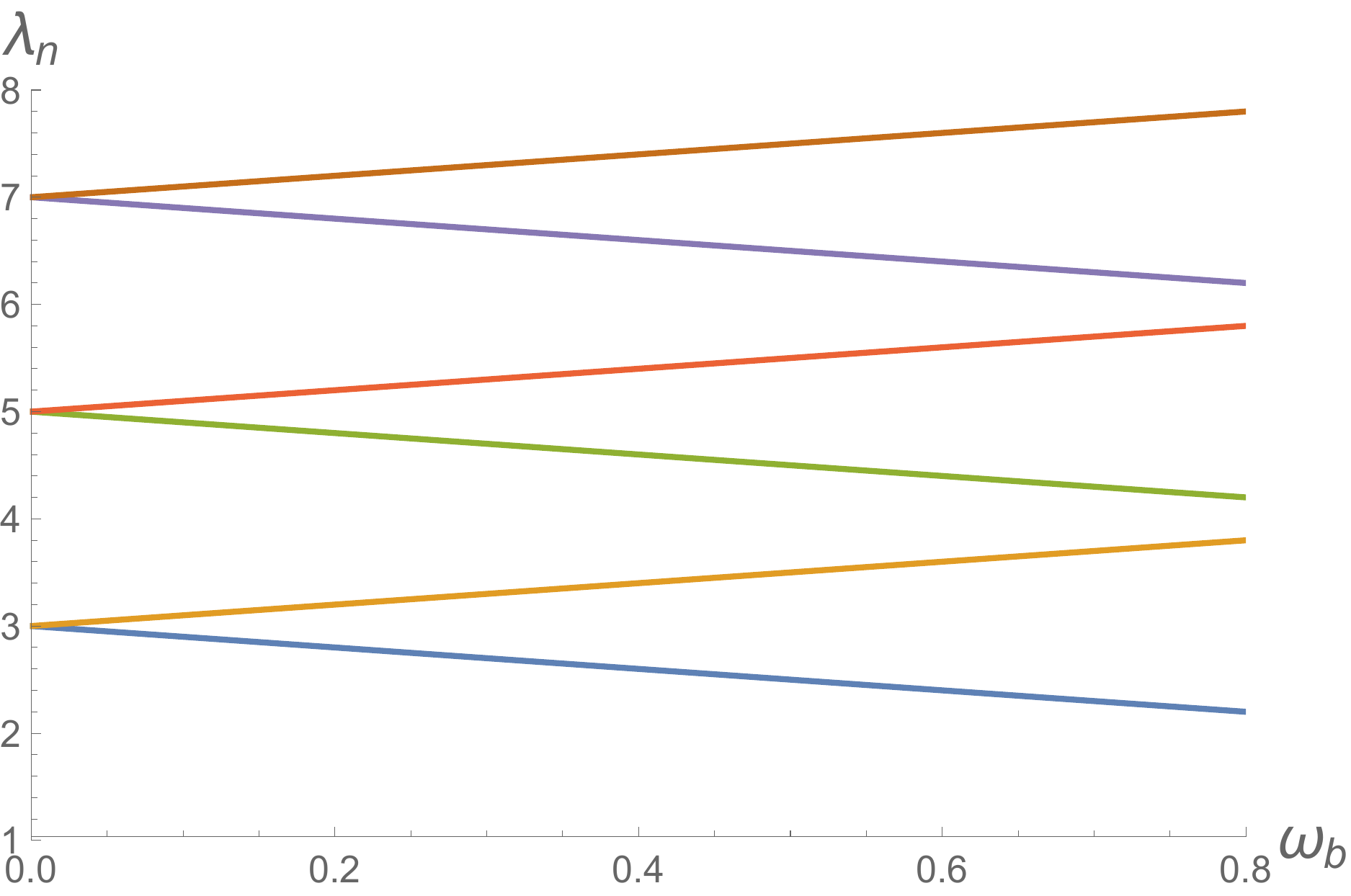}
\caption{\small Normal frequencies plotted against the boundary frequency of the source whose amplitude is $\phi_b=.001$.}
\label{lin}
\end{center}
\end{figure}

The repulsion phenomenon departs from the perturbative expectation about the mode behavior.
In fact a perturbative analysis at $\omega_b \ll 1$ suggests that the eigenfrequencies $\lambda_n$ behave as 
$\lambda_n = \lambda_n^*\pm \omega_b$, where $\lambda_n^*$ are the eigenfrequencies of AdS.
This behavior is maintained also at higher values of $\omega_b$ as the numerics shows, see Fig.\,\ref{lin}.
Nevertheless, the linear behavior of the modes has to break down as two modes approach in order to be compatible 
with a repulsion.

\section{Real periodic solutions in AdS$_4$.}
\label{realmethods}

The action is
\begin{equation}
S= \frac{1}{2\kappa^2} \int d^{4}x \sqrt{-g}\left( R - 2\Lambda\right)   -\int d^{4}x\sqrt{-g}\left(  \frac{1}{2} \partial_\mu\phi  \partial^\mu\phi  + V(\phi)\right) 
\end{equation}
where we will take as before $\kappa^2 = 8\pi G=1$,  $\Lambda = -6/l^2$. The isotropic ansatz for the metric of $AdS_{d+1}$ can be expressed as follows

\be
ds^2 = \frac{l^2}{\cos^2 x}\left( - f e^{-2\delta} dt^2+ f^{-1} dx^2 + \sin^2 x \, d \Omega_{d-1}^2\right) . \label{line1}
\ee

We are interested in time-periodic solutions with harmonic boundary conditions such that $\phi(t,\pi/2) = \rho_b \cos(\omega_b t)$. Continuing with the same notation as in the complex case we will use $\rho_o = \phi(0,0)$ and $\rho_b = \phi(0,\pi/2)$. The effective system of equations is conveniently expressed in terms of the following variables 
\begin{eqnarray}
F &=& f e^{-\delta} \nonumber\\
\Phi &\equiv& \phi'(t,x)   \nonumber\\
\Pi &\equiv&  \dot\phi /F  + \rho_b\omega_b \sin \omega_b t.  \nonumber
\end{eqnarray}
Working in the boundary gauge, we have that $\Pi(t,\pi/2)=0$, while $\Phi(t,\pi/2)=0$ due to the asymptotic near-boundary expansion. 
With these definitions, the equations of motion become 
\begin{eqnarray}
\dot{\Phi} &=& (F \Pi)' -  \rho_b\omega_b\sin \omega_b t \,  F'   \label{eqPhireal}\\
\dot{\Pi} &=& \frac{1}{\tan^{2} x}(\tan^{2} x F\Phi)' + \rho_b\omega_b^2 \cos \omega_b t \label{eqPireal}\\ 
\delta' &=&-\displaystyle   \sin x \cos x \left(  \Phi^2 + \Pi ^2  - 2 \rho_b\omega_b \sin \omega_b t \, \Pi+ \rho_b^2\omega_b^2 \sin^2 \omega_b t \right) \label{edeltareal}\\
F ' &=&  \frac{1+2\sin x}{\sin x\cos x} ( e^{-\delta} - F)  \label{eqFreal}
\end{eqnarray}

The periodicity of the solution \eqref{periodicansatz} instructs us to consider the most general Fourier series expansion. 
However, the structure of the equations of motion allows for a truncation to odd modes, which correspond to bifurcations emanating from single normal modes of AdS, 
\begin{equation}
{\Phi}=\sum_{k=0}^{\infty} {\Phi}_{k} (x)\cos[(2k+1)\omega_b t ] \qquad {\Pi}=\sum_{k=0}^{\infty}  {\Pi}_{k}(x) \sin[  (2k+1)\omega_b t  ].
\label{eq:periodic_field_time_structure}
\end{equation}
While $\delta$ and $F$ are obtained from $\Phi$ and $\Pi$ through \eqref{edeltareal} and \eqref{eqFreal}. 
Following the methods introduced in \cite{Maliborski:2016zlh}, we rescale time as $\tau = \omega_b t$ and use Chebyshev polynomials to expand the spatial dependence.
The collocation grids are associated to these two basis of functions, Fourier and Chebyshev, 
\begin{eqnarray}
\tau_n &=&  \frac{\pi}{2}\frac{2n-1}{2K+1}   ~~~ ~~~~~~~~~~~~ \hbox{with} ~~~ n=0,1,...,K-1 \nonumber \\
x_i &=& \frac{\pi}{2} \cos[\pi i /(2N+1)]  ~~~ \hbox{with} ~~~ i = 0,1,...,N\, .
\label{eq:apx_real_collocation_points}
\end{eqnarray}
Let us define $f_{ki} \equiv f_k(x_i)$.  Then, the discretized values are
\begin{equation}
{\Phi(\tau_n,x_i)}=\sum_{k=0}^{\infty} {\Phi}_{ki} \cos[(2k+1)\tau_n ] \qquad {\Pi (\tau_n,x_i)}=\sum_{k=0}^{\infty}  {\Pi}_{ki} \sin[  (2k+1)\tau_n  ].
\label{eq:periodic_field_time_structure_discr}
\end{equation}

Using the boundary conditions discussed previously ($\Phi(t,\pi/2) = \Pi(t,\pi/2) = 0$), the values of the fields at $x_0$ are restricted to $\Phi_{k0} = \Pi_{k0} = 0$. 
Hence, the unknowns to be fixed are
$\omega_b, \rho_b, \Phi_{ki}$ and $\Pi_{ki}$ for $k=0,...,K-1$ and $i=1,...,N$, while the equations are \eqref{eqPhireal} and \eqref{eqPireal} evaluated at each of 
these collocation points $(\tau_n,x_i)$. This gives $2KN$ equations for $2KN+2$ unknowns. 
We fix a numerical value for one of them, being it either $\omega_b$ or $\rho_b$, and add another equation of motion that sets the value of 
$\rho_o$. Finally, the number of equations and variables match, $2KN+1$, and the discretized system yields an algebraic nonlinear system of 
equations that can be solved using a Newton-Raphson algorithm. 

After obtaining time-periodic solutions we are interested in their linear stability. Consider linearized fluctuations of 
the form\footnote{hereafter we will find it convenient to work with the $\phi$ rather than $\Phi$} 
\begin{eqnarray}
\phi(t,x)  &=& \phi_p(t,x)  + \tilde\phi(t,x), \nonumber\\
\Pi(t,x)  &=& \Pi_p(t,x)  + \tilde\Pi(t,x), \nonumber\\
\delta(t,x) &=& \delta_p(t,x) + \tilde \delta(t,x), \nonumber\\
F(t,x) &=& F_p(t,x) + \tilde F(t,x), \nonumber
\end{eqnarray}
where the subindex $p$ stands for ``periodic solution" and fields with a tilde are the perturbations. 
Inserting these ansatz into the equations of motion, and at first order in the amplitude, we obtain the equations for the perturbations
\begin{eqnarray}
\tilde{\delta}' & = & -2\sin{x}\cos{x}\left(\Phi_{p}\partial_{x}\tilde{\phi} + \left(\Pi_{p}-\rho_b \omega_b \sin{\omega_b t}\right)\tilde{\Pi}\right)
\label{eq:apx_real_delta_pert}, \\
\tilde{F}' & = & -\frac{1+2\sin^{2}x}{\sin{x}\cos{x}}\left(\tilde{\delta} e^{-\delta_{p}} + \tilde{F}\right)
\label{eq:apx_real_F_pert}, \\
\dot{\tilde{\phi}} & = & F_p \tilde{\Pi} + \left(\Pi_{p}-\rho_b\omega_b\sin{\omega_b t}\right)\tilde{F}
\label{eq:apx_real_phi_pert},\\
\dot{\tilde{\Pi}} & = & \frac{1}{\tan^{2}x}\partial_{x}\left[\tan^{2}x \left(F_{p}\partial_{x}\tilde{\phi} + \Phi_{p}\tilde{F}\right)\right]
\label{eq:apx_real_Pi_pert}.
\end{eqnarray}
$\tilde{\delta}$ and $\tilde{F}$ can be seen as linear operators acting on $\tilde{\phi}$ and $\tilde{\Pi}$, with this point of view, equations (\ref{eq:apx_real_phi_pert}) and (\ref{eq:apx_real_Pi_pert}) are expressed in the following form 
\begin{equation}
\left( \begin{array}{c}\dot{\tilde\phi} \\ \dot{\tilde \Pi} \end{array} \right) = {\bf L}(t) \left( \begin{array}{cc} \tilde\phi \\ \tilde \Pi \end{array} \right), \, 
\label{eq:apx_real_L_system}
\end{equation}
where ${\bf L}(t)$ is a linear integro-differential operator (in $x$) constructed with the periodic fields. 
The periodicity of the background is inherited by the operator, ${\bf L}(t) = {\bf L}(t+T)$, 
$T$ being the period of $\phi_{p}$. For ${\bf x}(t)  = (\tilde\phi,\tilde\Pi)$, the Floquet theorem 
establishes the existence of a solution of the form
\begin{equation}
{\bf x} (t) = e^{\lambda t} {\bf P}(t)~~~~\hbox{with} ~~~~{\bf P}(t) = {\bf P}(t+T)\, .
\end{equation}
The structure of (\ref{eq:apx_real_delta_pert})-(\ref{eq:apx_real_Pi_pert}) allows a truncated version 
of the Fourier expansion for ${\bf P}(t)$ in odd modes which rule the transitions of stability that we have found. 
It is convenient to express ${\bf P}(t)$ in two parts
\begin{equation}
{\bf P}(t,x) = {\bf p}^{(1)}(t,x) + {\bf p}^{(2)}(t,x) = \left( \begin{array}{c}\tilde\phi^{(1)} \\ \tilde \Pi^{(1)} \end{array} \right) + \left( \begin{array}{c}\tilde\phi^{(2)} \\ \tilde \Pi^{(2)} \end{array} \right), 
\label{eq:P_general_Fourier_2}
\end{equation}
where the time dependence is distributed as follows 
\begin{eqnarray}
\tilde{\phi}^{(1)}  =  \sum_{k=0}^{\infty} \tilde{\phi}_{k}^{(1)}(x)\cos\left[(2k+1)\omega_b t\right], & & \tilde{\Pi}^{(1)} =  \sum_{k=0}^{\infty} \tilde{\Pi}_{k}^{(1)}(x)\sin\left[(2k+1)\omega_b t\right], 
\label{eq:eq_1_time_ansatz} \\
\tilde{\phi}^{(2)}  =  \sum_{k=0}^{\infty} \tilde{\phi}_{k}^{(2)}(x)\sin\left[(2k+1)\omega_b t\right], & &
\tilde{\Pi}^{(2)}  =  \sum_{k=0}^{\infty} \tilde{\Pi}_{k}^{(2)}(x)\cos\left[(2k+1)\omega_b t\right]. \label{eq:eq_4_time_ansatz}
\end{eqnarray}
Inserting this ansatz into (\ref{eq:eq_1_time_ansatz})-(\ref{eq:eq_4_time_ansatz}) and taking advantage of the linear independence of 
the trigonometric functions, equations 
split in the following form
\begin{eqnarray}
\lambda \tilde{\phi}^{(2)} & = & \tilde{F}^{(1)}\left(\Pi_{p}-\rho_b\omega\sin{\omega t}\right)+F_{p}\tilde{\Pi}^{(1)} - \dot{\tilde{\phi}}^{(1)} \label{eq:eq_1_system_eq}, \\
\lambda \tilde{\phi}^{(1)} & = & \tilde{F}^{(2)}\left(\Pi_{p}-\rho_b\omega\sin{\omega t}\right)+F_{p}\tilde{\Pi}^{(2)} - \dot{\tilde{\phi}}^{(2)}, \\
\lambda \tilde{\Pi}^{(2)} & = & \frac{1}{\tan^{2}x}\left[\tan^{2}x\tilde{F}^{(1)}\Phi_{p}+\tan^2x F_{p}\left(\tilde{\phi}^{(1)}\right)'\right]' - \dot{\tilde{\Pi}}^{(1)}, \\
\lambda \tilde{\Pi}^{(1)} & = & \frac{1}{\tan^{2}x}\left[\tan^{2}x\tilde{F}^{(2)}\Phi_{p}+\tan^2x F_{p}\left(\tilde{\phi}^{(2)}\right)'\right]' - \dot{\tilde{\Pi}}^{(2)}, 
\label{eq:eq_4_system_eq}
\end{eqnarray}
where we have defined $\tilde{F}^{(i)} \equiv \tilde{F}(\tilde{\phi}^{(i)},\tilde{\Pi}^{(i)})$ and used the time structure 
of the periodic fields (\ref{eq:periodic_field_time_structure}). 
This system of equations has the property that given a solution $\left(\lambda, {\bf p}^{(1)}, {\bf p}^{(2)}\right)$ another solution 
exists with $\left(-\lambda, {\bf p}^{(1)}, -{\bf p}^{(2)}\right)$, implying that an unstable mode exists if $Re(\lambda)\neq 0$. 
On the other hand, if all $\lambda \in i\mathbb{R}$, the periodic solution is linearly stable.  

Solving \eqref{eq:eq_1_system_eq} through \eqref{eq:eq_4_system_eq} is very similar to obtaining the time-periodic solutions $\phi_p$ 
and $\Pi_p$ themselves. So the same techniques explained in the previous section are in order here. The differences lie in the fact that, 
in this case, the ansatz (\ref{eq:P_general_Fourier_2}) has two more unknown functions (${\bf p}^{(2)}$) and the role 
of $\omega$ is played by $\lambda$. Namely, we discretize the problem as in \eqref{eq:apx_real_collocation_points} and set 
boundary conditions so as to study perturbations which don't modify the source ($\tilde{\phi}^{(i)}(t,\pi/2) = \tilde{\Pi}^{(i)}(t,\pi/2) =  0$). 
Finally, by adding an additional equation to fix the amplitude of $\tilde{\phi}^{(1)}(0,0) = 1$, we obtain a system of $4NK+1$ nonlinear equations 
for our $4NK+1$ unknowns, which can be solved using a Newton-Raphson algorithm.

\end{appendix}



\end{document}